{}
{}

\def\verdraft{3}
\def\verpublish{4}
\def\verarxiv{5}

\def\version{5}

\documentclass[12pt,a4paper]{article}

\usepackage{amsmath,amssymb}
\ifnum\version<2\usepackage{showkeys}\fi
\usepackage{graphicx}
\usepackage[bookmarks=true,hyperfigures=true]{hyperref}
\usepackage[usenames,dvipsnames]{color}
\usepackage[nosort]{cite}
\usepackage[bulletsep]{collref}
\usepackage[utf8]{inputenc}

\def\showkeyslabelformat#1{{\usefont\encodingdefault\familydefault%
 \seriesdefault\shapedefault\normalfont\tiny\ttfamily#1}}
\def\showkeysrefformat#1{{\normalfont\tiny\ttfamily#1}}
\makeatletter
\def\SK@@ref#1>#2\SK@{%
 {\@inlabelfalse\leavevmode\vbox to\z@{%
 \vss\SK@refcolor\rlap{\vrule\raise .75em%
  \hbox{\showkeysrefformat{#2}}}}}}
\makeatother

\usepackage[a4paper,text={160mm,247mm},centering]{geometry}

\RequirePackage{verbatim}

\makeatletter
\ifnum\version<\verarxiv\relax
\RequirePackage{shellesc}
\ifnum\ShellEscapeStatus>0\relax
\AddToHook{enddocument/afteraux}{\ShellEscape{bibtex \jobname}}
\fi
\newwrite\bibinl@out
\newenvironment{bibtex}[1][\jobname]{%
  \immediate\openout\bibinl@out #1.bib
  \immediate\write\bibinl@out{\@percentchar generated from `\jobname' starting line \the\inputlineno^^J}%
  \def\verbatim@processline{\immediate\write\bibinl@out{\the\verbatim@line}}%
  \@bsphack\let\do\@makeother\dospecials\catcode`\^^M\active\verbatim@start
}%
{\immediate\closeout\bibinl@out\@esphack}
\else
\newenvironment{bibtex}[1][\jobname]{\comment}{\endcomment}
\fi
\makeatother

\counterwithin{equation}{section}

\usepackage[font=small,labelfont=bf,width=0.85\textwidth]{caption}

\usepackage{eqnlines}
\eqnlinesset{number=on}
\eqnlinesset{numberline=center}
\eqnlinesset{punctall={,,.},punctcases={,}}

\ifnum\version=11
\usepackage[bitstream-charter]{mathdesign}
\fi

\usepackage{mathfixs}
\ProvideMathFix{autobold,greekcaps,frac,root}
\ProvideMathFix{multskip}
\ProvideMathFix{numsets,Order,res,const}

\ProvideMathFix{der,diff,hyp,tr,diag,reim,mapclass,span,defeq,eqdef}

\RequirePackage[extdef]{delimset}
\newcommand{\spn}{\brk[a]}
\newcommand{\polyspc}{\brk[s]}
\newcommand{\pwrser}{\brk[S]}
\renewcommand{\setcond}{\pair[c;]}

\ProvideMathFix{vfrac,rfrac,iunit*,econst}
\newcommand{\half}{\rfrac{1}{2}}
\newcommand{\ihalf}{\rfrac{\iunit}{2}}
\newcommand{\quarter}{\rfrac{1}{4}}

\newcommand{\alg}[1]{\mathfrak{#1}}
\newcommand{\grp}[1]{\mathrm{#1}}
\newcommand{\gen}[1]{\mathrm{#1}{}}

\newcommand{\mspalg}{\alg{h}}
\newcommand{\algdbl}{\alg{D}}

\newcommand{\genJ}{\gen{J}}
\newcommand{\genJL}{\genJ_\algleft}
\newcommand{\genJR}{\genJ_\algright}
\newcommand{\genL}{\gen{L}}

\newcommand{\genP}{\gen{P}}
\newcommand{\genQ}{\gen{Q}}
\newcommand{\genD}{\gen{D}}
\newcommand{\genC}{\gen{C}}
\newcommand{\genOD}{\genD^0}
\newcommand{\genOC}{\genC^0}
\newcommand{\genODL}[1][]{\genOD_{\genL\if?#1?\else,#1\fi}}
\newcommand{\genODP}[1][]{\genOD_{\genP\if?#1?\else,#1\fi}}
\newcommand{\genOCL}[1][]{\genOC_{\genL\if?#1?\else,#1\fi}}
\newcommand{\genOCP}[1][]{\genOC_{\genP\if?#1?\else,#1\fi}}
\newcommand{\genZD}{\genD}
\newcommand{\genZC}{\genC}
\newcommand{\genZDL}{\genZD_{\genL}}
\newcommand{\genZDP}{\genZD_{\genP}}
\newcommand{\genZCL}[1][]{\genZC_{\genL\if?#1?\else,#1\fi}}
\newcommand{\genZCP}[1][]{\genZC_{\genP\if?#1?\else,#1\fi}}
\newcommand{\genRDL}[1][]{\genD_{\genL\if?#1?\else,#1\fi}}
\newcommand{\genRDP}[1][]{\genD_{\genP\if?#1?\else,#1\fi}}
\newcommand{\genRCL}[1][]{\genC_{\genL\if?#1?\else,#1\fi}}
\newcommand{\genRCP}[1][]{\genC_{\genP\if?#1?\else,#1\fi}}

\newcommand{\genJtw}{\hat{\genJ}{}}
\newcommand{\genLtw}{\hat{\genL}{}}
\newcommand{\genPtw}{\hat{\genP}{}}
\newcommand{\genZDtw}{\hat{\genD}{}}
\newcommand{\genZDLtw}[1][]{\genZDtw_{\genL\if?#1?\else,#1\fi}}
\newcommand{\casJJtw}{\genJtw^2}

\newcommand{\genRLo}{\genL}
\newcommand{\genRPo}{\genP}

\newcommand{\genRLt}{\tilde{\genL}}
\newcommand{\genRLom}{\genRLo{}'}
\newcommand{\genRLtm}{\genRLt{}'}
\newcommand{\genRPt}{\tilde{\genP}{}}
\newcommand{\genRQt}{\tilde{\genQ}{}}
\newcommand{\genRDLt}[1][]{\tilde{\genD}_{\genL\if?#1?\else,#1\fi}}
\newcommand{\genRDPt}[1][]{\tilde{\genD}_{\genP\if?#1?\else,#1\fi}}
\newcommand{\cclabel}[1]{[#1]}

\newcommand{\shiftpar}{{\hlcolor{red}u}}
\newcommand{\looppar}{{\hlcolor{red}\zeta}}
\newcommand{\zpar}{{\hlcolor{red}z}}

\newcommand{\cybe}{\pair[S,]}
\newcommand{\liebr}{\comm}
\newcommand{\cobra}{{\hlcolor{red}\delta}}
\newcommand{\biform}{\pair[r,]}
\newcommand{\knpair}{\pair[a,]}
\newcommand{\rmat}{{\hlcolor{red}r}}

\newcommand{\rmatmsp}{\rmat_{\mspalg}}
\newcommand{\rmatred}{\tilde{\rmat}}

\newcommand{\ad}{\mathrm{ad}}

\newcommand{\casJJ}{\genJ^2}

\newcommand{\casPP}{\genP^2}
\newcommand{\casLP}{\genL\mathord{\cdot}\genP}
\newcommand{\casPL}{\genP\mathord{\cdot}\genL}
\newcommand{\casQQ}{\genQ^2}

\newcommand{\slstruct}{{\hlcolor{red}f}}
\newcommand{\slinv}{{\hlcolor{red}c}}
\newcommand{\slpauli}{{\hlcolor{red}\tilde\sigma}}
\newcommand{\sleps}{{\hlcolor{red}\varepsilon}}
\newcommand{\spinup}{{\uparrow}}
\newcommand{\spindown}{{\downarrow}}
\newcommand{\spinupdown}{{\updownarrow}}
\newcommand{\algleft}{\mathrm{A}}
\newcommand{\algright}{\mathrm{B}}

\newcommand{\mref}{{\hlcolor{red}\bar m}}
\newcommand{\rnorm}{{\hlcolor{red}\nu}}
\newcommand{\rnormL}{\rnorm_{\genL}}
\newcommand{\rnormP}{\rnorm_{\genP}}
\newcommand{\rnormRL}{\tilde{\rnorm}_{\genL}}
\newcommand{\rnormRP}{\tilde{\rnorm}_{\genP}}
\newcommand{\rtwist}{{\hlcolor{red}\xi}}
\newcommand{\rtwistL}{\rtwist_{\genL}}
\newcommand{\rtwistP}{\rtwist_{\genP}}
\newcommand{\redang}{{\hlcolor{red}\alpha}}
\newcommand{\redpar}{{\hlcolor{red}\beta}}

\newcommand{\reddir}{{\hlcolor{red}{V}}}
\newcommand{\redrot}{{\hlcolor{red}{W}}}

\newcommand{\trigq}{{\hlcolor{red}{\eta}}}

\newcommand{\minusg}{\mathbin{\rightarrow}}
\newcommand{\minusl}{\mathbin{\leftarrow}}
\newcommand{\minuslg}{\mathbin{\leftrightarrow}}
\newcommand{\minush}{\mathbin{\mapsto}}

\newcommand{\KNalg}{\mathcal{F}}
\newcommand{\KNfuns}{\mathcal{A}}
\newcommand{\KNders}{\mathcal{L}}
\newcommand{\KNbas}[3]{f^{#1}_{#2,#3}}
\newcommand{\KNAbas}[2]{A_{#1,\cclabel{#2}}}
\newcommand{\RSurf}{\mathrm{\Sigma}}
\newcommand{\ord}{\operatorname{ord}}

\providecommand{\href}[2]{#2}

\makeatletter
\def\mr@ignsp#1 {\ifx\:#1\@empty\else #1\expandafter\mr@ignsp\fi}%
\newcommand{\multiref}[1]{\begingroup
\xdef\mr@no@sparg{\expandafter\mr@ignsp#1 \: }%
\def\mr@comma{}%
\@for\mr@refs:=\mr@no@sparg\do{\mr@comma\def\mr@comma{,}\ref{\mr@refs}}%
\endgroup}
\renewcommand{\eqref}[1]{(\multiref{#1})}
\makeatother

\makeatletter
\newcommand{\namedref}[2]{\hyperref[#2]{#1~\ref*{#2}}}
\newcommand{\secref}{\@ifstar{\namedref{Section}}{\namedref{Sec.}}}
\newcommand{\appref}{\@ifstar{\namedref{Appendix}}{\namedref{App.}}}
\newcommand{\tabref}{\@ifstar{\namedref{Table}}{\namedref{Tab.}}}
\newcommand{\figref}{\@ifstar{\namedref{Figure}}{\namedref{Fig.}}}
\makeatother

\edef\thebibliography{\unexpanded
  {\phantomsection\addcontentsline{toc}{section}{\refname}}
  \unexpanded\expandafter{\thebibliography}}
\edef\tableofcontents{\unexpanded
  {\phantomsection\addcontentsline{toc}{section}{\contentsname}}
  \unexpanded\expandafter{\tableofcontents}}

\providecommand{\hypersetup}[1]{}
\providecommand{\texorpdfstring}[2]{#1}

\hypersetup{plainpages=false}
\hypersetup{pdfpagemode=UseNone}
\hypersetup{bookmarksnumbered=true}
\hypersetup{pdfstartview=FitH}
\hypersetup{colorlinks=false}
\hypersetup{citebordercolor={.5 1 .5}}
\hypersetup{urlbordercolor={.5 1 1}}
\hypersetup{linkbordercolor={1 .7 .7}}

\ifnum\version<\verpublish\PassOptionsToPackage{draft}{metastr}\fi
\RequirePackage{metastr}

\ifnum\version<\verdraft\metasetterm{draft}{EDITING}\fi

\usepackage{ifpdf}
\ifpdf\else\RequirePackage[active]{srcltx}\fi
\RequirePackage{color}

\newcommand{\remark}[2][]{}

\newcommand{\hlcolor}{\color}
\ifnum\version<\verpublish\else
\renewcommand{\hlcolor}[2][]{}
\fi

\ifnum\version<\verpublish
\renewcommand{\remark}[2][]{\ifmmode\hbox\fi{%
  \def\emph{\textsl}\def\textbullet{$\bullet$}%
  \def\tmparga{#1}\def\tmpargb{draft}\ifx\tmparga\tmpargb%
  \color[rgb]{0.5,0,0}\normalfont\sffamily\hspace{1ex} #2\hspace{1ex}\fi}}
\fi

\ifnum\version<\verdraft
\renewcommand{\remark}[2][]{\ifmmode\hbox\fi{\normalfont\sffamily\hspace{1ex}%
  \def\emph{\textsl}\def\textbullet{$\bullet$}%
  \def\tmparga{#1}%
  \def\tmpargb{NB}\ifx\tmparga\tmpargb\color[rgb]{0,0,0.6}\fi%
  \def\tmpargb{EI}\ifx\tmparga\tmpargb\color[rgb]{0,0.5,0}\fi%
  \def\tmpargb{HM}\ifx\tmparga\tmpargb\color[rgb]{0,0.3,0.3}\fi%
  \def\tmpargb{}\ifx\tmparga\tmpargb\color{red}\fi%
  \def\tmpargb{}\ifx\tmparga\tmpargb\else \textbf{#1:} \fi%
  #2\hspace{1ex}}}
\fi

\metaset{title}{Multi-Point Affine Bialgebras\texorpdfstring{\\}{ }%
for AdS/CFT Integrability}
\metaset{author}{Niklas Beisert, Egor Im, Hannes Malcha}

\begin{document}

\ifnum\version>0
\pdfbookmark[1]{Title Page}{title}
\thispagestyle{empty}

\begingroup\raggedleft\footnotesize\ttfamily
\par\endgroup

\vspace*{2cm}
\begin{center}%
\begingroup\Large\bfseries\metapick[print]{title}\par\endgroup
\vspace{1cm}

\begingroup\scshape
\metapick[print]{author}
\endgroup
\vspace{5mm}

\textit{Institut für Theoretische Physik,\\
Eidgenössische Technische Hochschule Zürich,\\
Wolfgang-Pauli-Strasse 27, 8093 Zürich, Switzerland}
\vspace{0.1cm}

\begingroup\ttfamily\small
\verb+{+nbeisert,egorim\verb+}+@itp.phys.ethz.ch\par
\endgroup
\vspace{5mm}

\vfill

\metaif[]{draft}{
\vspace{5mm}
\begingroup\bfseries\Large \metaterm{draft}\endgroup
\par
\vspace{5mm}
\vfill
}{}

\textbf{Abstract}\vspace{5mm}

\begin{minipage}{12.5cm}
In this work, we revisit the analytical structure
of the classical affine Lie bialgebra underlying the
integrable structures of AdS/CFT worldsheet scattering
and the one-dimensional Hubbard model.
Earlier work showed that the algebraic relations implied
two singularity locations in the evaluation parameter plane
besides the common expansion points $\infty$ and $0$ for Laurent polynomials
forming the basis for loop and affine algebras.
Our aim is to formulate these singularities rigorously using meromorphic
functions rather than by expansion into geometric series.
\unskip\smallskip

We thus carefully analyse the analytical structure of the affine bialgebra
relations in the complex plane paying attention to potential restrictions and
extensions due to the additional singularities.
As a first attempt with some shortcomings, we reformulate the affine bialgebra
relations with the extra singularities serving as the expansion points for
Laurent polynomials.
We finally overcome the remaining issues by proposing three-point and four-point
affine algebras for the rational and trigonometric cases.
With a suitable choice of basis for meromorphic functions on the Riemann sphere
with several punctures based on Krichever--Novikov algebras,
we show that the multi-point affine algebras form consistent quasi-triangular
Lie bialgebras.
\end{minipage}

\vspace*{2cm}

\end{center}

\ifnum\version>1
\newpage
\tableofcontents
\fi

\newpage
\fi

\section{Introduction}
\label{sec:intro}

In this work, we continue the study of the classical affine Lie bialgebra
underlying integrability in the one-dimensional Hubbard model and the
factorised worldsheet scattering encountered in AdS/CFT \cite{Beisert:2006qh},
see \cite{Essler:2005aa,Beisert:2010jr} for reviews of the integrable structures
in these models.
Integrability in models of this kind is characterised by a quantum group,
which is a deformation of the universal enveloping algebra of an affine
Kac--Moody algebra \cite{Drinfel'd:1985, Drinfel'd:1988}.
In particular, these algebras retain the Hopf algebra
structure that is naturally present in universal enveloping algebras.
Furthermore, the Hopf algebra is quasi-triangular, i.e.\ it has a
universal R-matrix satisfying the Yang--Baxter equation.
When evaluated in a representation,
the universal R-matrix yields the S-matrix
for scattering on the string worldsheet.
See \cite{Shastry:1986bb,Beisert:2005tm}
for the relevant R- and S-matrices in the above models.

In the present case, the universal R-matrix is still not known
despite the many efforts that have been devoted towards its construction.
On the one hand, the difficulty of this task is comprehensible
given that the relevant representations of the R-matrix
have involved combinatorial and analytical structures
with several unconventional features for integrable systems
such as an R-matrix that deviates from the standard difference form.
On the other hand, it would be very desirable to complete the task,
as the full universal R-matrix is expected to
expose the full structure of the overall phase factor
\cite{Janik:2006dc,Beisert:2006ez,Beisert:2006ib,Dorey:2007xn}.

Towards reaching this goal, the classical limit proves useful:
In the classical limit, the quantum algebra reduces to
a quasi-triangular classical Lie bialgebra.
In the particular context, the classical limit was introduced and formulated in
\cite{Klose:2006zd,Torrielli:2007mc,Moriyama:2007jt,Beisert:2007ty}.
The underlying algebra turned out to be
a special deformation of the $\alg{u}(2|2)$ loop superalgebra.
In \cite{Beisert:2022vnc} the Lie bialgebra of the deformed $\alg{u}(2|2)$ loop
superalgebra was obtained from the bialgebra of the semi-simple loop
superalgebra $\alg{d}(2,1;\epsilon) \times \alg{sl}(2)$ in a novel contraction
and reduction procedure.
In the contraction, the algebra $\alg{d}(2,1;\epsilon) \times \alg{sl}(2)$
is evaluated for a particular $\epsilon$-dependent parametrisation
of the generators.
In the limit $\epsilon \to 0$, the resulting algebra
is a 3D Poincaré superalgebra
$\alg{sl}(2)\ltimes\alg{psu}(2|2)\ltimes\Real^{2,1}$.
The subsequent reduction consists in restricting to a one-dimensional
subalgebra of the Lorentz algebra $\alg{sl}(2)$ and dividing out a
two-dimensional ideal of the three momentum generators $\Real^{2,1}$,
leading to the appropriate space of the algebra $\alg{u}(2|2)$.

The contraction and reduction procedure is a remarkable result for several
reasons.
Most importantly, it is compatible with the quasi-triangular Lie bialgebra
structure.
In other words, it gives rise to the classical r-matrix
for the deformed $\alg{u}(2|2)$ loop superalgebra
via contraction and reduction of the classical r-matrix
for the $\alg{d}(2,1;\epsilon) \times \alg{sl}(2)$ loop superalgebra.
Also, the representation theory of the deformed $\alg{u}(2|2)$ loop superalgebra
emerges from the well-known representation theory of $\alg{sl}(2)$
combined with well-understood representations of $\alg{d}(2,1;\epsilon)$.
Finally, the contraction and reduction procedure opens a path to obtaining
the quantum group of the deformed $\alg{u}(2|2)$ loop superalgebra by lifting
the procedure to the quantum level.
The universal R-matrix of the quantum group of the deformed $\alg{u}(2|2)$
loop superalgebra would then follow from the universal R-matrix
of $\alg{d}(2,1;\epsilon) \times \alg{sl}(2)$ via contraction and reduction.
While the universal R-matrix of $\alg{sl}(2)$
is already well known \cite{Drinfel'd:1985, Drinfel'd:1988},
the universal R-matrix of $\alg{d}(2,1;\epsilon)$
follows the standard construction procedures for simple algebras and
superalgebras
\cite{1995q.alg.....6005E,1997q.alg.....1038E,1996q.alg....10030E}.

Before addressing the elaborate task of quantisation,
it makes good sense to consider the affinisation
of the loop Lie bialgebra of \cite{Beisert:2022vnc}.
The affine Lie bialgebra for the one-dimensional Hubbard model
and for AdS/CFT worldsheet scattering has been discussed in
\cite{Beisert:2010kk,Beisert:2024rke}.
Besides the loop algebra over the ring of Laurent polynomials
$\Complex\polyspc{\shiftpar,\shiftpar^{-1}}$ in a formal variable $\shiftpar$,
the affine bialgebra contains a non-trivial central charge element $\genC$
and a derivation $\genD$ (or, more precisely, a pair of each
\cite{Beisert:2024rke}).
The affine derivation acts as an additional symmetry that
can be associated to measuring the deviation of the classical r-matrix from the
difference form.
The affine derivation can also be related with the classical limit of the
two-dimensional q-deformed Poincaré boost generator
\cite{Young:2007wd,Borsato:2017lpf}.
Furthermore, the phase of the classical r-matrix receives some constraints,
since the derivation acts directly on the loop variable $\shiftpar$ of the
algebra.
The physical interpretation of the central charge(s) is not yet known;
so far, they have been projected out in all applications.

The contraction for the affine Lie bialgebra is largely analogous
to the one for the loop algebra, but there is one additional complication:
The reduction requires the affine derivation to be supplemented
by particular loop algebra generators, which introduce extra poles at
$\shiftpar = \pm \redpar$ in the affine algebra.
In \cite{Beisert:2024rke} it was proposed to incorporate these extra poles
in the affine algebra by expanding the underlying ring of
polynomials $\Complex\polyspc{\shiftpar,\shiftpar^{-1}}$ to a ring of formal
series $\Complex\pwrser{\shiftpar,\shiftpar^{-1}}$.
While the expansion yields a consistent formulation of the
quasi-triangular affine bialgebra,
it introduces geometric series expressions in both the Lie algebra
and the coalgebra.
Summing up the geometric series, one recovers the above poles,
but in addition, also certain distributional terms
(see \cite{Beisert:2010kk}),
which play an essential role for the quasi-triangular structures
of the bialgebra.
The nature of the infinite series contributions
and/or the extra singularities at $\shiftpar = \pm \redpar$
remains to be understood better.
In particular, it is not clear how to properly quantise these contributions,
i.e.\ how to lift the combinatorics in the bialgebra relations
towards formal deformation expansions in $\hbar$.

\medskip

In this work, we propose a more elegant formulation of the affine Lie
bialgebra of \cite{Beisert:2024rke} that comes without
the need for infinite series expansions
and may pave the way towards quantisation.
The key to this formulation is a Möbius transformation of the loop variable
$\shiftpar$ in the affine algebra to a new variable $\looppar$ such that the
pair of singular points $\shiftpar = \pm \redpar$ is mapped to
$\looppar=0,\infty$, respectively.
Upon a subsequent reparametrisation of two generators,
the Lie bialgebra is completely free from unwanted poles.
Furthermore, the transformation provides a much clearer common ground between
the rational and trigonometric forms of the affine Lie bialgebra.

We further explore the idea \cite{Beisert:2024rke} of addressing the additional
poles via a multi-point Lie algebra as formulated by Krichever and Novikov
\cite{krichever1987algebras} and further developed by Schlichenmaier
\cite{Schlichenmaier+2014}.
This class of algebras extends the relevant ring of functions
from Laurent polynomials for ordinary affine algebras
to meromorphic functions on a Riemann sphere with multiple punctures,
i.e.\ to rational functions with a prescribed set of pole locations.
We propose a formulation of the quasi-triangular bialgebra structure
within the multi-point algebra, both for a rational and a trigonometric
r-matrix.
This formulation allows one to consistently treat the additional singularities
together with the affine extensions as a quasi-triangular Lie bialgebra.
While the approach will eventually require the extension
of the notion of a multi-point algebra to the quantum realm,
which may turn out to be cumbersome as well,
this novel application of Krichever--Novikov algebras
is certainly an interesting concept to study
for the field of integrable models.

\medskip

This work is organised as follows:
In \secref{sec:poincare} we introduce the affine Poincaré superalgebra.
In \secref{sec:rational} we rederive and expose
the rational affine classical Lie bialgebra
relevant to the one-dimensional Hubbard model and the AdS/CFT worldsheet
scattering.
In \secref{sec:analysis} we analyse and discuss several aspects of the
resulting affine classical Lie bialgebra.
In particular, we discuss the pole singularities,
the quasi-triangular structure, and compare our coalgebra relations to the
results of \cite{Borsato:2017lpf}.
Most importantly, we comment on the extension to a so-called three-point affine
algebra
which can simultaneously accommodate all relevant singularities and the affine
extensions.
In \secref{sec:trigonometric} we discuss the derivation
of the trigonometric r-matrix that is relevant to integrable q-deformations.
The final \secref{sec:conclusions} contains a brief conclusion.
The work is completed by three appendices
\ref{app:dual-jacobi}, \ref{app:krichever-novikov}, and \ref{app:triang_decomp}
discussing an instance of the dual Jacobi identity as well as
technical details of Krichever--Novikov algebras and
the quasi-triangular structure.

\section{Affine Poincaré Superalgebra}
\label{sec:poincare}

We start by recollecting some foundations for the 3D affine Poincaré
superalgebra based on the maximally extended $\alg{psu}(2|2)$ algebra
\cite{Nahm:1977tg}
\[@{eq:h}
\mspalg \defeq \alg{sl}(2)\ltimes\alg{psu}(2|2)\ltimes\Real^{2,1}
\]
In this section, we discuss the various components of the algebra,
and we spell out their relations and our conventions.

\subsection{Algebra \texorpdfstring{$\alg{sl}(2)$}{sl(2)}}

The algebra $\alg{sl}(2)$ is spanned by three generators $\genJ^a$ with
$a=0,\pm$.
We write the algebra relations equivalently using structure constants
or Cartan raising and lowering generators
\[@{eq:Jcomm}
\comm{\genJ^a}{\genJ^b}=\iunit \slstruct^{ab}{}_c\genJ^c
\eqnjoin{\iff}
\comm{\genJ^0}{\genJ^\pm}=\pm\genJ^\pm
\&
\comm{\genJ^+}{\genJ^-}=-2\genJ^0
\]
The algebra has a quadratic invariant tensor
for which we choose the normalisation
\[@{eq:Jcas}
\casJJ =
\slinv_{ab}(\genJ^a\otimes\genJ^b)
\defeq
-\genJ^0\otimes\genJ^0
+ \half\genJ^+\otimes\genJ^-
+ \half\genJ^-\otimes\genJ^+
\]
The above relations imply the following non-zero values
for the structure constants $\slstruct^{ab}{}_c$, the invariant tensor
$\slinv_{ab}$, and their respective duals
$\slstruct^a{}_{bc}\defeq\slinv_{bd}\slstruct^{ad}{}_{c}$ and $\slinv^{ab}$
(with $\slinv^{ac}\slinv_{cb}={\delta^a}_b$)
\<@{eq:sl2Constants}
\slstruct^{0\pm}{}_\pm \&= \mp\iunit
\&
\slstruct^{\pm\mp}{}_0 \&= \pm 2\iunit
\&
\slinv_{00} \&= -1
\&
\slinv_{\pm\mp} \&= \half
\\
\slstruct^{0}{}_{\mp\pm} \&= \mp \ihalf
\&
\slstruct^{\pm}{}_{\pm0} \&= \pm \iunit
\&
\slinv^{00} \&= -1
\&
\slinv^{\pm\mp} \&= 2
\>
%

\subsection{Poincaré Algebra and Superalgebra}
\label{sec:poincarealg}

Next, we formulate the 3D Poincaré algebra $\alg{iso}(2,1)$
and its maximally supersymmetric extension $\mspalg$.
The Poincaré algebra in three dimensions is spanned by
a triplet of Lorentz rotation generators $\genL^a$ and a triplet
of translation generators $\genP^a$.
Their algebra relations generalise the $\alg{sl}(2)$ relations
where we use the same conventions as above
\<
\comm{\genL^a}{\genL^b} \&= \iunit \slstruct^{ab}{}_c \genL^c
\&
\comm{\genL^a}{\genP^b} \&= \comm{\genP^a}{\genL^b}
= \iunit \slstruct^{ab}{}_c \genP^c
\&
\comm{\genP^a}{\genP^b} \&= 0
\>
This algebra has two independent invariant quadratic combinations,
namely the Pauli--Lubanski scalar $\casLP+\casPL$ and the mass invariant
$\casPP$ with
\<@{eq:poincareinvariant}
\casLP \&= \slinv_{ab}(\genL^a\otimes\genP^b)
\&
\casPL \&= \slinv_{ab}(\genP^a\otimes\genL^b)
\&
\casPP \&= \slinv_{ab}(\genP^a\otimes\genP^b)
\>
The maximally extended 3D Poincaré superalgebra $\mspalg$,
defined in \eqref{eq:h},
has an additional set of eight supercharges $\genQ^{j,lr}$ with
$j,l,r=\spinup,\spindown$
as well as two additional triplets of $\alg{su}(2)_{\algleft,\algright}$
generators $\genJ^a_{\algleft,\algright}$ with $a=0,\pm$.
The supercharges transform in spin-$\vfrac{1}{2}$ representations
of the Lorentz algebra $\alg{sl}(2)$ and the internal symmetries
$\alg{su}(2)_{\algleft,\algright}$
\<
\comm{\genL^a}{\genQ^{j, l r}}
\&=
- \half \slinv^{ab} (\slpauli_b)^j{}_k \genQ^{k, l r}
\\
\comm{\genJ_\algleft^a}{\genQ^{j, l r}}
\&=
- \half \slinv^{ab} (\slpauli_b)^l{}_m \genQ^{j, m r}
\\
\comm{\genJ_\algright^a}{\genQ^{j, l r}}
\&=
- \half \slinv^{ab} (\slpauli_b)^r{}_p \genQ^{j, l p}
\>
The $\slpauli$ are a set of Pauli matrices adjusted to
the signature of $\alg{sl}(2)$ and $\sleps$
is the anti-symmetric tensor
\<@{eq:pauli}
\slpauli_0 \&= \<* 1 & 0 \\ 0 & -1 \>
\&
\slpauli_+ \&= \<* 0 & \iunit\\ 0 & 0 \>
\&
\slpauli_- \&= \<* 0 & 0\\ \iunit & 0 \>
\&
\sleps \&= \<* 0 & +1 \\ -1 & 0 \>
\>
The (symmetric) Lie bracket between the supercharges is
\[
\acomm{\genQ^{j, l r}}{\genQ^{k, m p}}
= - \frac{1}{\mref}(\slpauli_a \sleps)^{jk} \sleps^{lm}
\sleps^{rp} \.\genP^a
- \sleps^{jk} (\slpauli_a \sleps)^{lm} \sleps^{rp}
\.\genJ^a_\algleft
+ \sleps^{jk} \sleps^{lm} (\slpauli_a \sleps)^{rp}
\.\genJ^a_\algright
\]
Here $\mref$ is a reference mass scale, which could be absorbed by rescaling
the momentum generators.
The superalgebra has two invariant quadratic forms,
namely $\casPP$ and the following supersymmetry extension of $\casLP+\casPL$
\[@{eq:superpoincareinvariant}
\casJJ_{\mspalg}=\casLP + \casPL -
\mref(\casQQ - \casJJ_{\algleft} + \casJJ_{\algright})
\],
where $\casJJ_{\algleft,\algright}$ denote the invariant quadratic forms
for the even subalgebras $\alg{su}(2)_{\algleft,\algright}$, respectively,
and where we have defined $\casQQ$ as the following quadratic combination of the
supercharges
\[
\casQQ \defeq \half \sleps_{jk} \sleps_{lm} \sleps_{rp}
(\genQ^{j, lr}\otimes \genQ^{k, mp})
\]
%

\subsection{Affine Extension}

We now lift the (super)algebra $\mspalg$ to an extended affine algebra.
The loop algebra $\mspalg \otimes \Complex\polyspc{\looppar,\looppar^{-1}}$
is spanned by the Laurent monomials
$\genJ^a_n\defeq\looppar^n\genJ^a$, $n\in\Integer$, in the formal loop variable
$\looppar$, where the $\genJ^a$ denote a basis for the underlying Lie algebra
$\mspalg$.
We shall use Laurent polynomials $a(\looppar)$ in $\looppar$
to express loop algebra elements $a(\looppar)\genJ^a$ and their relations.

In the affine algebra, the algebra relations of the loop elements
receive contributions from two sets of central charges
\unskip\footnote{The two central charges relate to the existence of two
independent quadratic invariant structures.}
$\genOCL[\cclabel{\kappa}]$ and $\genOCP[\cclabel{\kappa}]$
\<@{eq:algebra1}
\comm!{a(\looppar)\genL^a}{b(\looppar)\genL^b}
\&=
\iunit \slstruct^{ab}{}_{c} \. a(\looppar)b(\looppar) \genL^c
+\frac{1}{2\pi \iunit}\oint_{\cclabel{\kappa}} a(\looppar)\diff{b(\looppar)}
\. \slinv^{ab} \.\genOCL[\cclabel{\kappa}]
\\
\comm!{a(\looppar)\genL^a}{b(\looppar)\genP^b}
\&=
\iunit \slstruct^{ab}{}_{c}  \. a(\looppar)b(\looppar) \genP^c
+\frac{1}{2\pi \iunit}\oint_{\cclabel{\kappa}} a(\looppar)\diff{b(\looppar)}
\. \slinv^{ab} \.\genOCP[\cclabel{\kappa}]
\\
\comm!{a(\looppar) \genJL^a}{b(\looppar) \genJL^b}
\&=
\iunit \slstruct^{ab}{}_{c} \. a(\looppar)b(\looppar) \genJL^c
+ \frac{1}{\mref} \frac{1}{2\pi \iunit}
\oint_{\cclabel{\kappa}} a(\looppar) \diff{b(\looppar)} \slinv^{ab}
\genOCP[\cclabel{\kappa}]
\\
\comm!{a(\looppar) \genJR^a}{b(\looppar) \genJR^b}
\&=
\iunit \slstruct^{ab}{}_{c} \. a(\looppar)b(\looppar) \genJR^c
- \frac{1}{\mref} \frac{1}{2\pi \iunit}
\oint_{\cclabel{\kappa}} a(\looppar) \diff{b(\looppar)} \slinv^{ab}
\genOCP[\cclabel{\kappa}]
\\
\acomm!{a(\looppar) \genQ^{j,lr}}{b(\looppar) \genQ^{k,mp}}
\&=
a(\looppar) b(\looppar) \acomm!{\genQ^{j,lr}}{\genQ^{k,mp}}
+ \frac{2}{\mref}
\frac{1}{2\pi \iunit} \oint_{\cclabel{\kappa}} a(\looppar) \diff{b(\looppar)}
\sleps^{jk} \sleps^{lm} \sleps^{rp} \genOCP[\cclabel{\kappa}]
\>
Here, the contour integrals compute residues
at certain points abstractly labelled by $\cclabel{\kappa}$,
usually $\looppar=0$ and/or $\looppar=\infty$,
but, a priori, any set of locations could be accommodated.
We implicitly sum over the considered pole locations $\cclabel{\kappa}$,
which are associated to the central charges $\genOCL[\cclabel{\kappa}]$
and $\genOCP[\cclabel{\kappa}]$.

The affine algebra is completed by two towers of
derivation generators $\looppar^n\genODL$ and $\looppar^n\genODP$.
Their non-trivial algebra relations are summarised in functional form as
\<@{eq:algebra2}
\comm!{a(\looppar)\genODL}{b(\looppar)\genOD_{\gen{X}}}
\&=
\looppar \brk!{a(\looppar)b'(\looppar)-a'(\looppar)b(\looppar)} \genOD_{\gen{X}}
\eqnsep~ \text{for all }\genOD_{\gen{X}} \in \set{\genODL,\genODP}
\\
\comm!{a(\looppar)\genODL}{b(\looppar)\gen{X}}
\&=
\looppar a(\looppar)b'(\looppar) \gen{X}
\eqnsep~ \text{for all } \gen{X}
\in \set{\genL^a, \genP^a, \genJL^a, \genJR^a, \genQ^{j,lr}}
\\
\comm!{a(\looppar)\genODP}{b(\looppar)\genL^a}
\&=
\looppar a(\looppar)b'(\looppar) \genP^a
\>
%

\section{Rational Lie Bialgebra}
\label{sec:rational}

In this section, we derive the classical affine Lie bialgebra relevant for the
one-\hspace{0pt}dimensional Hubbard model and the AdS/CFT worldsheet scattering.
Essentially, we follow the reduction procedure introduced in
\cite{Beisert:2022vnc, Beisert:2024rke}.
However, additionally, we introduce a specific reparametrisation
of the loop parameter.
Subsequently, we derive the rational r-matrix and the coalgebra relations
of the resulting deformed $\alg{u}(2|2)^{(1)}$ algebra.
We will analyse and discuss aspects of the resulting expressions
in the subsequent \secref{sec:analysis}.

\subsection{Triangular Bialgebra}
\label{sec:triangular-bialgebra}

A Lie algebra $\alg{g}$ can be supplemented by a classical r-matrix $\rmat$
in order to turn it into a so-called quasi-triangular Lie bialgebra.
The classical r-matrix is an element of $\alg{g} \otimes \alg{g}$ that satisfies
the classical Yang--Baxter equation (cYBE)
\[
\cybe{\rmat}{\rmat}
\defeq
\liebr{\rmat_{12}}{\rmat_{13}}
+
\liebr{\rmat_{12}}{\rmat_{23}}
+
\liebr{\rmat_{13}}{\rmat_{23}}
=
0
\]
It gives rise to the Lie cobracket
$\cobra \colon \alg{g} \to \alg{g} \otimes \alg{g}$ via
the coboundary relation
\[@{eq:cobra}
\cobra(X) = \comm{X_1 + X_2}{\rmat}
\],
with the standard notation $X_1 = X \otimes 1$ and $X_2 = 1 \otimes X$.
For the cobracket to be anti-symmetric,
the symmetric part of the r-matrix must be a quadratic invariant element of $\alg{g}$.
That is, for any $X \in \alg{g}$
\[
\comm{X_1 + X_2}{\rmat_{12} + \rmat_{21}} = 0
\]
Moreover, the cobracket satisfies the dual Jacobi identity by virtue of the
cYBE.
Altogether, the cobracket extends the Lie algebra to a proper Lie bialgebra.

The rational r-matrix of the affine Poincaré superalgebra $\mspalg$ reads
\cite{Beisert:2022vnc}
\[@{eq:r-rational}
\rmatmsp(\looppar_1,\looppar_2)
=
-\frac{\rnormL\casJJ_{\mspalg}}{\shiftpar(\looppar_1)-\shiftpar(\looppar_2)}
+ \rtwistL\. \genL^0 \wedge \genP^+
-\frac{\rnormP\casPP}{\shiftpar(\looppar_1)-\shiftpar(\looppar_2)}
\]
The quadratic invariants $\casJJ_{\mspalg}$ and $\casPP$
are given in \eqref{eq:superpoincareinvariant} and \eqref{eq:poincareinvariant}.
Here, $\rtwistL$ parametrises an admissible null twist
which we will need later for the algebraic reduction.
The factors $\rnormL$ and $\rnormP$ are arbitrary parameters.
The function $\shiftpar(\looppar)$ represents
a map $\looppar\mapsto\shiftpar(\looppar)$ to a variable $\shiftpar$
in which the r-matrix takes a difference form.
Conventionally, one assumes $\shiftpar=\looppar$,
but in this work, we will be interested in a non-trivial transformation
$\looppar\mapsto\shiftpar(\looppar)$.
In this way, we will avoid the introduction of formal
Laurent series as in \cite{Beisert:2024rke}.

\subsection{Reparametrisation}
\label{sec:reparametrisation}

The algebraic reduction leading to the rational AdS/CFT r-matrix
starts with restricting the Lorentz rotations to a particular
$\looppar$-dependent direction
\[@{eq:Lofu}
\genRLo
= \redpar^{-1}\shiftpar(\looppar) \genL^0
- \half \econst^{-\iunit \redang} \genL^+
- \half \econst^{+\iunit \redang} \genL^-
\]
Consequently, the triplet of momentum generators is projected
to the same direction (up to affine central charge contributions)
\[@{eq:Pofu}
\genP^0 = \redpar^{-1}\shiftpar(\looppar) \genRPo
\&
\genP^\pm = \econst^{\pm\iunit \redang} \genRPo + \ldots
\]
This direction becomes light-like at the special locations
$\shiftpar(\looppar)=\pm\redpar$, which ultimately leads
to singularities in the definition of the appropriate affine derivation.

In order to clarify the nature of these singularities,
we shall align them with the points $\looppar=0,\infty$,
which are conventionally used as expansion points for the affine algebra.
This is conveniently achieved by the Möbius transformation
$\looppar\mapsto\shiftpar(\looppar)$,
which maps the special points $\shiftpar(\looppar)=\pm\redpar$
to the points $\looppar=0,\infty$, respectively
\[@{eq:repar}
\shiftpar(\looppar)=-\redpar\frac{\looppar+1}{\looppar-1}
\&
\looppar(\shiftpar)=\frac{\shiftpar-\redpar}{\shiftpar+\redpar}
\]
The functional formulation of the affine algebra is independent of a particular
choice of coordinates $\looppar$ vs.\ $\shiftpar$. In other words,
the algebraic relations take the same form when expressed in terms of
$\shiftpar$ rather than $\looppar$.
The coalgebra structures, however, make a concrete reference to the coordinates.
Here, we are interested in the algebra for the AdS/CFT r-matrix
\eqref{eq:r-rational} whose functional dependency is
$1/(\shiftpar_1-\shiftpar_2)$ with $\shiftpar$ as used in \eqref{eq:Lofu}.
Note that the point $\shiftpar=0$ is distinguished as a zero of this function.
The functional dependency of the r-matrix
translates to a meromorphic function in $\looppar$
which has the same difference form in the denominator $\looppar_1-\looppar_2$
but is dressed by additional factors in the numerator
\[
\frac{1}{\shiftpar(\looppar_1)-\shiftpar(\looppar_2)}
= \frac{(\looppar_1-1)(\looppar_2-1)}{2\redpar(\looppar_1-\looppar_2)}
\]
These clearly single out the point $\looppar=1$ translating to the point
$\shiftpar=0$.
With transformed parameters, the rational r-matrix \eqref{eq:r-rational}
for the affine Poincaré algebra takes the form
\[@{eq:r-rat-repara}
\rmatmsp(\looppar_1, \looppar_2)
=
-\frac{\rnormL}{2\redpar}
\frac{(\looppar_1-1)(\looppar_2-1)}{\looppar_1-\looppar_2}\casJJ_{\mspalg}
+ \rtwistL\.  \genL^0 \wedge \genP^+
-\frac{\rnormP}{2\redpar}
\frac{(\looppar_1-1)(\looppar_2-1)}{\looppar_1-\looppar_2}\casPP
\]
We note that this r-matrix does not depend on the difference
$\looppar_1-\looppar_2$ alone, but also on $\looppar_1$ and $\looppar_2$
individually.
Nevertheless, it preserves the difference form in the original variables
$\shiftpar_1$ and $\shiftpar_2$.
The difference form is expressed by invariance under shifts generated by the
derivative
\[
\frac{\partial}{\partial \shiftpar(\looppar)}
=
\frac{(\looppar-1)^2}{2\redpar}
\frac{\partial}{\partial\looppar}
\]
Hence, we introduce a pair of transformed affine derivations
\[@{eq:der-repara}
\genZDL \defeq \frac{(\looppar-1)^2}{2\redpar\looppar} \genODL
\&
\genZDP \defeq \frac{(\looppar-1)^2}{2\redpar\looppar} \genODP
\],
where the affine derivations $\genOD$ represent the logarithmic derivative
$\looppar\partial/\partial\looppar$ in $\looppar$, see \eqref{eq:algebra2}.

\subsection{Algebra Reduction}
\label{sec:reduction}

For the reduction, we define a loop subalgebra of the rotation generators
$a(\looppar)\genL^a$.
We first apply the coordinate transformation to the above expression
\eqref{eq:Lofu}
\[
\genRLo
= -\frac{\looppar+1}{\looppar-1} \genL^0
- \half \econst^{-\iunit \redang} \genL^+
- \half \econst^{+\iunit \redang} \genL^-
\]
It is furthermore convenient to redefine the generator
$\genRLt \defeq -(\looppar-1)\genRLo$ in order to eliminate the denominator
$\looppar-1$.
\unskip\footnote{We note that
$\genRLt\polyspc{\looppar,\looppar^{-1}} \subset
\genRLo\polyspc{\looppar,\looppar^{-1}}$
but formally $\genRLo\notin\genRLt\polyspc{\looppar,\looppar^{-1}}$.
This restriction can be viewed as irrelevant because
$\genRLo$ does not represent an original symmetry within some physics model,
but it was identified as a symmetry in the scattering picture of an r-matrix.
Hence, it is sufficient to consider $a(\looppar)\genRLt$
as the symmetries associated to the r-matrix in question.}
The resulting loop algebra of rotations
is spanned by $a(\looppar)\genRLt$ with the base generator
\[@{eq:L-reduction}
\genRLt
= (\looppar+1) \genL^0
+ \half \econst^{-\iunit \redang} (\looppar-1) \genL^+
+ \half \econst^{+\iunit \redang} (\looppar-1) \genL^-
= \slinv_{ab} \reddir^a(\looppar)  \genL^b
\]~
and the $\looppar$-dependent vector
\[@{eq:V-vector}
\reddir(\looppar) =
\<*
\reddir^0(\looppar) \\ \reddir^+(\looppar) \\ \reddir^-(\looppar)
\>
\defeq
\<*
-(\looppar+1) \\ \econst^{+\iunit \redang} (\looppar-1) \\
\econst^{-\iunit \redang}(\looppar-1)
\>
\]

The restriction to the subalgebra activates two ideals among the momentum
generators $a(\looppar)\genP^a$ (see \cite{Beisert:2024rke} for details).
Dividing out these ideals, the momentum generators can be expressed in terms
of a single base generator $\genRPt$
\[@{eq:P-reduction}
a(\looppar)\genP^a = - a(\looppar) \reddir^a(\looppar) \genRPt
+
\frac{\redrot^a}{2\pi \iunit}\oint_{\cclabel{\kappa}}
\frac{\diff{\looppar}}{2\looppar}a(\looppar)
\genOCP[\cclabel{\kappa}]
\]
Note that $\genRPt$ has been rescaled with regard to $\genRPo$ defined in
\eqref{eq:Pofu} as $\genRPt\defeq-\genRPo/(\looppar-1)$.
The vector $\redrot$ describes some additional dependence on the central charges,
which is needed for consistency with the algebra relations \eqref{eq:algebra1}
\[@{eq:W-vector}
\redrot
=
\<* \redrot^0\\ \redrot^+\\ \redrot^- \>
\defeq
\<*
0 \\
+\econst^{+\iunit\redang} \\
-\econst^{-\iunit\redang}
\>
\]

For the subalgebra to close properly,
the affine derivations must be adjusted as follows:
\<@{eq:D-reduction}
\genRDLt[]
\&=
\genZDL-
\frac{(\looppar-1)^2}{4\redpar\looppar}
\slinv_{ab}\redrot^a \genL^b
=
\frac{(\looppar-1)^2}{2\redpar\looppar}
\brk!{\genODL - \half \slinv_{ab}\redrot^a \genL^b}
\\
\genRDPt[]
\&=
\genZDP=
\frac{(\looppar-1)^2}{2\redpar\looppar} \genODP
\>
Finally, we identify the central charges before and after the reduction
without any modifications
\[
\genOCP[\cclabel{\kappa}] = \genRCP[\cclabel{\kappa}]
\&
\genOCL[\cclabel{\kappa}] = \genRCL[\cclabel{\kappa}]
\].
In this setting, the algebra relations \eqref{eq:algebra1} reduce to
\<@{eq:algebra-reduced}
\comm!{a(\looppar)\genRLt}{b(\looppar)\genRPt}
\&=
-\frac{1}{2\pi \iunit}\oint_{\cclabel{\kappa}} \diff\looppar a(\looppar)
\brk[s]*{b'(\looppar)
-\frac{b(\looppar)}{2\looppar}} \genRCP[\cclabel{\kappa}]
\\
\comm!{a(\looppar)\genRLt}{b(\looppar)\genRLt}
\&=
-\frac{1}{2\pi \iunit}\oint_{\cclabel{\kappa}}
\diff\looppar 4\looppar a(\looppar)
\brk[s]*{b'(\looppar)+\frac{b(\looppar)}{2\looppar}} \genRCL[\cclabel{\kappa}]
\\
\comm!{a(\looppar)\genRLt}{b(\looppar)\genQ^{j,lr}}
\&=
-\half a(\looppar) b(\looppar)
\brk!{\reddir^a(\looppar)\slpauli_a}{}^j{}_{k}\. \genQ^{k,lr}
\\
\acomm!{a(\looppar)\genQ^{j,lr}}{b(\looppar)\genQ^{k,mp}}
\&=
 \mref^{-1} a(\looppar) b(\looppar)
 \brk!{\reddir^a(\looppar)\slpauli_a \sleps}{}^{jk} \sleps^{lm}
\sleps^{rp} \.  \genRPt
\\;
+ a(\looppar) b(\looppar) \sleps^{jk}
\brk[s]!{
-(\slpauli_a \sleps)^{lm} \sleps^{rp} \.\genJ^a_\algleft
+ \sleps^{lm} (\slpauli_a \sleps)^{rp} \.\genJ^a_\algright
}
\\;
+ \frac{2}{\mref}
\frac{1}{2\pi \iunit} \oint_{\cclabel{\kappa}} \diff{\looppar} a(\looppar)
\brk[s]*{b'(\looppar)\sleps^{jk} - \frac{b(\looppar)}{4\looppar}
(\redrot^a\slpauli_a\sleps)^{jk}}
\sleps^{lm} \sleps^{rp} \genRCP[\cclabel{\kappa}]
\>
The relations involving the internal symmetry generators
$\genJ_{\algleft}^a$, $\genJ_{\algright}^a$ spelled out in
\secref{sec:poincarealg} remain unaffected.
The algebra relations \eqref{eq:algebra2} involving the derivation become
\<@{eq:deralgebra-reduced}
\comm!{a(\looppar)\genRDLt[]}{b(\looppar)\genRDLt[]}
\&=
\frac{(\looppar-1)^2}{2\redpar}
\brk[s]!{a(\looppar)b'(\looppar)-a'(\looppar)b(\looppar)}\genRDLt[]
\\
\comm!{a(\looppar)\genRDLt[]}{b(\looppar)\genRDPt[]}
\&=
\frac{(\looppar-1)^2}{2\redpar}
\brk[s]!{a(\looppar)b'(\looppar)-a'(\looppar)b(\looppar)}\genRDPt[]
\\
\comm!{a(\looppar)\genRDLt[]}{b(\looppar)\genRLt}
\&=
\frac{(\looppar-1)^2}{2\redpar}
a(\looppar)\brk[s]*{b'(\looppar)+\frac{b(\looppar)}{2\looppar}}\genRLt
\\
\comm!{a(\looppar)\genRDLt[]}{b(\looppar)\genRPt}
\&=
\frac{(\looppar-1)^2}{2\redpar}
a(\looppar)\brk[s]*{b'(\looppar)-\frac{b(\looppar)}{2\looppar}}\genRPt
\\
\comm!{a(\looppar)\genRDLt[]}{b(\looppar)\genJL^a}
\&=
\frac{(\looppar-1)^2}{2\redpar}
a(\looppar)b'(\looppar)\genJL^a
\\
\comm!{a(\looppar)\genRDLt[]}{b(\looppar)\genJR^a}
\&=
\frac{(\looppar-1)^2}{2\redpar}
a(\looppar)b'(\looppar)\genJR^a
\\
\comm!{a(\looppar)\genRDLt[]}{b(\looppar)\genQ^{j,lr}}
\&=
\frac{(\looppar-1)^2}{2\redpar}
a(\looppar) \brk[s]*{b'(\looppar) \delta^j{}_k
+ \frac{b(\looppar)}{4 \looppar}
(\redrot^a\slpauli_a)^j{}_k}\genQ^{k,lr}
\\
\comm!{a(\looppar)\genRDPt[]}{b(\looppar)\genRLt}
\&=
\frac{(\looppar-1)^2}{2\redpar}
4\looppar a(\looppar)\brk[s]*{b'(\looppar)+\frac{b(\looppar)}{2\looppar}}\genRPt
\>
Here, the vectors $\reddir^a$ and $\redrot^a$ defined in
\eqref{eq:V-vector} and \eqref{eq:W-vector}, respectively, are converted to
matrix form via the triplet of matrices $\slpauli_a$ in \eqref{eq:pauli}
\[
\reddir^a(\looppar)\slpauli_a=
\<*
-(\looppar+1) & \iunit\econst^{+\iunit\redang}(\looppar-1) \\
\iunit\econst^{-\iunit\redang}(\looppar-1) & +(\looppar+1)
\>
\&
\redrot^a\slpauli_a=
\<*
0 & \iunit\econst^{+\iunit\redang} \\ -\iunit\econst^{-\iunit\redang} & 0
\>
\]
Notice that the reduction only affects the $\alg{sl}(2)$ part of the algebra.

\subsection{Reduced r-Matrix}
\label{sec:r-matrix}

Using the reduced generators \eqref{eq:L-reduction} and \eqref{eq:P-reduction},
the r-matrix \eqref{eq:r-rat-repara} becomes
\<@{eq:r-rat-red-long}
\rmatmsp(\looppar_1, \looppar_2)
\&=
\frac{\rnormL}{2\redpar}
\frac{(\looppar_2-1)^2}{\looppar_1-\looppar_2}\genRLt\otimes\genRPt
+ \frac{\rnormL}{2\redpar}
\frac{(\looppar_1-1)^2}{\looppar_1-\looppar_2}\genRPt\otimes\genRLt
\\;
+\frac{\rnormP}{\redpar}
\frac{(\looppar_1+\looppar_2)(\looppar_1-1)(\looppar_2-1)}
{\looppar_1-\looppar_2}
\genRPt\otimes\genRPt
\\;
+ \frac{\rnormL\mref}{2\redpar} \frac{(\looppar_1-1)(\looppar_2-1)}
{\looppar_1-\looppar_2}
 \brk!{\casQQ - \casJJ_{\algleft} + \casJJ_{\algright}}
\\;
+
\brk[s]*{ \rtwistL \econst^{ \iunit \redang} - \frac{\rnormL}{\redpar} }
\brk!{ - (\looppar_2-1) \genL^0 \otimes \genRPt
       + (\looppar_1-1) \genRPt \otimes \genL^0 }
\\;
+ \rtwistL \econst^{+\iunit\redang} \frac{1}{2\pi \iunit}
 \oint_{\cclabel{\kappa}} \frac{\diff{\looppar'}} {2\looppar'}
 \genL^0 \wedge \genRCP[\cclabel{\kappa}]
\\;
+\frac{\rnormL}{4\redpar}
 \frac{1}{2\pi \iunit} \oint_{\cclabel{\kappa}}
 \frac{\diff{\looppar'}}{2\looppar'}
 \frac{(\looppar-1)(\looppar'-1) }{\looppar-\looppar'}
 ( \econst^{- \iunit \redang}  \genL^+ - \econst^{+\iunit \redang}  \genL^-)
 \wedge \genRCP[\cclabel{\kappa}]
\\;
+\frac{\rnormP}{2\redpar}   \frac{1}{(2\pi \iunit)^2}
\oint_{\cclabel{\kappa}} \oint_{\cclabel{\lambda}}
\frac{\diff{\looppar'}}{2\looppar'}
\frac{\diff{\looppar''}}{2\looppar''}
\frac{(\looppar'-1)(\looppar''-1)}{\looppar'-\looppar''}
( \genRCP[\cclabel{\kappa}] \otimes \genRCP[\cclabel{\lambda}] )
\>
Here, only the first three lines and the last one are fully expressed within the
reduced algebra.
For the reduction to extend to the coalgebra, the terms on lines four through
six must vanish from the r-matrix.
The fourth line can be removed readily by tuning the twist phase as
$\rtwistL=\econst^{-\iunit \redang}\lfrac{\rnormL}{\redpar}$.

The residue integrals a priori have non-vanishing contributions
for integration contours winding around $\looppar = 0$ and $\looppar = \infty$.
Moreover, one has to be careful about the point $\looppar_1=\looppar_2$
in the evaluation of the integrals.
In the following, we assume that $\looppar_1$ and $\looppar_2$
are always well separated
and we remove all charges at $\looppar = 0$ and $\looppar = \infty$,
i.e.\ we set $\genRCP[\cclabel{0}] = \genRCP[\cclabel{\infty}] = 0$.
This removes the last three lines in \eqref{eq:r-rat-red-long}.
In \secref{sec:rational-extra} we explore more general central
charge configurations.

We note that a proper reduction of the r-matrix does
not strictly force us to set both central charges to zero:
Instead, it suffices to restrict them to be equal,
$\genRCP[\cclabel{0}] = \genRCP[\cclabel{\infty}]$.
In that case, the residues at $\looppar = 0$ and $\looppar = \infty$
combine to yield a functional dependency of
$\looppar^{-1}(\looppar-1)^2$ times the algebraic structure $\redrot_a\genL^a$.
Such a term can be absorbed by adding the twist proportional to
$\genZDL\wedge \genRCP[\cclabel{\infty}]$
to the r-matrix. The residue and the twist then nicely combine
to a term proportional to $\genRDLt\wedge \genRCP[\cclabel{\infty}]$
via the reduction of the affine derivation \eqref{eq:D-reduction}.
As the term belongs to the reduced algebra, the r-matrix reduces properly.
Nevertheless, this generalisation is rather trivial.
The algebra relations \eqref{eq:algebra-reduced}
produce central charges only as residues.
Assuming that the functional dependencies of the loop algebra generators
have no poles away from $\looppar=0$ and $\looppar=\infty$,
the residues at these two points must be equal but opposite.
Hence, the algebra only produces the combination
$\genRCP[\cclabel{0}] - \genRCP[\cclabel{\infty}]=0$,
and $\genRCP[\cclabel{0}] = \genRCP[\cclabel{\infty}]$
can be considered as an external central charge.
We will thus disregard this possibility.

Subsequently, we arrive at the following reduced r-matrix
\<@{eq:r-rat-red}
\rmatred(\looppar_1, \looppar_2)
\&=
\frac{\rnormL}{2\redpar}
\frac{(\looppar_2-1)^2}{\looppar_1-\looppar_2}\genRLt\otimes\genRPt
+ \frac{\rnormL}{2\redpar}
\frac{(\looppar_1-1)^2}{\looppar_1-\looppar_2}\genRPt\otimes\genRLt
\\;
+\frac{\rnormP}{\redpar}
\frac{(\looppar_1+\looppar_2)(\looppar_1-1)(\looppar_2-1)}
{\looppar_1-\looppar_2}
\genRPt\otimes\genRPt
\\;
+ \frac{\rnormL\mref}{2\redpar} \frac{(\looppar_1-1)(\looppar_2-1)}
{\looppar_1-\looppar_2}
 \brk!{\casQQ - \casJJ_{\algleft} + \casJJ_{\algright}}
\>
It is easy enough to verify that the reduced r-matrix
\eqref{eq:r-rat-red} satisfies the classical Yang--Baxter equation.
For illustration, we display the subset of resulting terms involving
residue integrals leading to terms proportional
to $\genRCL[] \otimes \genRPt \otimes \genRPt$
and $\genRCP[] \otimes \genRPt \otimes \genRPt$
\<
\cybe{\rmatred}{\rmatred}
\&=
\liebr!{\rmatred(\looppar_1,\looppar_2)}{\rmatred(\looppar_1,\looppar_3)}
+\liebr!{\rmatred(\looppar_1,\looppar_2)}{\rmatred(\looppar_2,\looppar_3)}
+\liebr!{\rmatred(\looppar_1,\looppar_3)}{\rmatred(\looppar_2,\looppar_3)}
\\=
  \frac{\rnormL^2}{2\redpar^2} \frac{1}{2\pi \iunit} \oint_{\cclabel{\kappa}}
  \diff{\looppar_1}
\frac{(\looppar_2-1)^2(\looppar_3-1)^2(\looppar_1+\looppar_3)}
{(\looppar_1-\looppar_2)(\looppar_1-\looppar_3)^2}
\genRCL[\cclabel{\kappa}] \otimes \genRPt \otimes \genRPt
\\;
+ \frac{\rnormL \rnormP}{4\redpar^2} \frac{\looppar_2\looppar_3}{2\pi \iunit}
\oint_{\cclabel{\kappa}} \diff{\looppar_1}
\frac{(\looppar_1-1)(\looppar_2-1)(\looppar_3-1)^2}
{2\looppar_1(\looppar_1-\looppar_3)}
\genRCP[\cclabel{\kappa}] \otimes \genRPt \otimes \genRPt
\\;
- \frac{\rnormL \rnormP}{4\redpar^2} \frac{\looppar_2\looppar_3}{2\pi \iunit}
\oint_{\cclabel{\kappa}} \diff{\looppar_1}
\frac{(\looppar_1+\looppar_2)(\looppar_2-1)^2(\looppar_3-1)^2}
{(\looppar_1-\looppar_2)^2(\looppar_1-\looppar_3)}
\genRCP[\cclabel{\kappa}] \otimes \genRPt \otimes \genRPt
\\;
- \frac{\rnormL \rnormP}{4\redpar^2} \frac{\looppar_2\looppar_3}{2\pi \iunit}
\oint_{\cclabel{\kappa}} \diff{\looppar_1}
\frac{(\looppar_1-1)(\looppar_2-1)^2(\looppar_3-1)}
{2\looppar_1(\looppar_1-\looppar_2)}
\genRCP[\cclabel{\kappa}] \otimes \genRPt \otimes \genRPt
\\;
+ \frac{\rnormL \rnormP}{4\redpar^2} \frac{\looppar_2\looppar_3}{2\pi \iunit}
\oint_{\cclabel{\kappa}} \diff{\looppar_1}
\frac{(\looppar_1+\looppar_3)(\looppar_2-1)^2(\looppar_3-1)^2}
{(\looppar_1-\looppar_2)(\looppar_1-\looppar_3)^2}
\genRCP[\cclabel{\kappa}] \otimes \genRPt \otimes \genRPt
\\;
+ \ldots
\\= 0
\>
Once again $\genRCP[\cclabel{0}] = \genRCP[\cclabel{\infty}] =0$ is required to
cancel all such terms.

\subsection{Coalgebra Reduction}
\label{sec:coalgebra}

In the following, we compute and spell out the coalgebra relations
for analysis in \secref{sec:analysis} and future reference.
Concretely, we apply the coboundary relation \eqref{eq:cobra} to the r-matrix
\eqref{eq:r-rat-red} using the algebra relations in \secref{sec:reduction}.

The resulting cobrackets of the derivations are
\<@{eq:coalgebra1}
\cobra\brk!{a(\looppar)\genRDLt[]}
\&=
\frac{1}{\looppar_1-\looppar_2}
\brk[s]*{\frac{(\looppar_1-1)^2}{\looppar_1}a(\looppar_1)
- \frac{(\looppar_2-1)^2}{\looppar_2}a(\looppar_2)}
\\;\quad\cdot
\brk[s]^3[
-\frac{\rnormL}{8\redpar^2}
\frac{\looppar_1\looppar_2+1}{\looppar_1\looppar_2-1}
\brk!{(\looppar_2-1)^2 \genRLt\otimes\genRPt
- (\looppar_1-1)^2 \genRPt\otimes\genRLt}
\\;\qquad
+\frac{\rnormP}{4\redpar^2}(\looppar_1-1)(\looppar_2-1)(\looppar_1-\looppar_2)
\genRPt\otimes\genRPt
\\;\qquad
-\frac{\rnormL \mref }{32 \redpar^2} (\looppar_1-1)(\looppar_2-1)
(\redrot^a \sleps\slpauli_a)_{jk}\sleps_{lm} \sleps_{rp}
\genQ^{j,lr}\otimes \genQ^{k,mp}
\brk]
\\;
+(\looppar_1-1)^2(\looppar_2-1)^2
\frac{a(\looppar_1)-a(\looppar_2)}{(\looppar_1-\looppar_2)^2}
\\;\quad\cdot
\brk[s]^3[
-\frac{\rnormL}{4\redpar^2}
\frac{1}{\looppar_1\looppar_2-1}
\brk!{(\looppar_2^2-1)
\genRLt\otimes\genRPt+(\looppar_1^2-1) \genRPt\otimes\genRLt}
\\;\qquad
-\frac{\rnormP}{2\redpar^2}
(\looppar_1+\looppar_2)
\genRPt\otimes\genRPt
- \frac{\rnormL \mref}{4 \redpar^2}
\brk{\casQQ - \casJJ_\algleft + \casJJ_\algright}
\brk]
\\
\cobra\brk!{a(\looppar)\genRDPt[]}
\&=
-\frac{\rnormL}{2\redpar^2}
(\looppar_1-1)^2 (\looppar_2-1)^2 (\looppar_1+\looppar_2)
\frac{ a(\looppar_1)-a(\looppar_2)}{(\looppar_1-\looppar_2)^2}
\genRPt\otimes\genRPt
\>
The cobrackets of the Poincaré generators are
\<@{eq:coalgebra2}
\cobra\brk!{a(\looppar)\genRLt}
\&=
\frac{\rnormL \mref}{8 \redpar}
\frac{(\looppar_1-1)(\looppar_2-1)}
{\looppar_1 - \looppar_2}
\brk!{\reddir^a(\looppar_1) a(\looppar_1) - \reddir^a(\looppar_2) a(\looppar_2)}
\\;\qquad\cdot
(\sleps \slpauli_a)_{jk} \sleps_{lm} \sleps_{rp}
\genQ^{j,lr}\otimes \genQ^{k,mp}
\\;
+
\frac{1}{2 \pi \iunit}\oint_{\cclabel{\kappa}}
\frac{\diff{\looppar'}}{\looppar-\looppar'}
\brk[s]*{a'(\looppar')+\frac{a(\looppar')}{2\looppar'}}
\brk[s]^3[
\frac{\rnormL}{2\redpar}
(\looppar'-1)^2
\genRLt \wedge \genRCP[\cclabel{\kappa}]
\\;\qquad\qquad
+
\frac{2\rnormL}{\redpar}
(\looppar-1)^2\looppar'
\genRPt \wedge \genRCL[\cclabel{\kappa}]
+
\frac{\rnormP}{\redpar}
(\looppar-1)(\looppar'-1) (\looppar+\looppar')
\genRPt \wedge \genRCP[\cclabel{\kappa}]
\brk]
\\
\cobra\brk!{a(\looppar)\genRPt}
\&=
\frac{\rnormL}{2\redpar} \frac{1}{2\pi\iunit} \oint_{\cclabel{\kappa}}
\diff{\looppar'}
\frac{(\looppar-1)^2}{\looppar-\looppar'}
\brk[s]*{a'(\looppar')-\frac{a(\looppar')}{2\looppar'}}
\genRPt \wedge \genRCP[\cclabel{\kappa}]
\>
For the internal symmetry generators, the cobrackets are
\<@{eq:coalgebra3}
\cobra\brk!{a(\looppar)\genJL^a}
\&=
(\looppar_1-1)(\looppar_2-1)
\frac{a(\looppar_1)-a(\looppar_2)}{\looppar_1-\looppar_2}
\\;\quad\cdot
\brk[s]^3[
+\frac{\iunit\rnormL \mref}{2 \redpar} \slstruct^{a}{}_{bc} \.
\genJL^b \otimes \genJL^c
+\frac{\rnormL \mref}{8\redpar}
\slinv^{ab}\sleps_{jk} (\sleps\slpauli_b)_{lm} \sleps_{rp}
\genQ^{j,lr} \otimes \genQ^{k,mp}
\brk]
\\;
+ \frac{\rnormL}{2\redpar} \frac{1}{2\pi \iunit} \oint_{\cclabel{\kappa}}
\diff{\looppar'}
\frac{ (\looppar-1)(\looppar'-1)}{\looppar-\looppar'} a'(\looppar')
 \genJL^a \wedge \genRCP[\cclabel{\kappa}]
\\
\cobra\brk!{a(\looppar)\genJR^a}
\&=
(\looppar_1-1)(\looppar_2-1)
\frac{a(\looppar_1)-a(\looppar_2)}{\looppar_1-\looppar_2}
\\;\quad\cdot
\brk[s]^3[
-\frac{\iunit\rnormL \mref}{2 \redpar} \slstruct^{a}{}_{bc}
\genJR^b \otimes \genJR^c
+\frac{\rnormL \mref}{8\redpar}
\slinv^{ab}\sleps_{jk}  \sleps_{lm}(\sleps\slpauli_b)_{rp}
\genQ^{j,lr} \otimes \genQ^{k,mp}
\brk]
\\;
+ \frac{\rnormL}{2\redpar} \frac{1}{2\pi \iunit} \oint_{\cclabel{\kappa}}
\diff \looppar'
\frac{ (\looppar-1)(\looppar'-1)}{\looppar-\looppar'} a'(\looppar')
 \genJR^a \wedge  \genRCP[\cclabel{\kappa}]
\>
For the supercharges, the cobrackets are
\<@{eq:coalgebra4}
\cobra\brk!{a(\looppar)\genQ^{j,lr}}
\&=
-\frac{\rnormL}{4 \redpar}
\frac{1}{\looppar_1-\looppar_2}
\brk!{
    (\looppar_2-1)\reddir^a(\looppar_1) a(\looppar_1)
    - (\looppar_1-1)\reddir^a(\looppar_2) a(\looppar_2)
}
\\;\qquad\cdot
(\slpauli_a )^j{}_k
\brk[s]!{
  (\looppar_1-1) \genRPt \otimes \genQ^{k,lr}
  - (\looppar_2-1) \genQ^{k,lr} \otimes \genRPt
}
\\;
+
\frac{\rnormL \mref}{4 \redpar}
(\looppar_1-1)(\looppar_2-1)
\frac{a(\looppar_1)-a(\looppar_2)}{\looppar_1-\looppar_2}
\\;\qquad\cdot
\brk[s]*{ (\slpauli_a )^{l}{}_k
\.\genJ^a_\algleft  \wedge \genQ^{j,kr}
- (\slpauli_a )^r{}_k
\.\genJ^a_\algright \wedge \genQ^{j,lk} }
\\;
+
\frac{\rnormL}{2\redpar}
\frac{1}{2\pi \iunit} \oint_{\cclabel{\kappa}} \diff{ \looppar'}
\frac{(\looppar-1)(\looppar'-1)}{\looppar-\looppar'}
\brk[s]*{ a'(\looppar')\delta^j{}_k
+
\frac{a(\looppar')}{4\looppar'}(\redrot^a\slpauli_a)^j{}_k }
\genQ^{k,lr} \wedge \genRCP[\cclabel{\kappa}]
\>
In \appref{app:dual-jacobi}, we verify explicitly
that the dual Jacobi identity for $\genRDLt[]$ holds
based on the above expressions
(disregarding that they were derived from an r-matrix satisfying the cYBE).

\section{Analysis and Discussion}
\label{sec:analysis}

In this section, we analyse and discuss
several aspects of the structures and transformations
derived in the previous \secref{sec:rational}.

\subsection{Grading and Supercharge Eigenvectors}
\label{sec:grading}

The algebra relations of $\genRDLt[]$ in \eqref{eq:deralgebra-reduced}
take a very regular form, hinting at a particular pattern of charges of the
basic generators.
The two supercharges, however, appear in a mixed form represented through the
$2\times 2$ matrix $\redrot^a\slpauli_a$.
Let us therefore change to a basis of eigenvectors under this matrix.
The eigenvectors are given by
\[
\<
\redrot^a\slpauli_a\psi_\wedge \&= +\psi_\wedge \\
\redrot^a\slpauli_a\psi_\vee \&= -\psi_\vee
\>
\&~
\psi_\wedge=
\<*
1\\
-\iunit \econst^{-\iunit\redang}
\>
\&
\psi_\vee=
\<*
-\iunit\econst^{\iunit\redang}\\
1
\>
\]
We can thus expand the odd generators in this basis as
\[
\genQ^{j,lr} = \psi_\vee^j \genRQt^{\vee,lr} + \psi_\wedge^j \genRQt^{\wedge,lr}
\],
and find the following algebra relations with the derivation $\genRDLt[]$
\<
\comm!{a(\looppar)\genRDLt[]}{b(\looppar)\genRQt^{\wedge,lr}}
\&=
\frac{(\looppar-1)^2}{2\redpar}a(\looppar) \brk[s]*{b'(\looppar)
+
\frac{b(\looppar)}{4 \looppar}}
\genRQt^{\wedge,lr}
\\
\comm!{a(\looppar)\genRDLt[]}{b(\looppar)\genRQt^{\vee,lr}}
\&=
\frac{(\looppar-1)^2}{2\redpar} a(\looppar) \brk[s]*{b'(\looppar)
-
\frac{b(\looppar)}{4 \looppar}}
\genRQt^{\vee,lr}
\>.
This demonstrates that the loop algebra structure is $\Integer_4$-twisted
with the generators $\genRQt^\wedge,\genRQt^\vee$ at grading $\pm 1$,
respectively, $\genRLt,\genRPt$ at grading $2$ and $\genJL,\genJR$ at grading
$0$.
In fact, this twisted loop algebra is based on the superalgebra $\alg{u}(2|2)$,
see \cite{Beisert:2007ty}.
For the algebra relations of $\genRQt$ with $\genRLt$,
we express the matrix $\reddir^a(\looppar)\slpauli_a$ in the eigenvector basis
\[
\reddir^a(\looppar)\slpauli_a \psi_\wedge
\&= -2\iunit\econst^{-\iunit\redang} \psi_\vee
\&
\reddir^a(\looppar)\slpauli_a \psi_\vee
\&= 2\iunit\econst^{\iunit\redang}\looppar\psi_\wedge
\]
This shows explicitly that the generator $\genRLt$ with grading $2$ alternates
between the two types of $\genRQt$ generators with gradings $\pm1$
\<
\comm!{a(\looppar)\genRLt}{b(\looppar)\genRQt^{\wedge,lr}}
\&=
-\iunit\econst^{\iunit\redang} a(\looppar) b(\looppar)\looppar \genRQt^{\vee,lr}
\\
\comm!{a(\looppar)\genRLt}{b(\looppar)\genRQt^{\vee,lr}}
\&=
\iunit\econst^{-\iunit\redang} a(\looppar) b(\looppar)\genRQt^{\wedge,lr}
\>
Likewise, the algebra of supercharges respects the grading
\<
\acomm!{a(\looppar)\genRQt^{\wedge,lr}}{b(\looppar)\genRQt^{\vee,mp}}
\&=
\half a(\looppar) b(\looppar)
\brk[s]!{
-(\slpauli_a \sleps)^{lm} \sleps^{rp} \.\genJ^a_\algleft
+ \sleps^{lm} (\slpauli_a \sleps)^{rp} \.\genJ^a_\algright
}
\\;
+\frac{1}{\mref}
\frac{1}{2\pi \iunit} \oint_{\cclabel{\kappa}} \diff{\looppar} a(\looppar)
\brk[s]*{b'(\looppar) - \frac{b(\looppar)}{4\looppar} }
\sleps^{lm} \sleps^{rp} \genRCP[\cclabel{\kappa}]
\\
\acomm!{a(\looppar)\genRQt^{\wedge,lr}}{b(\looppar)\genRQt^{\wedge,mp}}
\&=
-\iunit\econst^{\iunit\redang}\mref^{-1} a(\looppar) b(\looppar)\looppar
\sleps^{lm} \sleps^{rp} \. \genRPt
\\
\acomm!{a(\looppar)\genRQt^{\vee,lr}}{b(\looppar)\genRQt^{\vee,mp}}
\&=
-\iunit\econst^{-\iunit\redang} \mref^{-1} a(\looppar) b(\looppar)
\sleps^{lm} \sleps^{rp} \.  \genRPt
\>
%

\subsection{Loop-Level Expansion and Central Charges}
\label{sec:rational-extra}

Quasi-triangular affine bialgebras are often formulated in terms
of loop levels, i.e.\ in terms of a basis of generators $\shiftpar^n \genJ$
with $n\in\Integer$.
In this picture, the formulation of the classical r-matrix
is problematic because it contains a pole at $\shiftpar_1=\shiftpar_2$
which cannot be expressed as a finite combination of basis elements.
Instead, it is expanded as a geometric series,
and there are two viable choices
which depend on which of the two variables
$\shiftpar_1$ and $\shiftpar_2$ is considered to have a smaller magnitude
\[
\sum_{n=0}^\infty \frac{\shiftpar_2^n}{\shiftpar_1^{n+1}}
\eqdef\frac{1}{\shiftpar_1\minusg \shiftpar_2}
\&
-\sum_{n=0}^\infty \frac{\shiftpar_1^n}{\shiftpar_2^{n+1}}
\eqdef
\frac{1}{\shiftpar_1\minusl \shiftpar_2}
\]
Here, the symbols `$\minusg$' and `$\minusl$' represent
the difference (`$-$') of two variables,
where the arrow points towards the variable of smaller
magnitude (compare to the inequalities `$>$' and `$<$', respectively).

With either of these two expansions,
the classical r-matrix loses the manifest anti-symmetric functional form,
but the classical Yang--Baxter equation is maintained
based on the identity
\[@{eq:cybeseries}
\frac{1}{\shiftpar_1\minusg \shiftpar_2}
\frac{1}{\shiftpar_1\minusg \shiftpar_3}
-\frac{1}{\shiftpar_1\minusg \shiftpar_2}
\frac{1}{\shiftpar_2\minusg \shiftpar_3}
+\frac{1}{\shiftpar_1\minusg \shiftpar_3}
\frac{1}{\shiftpar_2\minusg \shiftpar_3}=0
\]
Note that a consistent ordering of denominators
is essential for the identity to hold.
\unskip\footnote{A meromorphic function can be expanded by
$(\shiftpar_j-\shiftpar_k)/(\shiftpar_j\minuslg\shiftpar_k)$ as long as the
denominators permit a proper ordering for the magnitudes of $\shiftpar_j$'s.
In other words, there must not be closed loops in the flow described by the
arrows `$\minusg$' which would imply incompatible expansions of the underlying
geometric series.}
Furthermore, the treatment of central charges needs to be adjusted
because the series expansion effectively shifts some residues:

Consider the residues of a plain functional pole
\[
\frac{1}{2\pi\iunit}\oint_{\shiftpar''}
\frac{\diff{\shiftpar}}{\shiftpar-\shiftpar'}
=\<?[introtext=for]
0& \shiftpar''=0 \\ +1& \shiftpar''=\shiftpar' \\ -1 & \shiftpar''=\infty
\>!
\]
It has two opposite residues at $\shiftpar=\shiftpar'$
and at $\shiftpar=\infty$.
Conversely, the terms of the geometric series (before summation)
have residues at $\shiftpar=0$ and $\shiftpar=\infty$ only
\[
\frac{1}{2\pi\iunit}\oint_{\shiftpar''}
\frac{\diff{\shiftpar}}{\shiftpar\minusg\shiftpar'}
=\<?[introtext=for]
+1& \shiftpar''=0 \\ 0& \shiftpar''=\shiftpar' \\ -1 & \shiftpar''=\infty
\>!
\]
The opposite geometric series has no residues
\[
\frac{1}{2\pi\iunit}\oint_{\shiftpar''}
\frac{\diff{\shiftpar}}{\shiftpar\minusl\shiftpar'}=0
\]
The absence of residues follows from the expansion into non-negative powers
of $\shiftpar$.
The modification of the residue pattern can be explained in simple terms as
follows:
The geometric series ${\diff{\shiftpar}}/(\shiftpar\minusg\shiftpar')$
formally attributes its residue at $\shiftpar=\shiftpar'$ to the point
$\shiftpar=0$;
the opposite geometric series ${\diff{\shiftpar}}/(\shiftpar\minusl\shiftpar')$
formally attributes its residue at $\shiftpar=\shiftpar'$ to the point
$\shiftpar=\infty$ instead.
In both cases, the residue has to be evaluated together with any further factors
at $\shiftpar\minusg\shiftpar'$ before it is shifted to its new position.
This prescription is analogous to the treatment using distributional terms
introduced in \cite{Beisert:2010kk}.

In the present context, we want to generalise the above prescription
to accommodate the coordinate transformation $\shiftpar\mapsto\looppar$.
The Möbius transformation \eqref{eq:repar}
essentially maps the two expansion points $\shiftpar=\infty,0$
to two other points $\looppar=\looppar_{\pm}$,
in our case $\looppar_{\pm}=\pm 1$,
whose corresponding central charges we will label by $\cclabel{\pm}$.
We can thus define two alternative geometric series as follows:
\<@{eq:geometric}
\frac{1}{\looppar_1\minusg\looppar_2}
\&?\defeq
-(\looppar_+-\looppar_-)\sum_{n=0}^\infty
\frac{(\looppar_1-\looppar_+)^n}{(\looppar_1-\looppar_-)^{n+1}}
\frac{(\looppar_2-\looppar_-)^n}{(\looppar_2-\looppar_+)^{n+1}}
\\
\frac{1}{\looppar_1\minusl\looppar_2}
\&?\defeq
(\looppar_+-\looppar_-)\sum_{n=0}^\infty
\frac{(\looppar_1-\looppar_-)^n}{(\looppar_1-\looppar_+)^{n+1}}
\frac{(\looppar_2-\looppar_+)^n}{(\looppar_2-\looppar_-)^{n+1}}
\>
By construction, these have the following pattern of formal residues
\unskip\footnote{Any further residues at $\looppar'=\looppar_\pm$
encountered in the integrand will clearly contribute as well.}
\<
\frac{1}{2\pi\iunit}\oint_{\looppar_-}
\frac{\diff{\looppar'}}{\looppar'\minusg\looppar}\&=+1
\&
\frac{1}{2\pi\iunit}\oint_{\infty}
\frac{\diff{\looppar'}}{\looppar'\minusg\looppar}\&=-1
\\
\frac{1}{2\pi\iunit}\oint_{\looppar_+}
\frac{\diff{\looppar'}}{\looppar'\minusl\looppar}\&=+1
\&
\frac{1}{2\pi\iunit}\oint_{\infty}
\frac{\diff{\looppar'}}{\looppar'\minusl\looppar}\&=-1
\>

To illustrate the prescription, let us apply it to the r-matrix.
We start with the form \eqref{eq:r-rat-repara}
reduced to a plain rational r-matrix for a simple Lie algebra
\[@{eq:r-rat-repara-gen-DC}
\rmat(\looppar_1, \looppar_2)
=
\frac{(\looppar_1-1)(\looppar_2-1)}{\looppar_1\minusg\looppar_2}\casJJ
+\frac{(\looppar_1-1)^2}{\looppar_1}\genOD\otimes\genC_{\cclabel{+}}
-\frac{(\looppar_2-1)^2}{\looppar_2}\genC_{\cclabel{-}}\otimes\genOD
\]
Here, additional combinations of a central charge and the affine derivation
are required in the classical r-matrix
\unskip\footnote{Note that we add two independent central charges
$\genC_{\cclabel{\pm}}$.
In view of a triangular decomposition, see \secref{sec:triangular},
these two need to be attributed to two different Borel subalgebras.
This is consistent provided that each Borel subalgebra also has
an independent affine derivation serving as their duals.
Both affine derivations have the same algebra relations,
hence it makes sense to identify them in the combined algebra.
Analogously, only a linear combination of $\genC_{\cclabel{\pm}}$
appears in the combined algebra relations due to the fact
that the overall residue of any complex function must be zero.
The other combination of central charges generates an abelian subalgebra,
and it merely twists the coalgebra structures.}
to solve the Yang--Baxter equation
\<@{eq:cybe-rat-with-CD}
\cybe{\rmat}{\rmat}\&=\ldots
+\slinv_{ab}\frac{1}{2\pi\iunit}\oint_{\cclabel{-}}
\frac{(\looppar'-1)(\looppar_2-1)}{\looppar'\minusg\looppar_2}
\der\frac{(\looppar'-1)(\looppar_3-1)}{\looppar'\minusg\looppar_3}
\genC_{\cclabel{-}}\otimes \genJ^a\otimes \genJ^b
\\;
+\slinv_{ab}\frac{1}{2\pi\iunit}\oint_{\cclabel{\pm}}
\frac{(\looppar_1-1)(\looppar'-1)}{\looppar_1\minusg\looppar'}
\der\frac{(\looppar'-1)(\looppar_3-1)}{\looppar'\minusg\looppar_3}
\genJ^a\otimes\genC_{\cclabel{\pm}}\otimes  \genJ^b
\\;
+\slinv_{ab}\frac{1}{2\pi\iunit}\oint_{\cclabel{+}}
\frac{(\looppar_1-1)(\looppar'-1)}{\looppar_1\minusg\looppar'}
\der\frac{(\looppar_2-1)(\looppar'-1)}{\looppar_2\minusg\looppar'}
\genJ^a\otimes \genJ^b\otimes\genC_{\cclabel{+}}
\\;
+\frac{(\looppar_1-1)^2(\looppar_3-1)^2}{(\looppar_1\minusg\looppar_3)^2}
\slinv_{ab}\genJ^a\otimes
\brk!{-\genC_{\cclabel{+}}+\genC_{\cclabel{-}}}\otimes\genJ^b
\>
The ellipsis represents the terms independent of $\genC_{\cclabel{\pm}}$,
which all cancel independently.
The above prescription applied to the first and third line
yields potential contributions only for $\looppar'=\looppar_-$ and
$\looppar'=\looppar_+$, respectively.
The resulting two residues on both lines cancel.
Only the second integral contributes with (opposite) residues
from $\looppar'=\looppar_1$ and $\looppar'=\looppar_3$ shifted to the points
$\looppar'=\looppar_+$ and $\looppar'=\looppar_-$, respectively.
This contribution cancels with the last line,
hence the classical Yang--Baxter equation holds.

The rational r-matrix for the affine Poincaré algebra \eqref{eq:r-rat-repara}
must be completed with the analogue of the $\genD\otimes\genC$ terms
in \eqref{eq:r-rat-repara-gen-DC}
\<
\rmatmsp(\looppar_1, \looppar_2)
\&=\ldots
-\rnormL \brk[s]!{\genZDL\otimes
\genRCP[\cclabel{+}]-\genRCP[\cclabel{-}]\otimes\genZDL}
+\ldots
\>
It leads to additional contributions to the second but last line in
the reduction of the r-matrix in \eqref{eq:r-rat-red-long}
\<@{eq:r-rat-reduction-series}
\rmatmsp(\looppar_1, \looppar_2)
\&=
\ldots
-\rnormL
\brk[s]!{\genZDL\otimes \genRCP[\cclabel{+}]-\genRCP[\cclabel{-}]\otimes\genZDL}
\\;
-\frac{\rnormL}{2\redpar}
 \frac{1}{2\pi \iunit} \oint_{\cclabel{\kappa}}
 \frac{\diff{\looppar'}}{2\looppar'}
 \frac{(\looppar-1)(\looppar'-1) }{\looppar\minusg\looppar'}
 \slinv_{ab}\redrot^a\genL^b \otimes \genRCP[\cclabel{\kappa}]
\\;
+\frac{\rnormL}{2\redpar}
 \frac{1}{2\pi \iunit} \oint_{\cclabel{\kappa}}
 \frac{\diff{\looppar'}}{2\looppar'}
 \frac{(\looppar-1)(\looppar'-1) }{\looppar\minusl\looppar'}
 \genRCP[\cclabel{\kappa}]\otimes \slinv_{ab}\redrot^a\genL^b
+\ldots
\\=
\ldots
-\rnormL
 \brk[s]*{\genZDL-\frac{1}{4\redpar}\frac{(\looppar-1)^2}{\looppar}
 \slinv_{ab}\redrot^a\genL^b}\otimes\genRCP[\cclabel{+}]
\\;
+\rnormL
 \genRCP[\cclabel{-}]\otimes\brk[s]*{\genZDL-\frac{1}{4\redpar}
 \frac{(\looppar-1)^2}{\looppar} \slinv_{ab}\redrot^a\genL^b}
\\;
+\ldots
\\=
\ldots
-\rnormL \brk[s]!{\genRDLt\otimes \genRCP[\cclabel{+}]
-\genRCP[\cclabel{-}]\otimes\genRDLt}
+\ldots
\>
These contributions combine neatly with the residue prescription
to evaluate the second but last line in \eqref{eq:r-rat-red-long}
for the residues at $\looppar'=\looppar_\pm$
and yield the reduced affine derivations \eqref{eq:D-reduction}.

The r-matrix \eqref{eq:r-rat-red} with all relevant combinations
of affine derivations and central charges added takes the form
\<@{eq:r-rat-red-series}
\rmatred(\looppar_1, \looppar_2)
\&=
\frac{\rnormL}{2\redpar}
\frac{(\looppar_2-1)^2}{\looppar_1\minusg\looppar_2}\genRLt\otimes\genRPt
+ \frac{\rnormL}{2\redpar}
\frac{(\looppar_1-1)^2}{\looppar_1\minusg\looppar_2}\genRPt\otimes\genRLt
\\;
+\frac{\rnormP}{\redpar}
\frac{(\looppar_1+\looppar_2)(\looppar_1-1)(\looppar_2-1)}
{\looppar_1\minusg\looppar_2}
\genRPt\otimes\genRPt
\\;
+ \frac{\rnormL\mref}{2\redpar} \frac{(\looppar_1-1)(\looppar_2-1)}
{\looppar_1\minusg\looppar_2}
 \brk!{\casQQ - \casJJ_{\algleft} + \casJJ_{\algright}}
\\;
-\rnormL \brk[s]!{\genRDLt\otimes \genRCP[\cclabel{+}]
-\genRCP[\cclabel{-}]\otimes\genRDLt
+\genRDPt\otimes \genRCL[\cclabel{+}]-\genRCL[\cclabel{-}]\otimes\genRDPt}
\\;
-\rnormP \brk[s]!{\genRDPt\otimes \genRCP[\cclabel{+}]
-\genRCP[\cclabel{-}]\otimes\genRDPt}
\>
As the r-matrix reduces properly to the reduced algebra,
it is also guaranteed that the reduction
satisfies the classical Yang--Baxter equation.

The additional terms in the r-matrix largely do not modify the coalgebra
relations presented in \secref{sec:coalgebra}.
In fact, their contribution cancels out additional contributions
from the formal evaluation of residues of the geometric series.
Hence, the coalgebra relations \eqref{eq:coalgebra2}--\eqref{eq:coalgebra4}
remain unchanged such that residues from $\looppar=\looppar'$ effectively do not
contribute.
Only the coalgebra relations of the affine derivations \eqref{eq:coalgebra1}
must be adjusted by replacing the second-order poles
$1/(\looppar_1-\looppar_2)^2$  with the geometric series
$1/(\looppar_1\minusg\looppar_2)^2$
\<@{eq:coalgebra1loop}
\cobra\brk!{a(\looppar)\genRDLt[]}
\&=
\frac{1}{\looppar_1-\looppar_2}
\brk[s]*{\frac{(\looppar_1-1)^2}{\looppar_1}a(\looppar_1)
-\frac{(\looppar_2-1)^2}{\looppar_2}a(\looppar_2)}
\\;\quad\cdot
\brk[s]^3[
-\frac{\rnormL}{8\redpar^2}
\frac{\looppar_1\looppar_2+1}{\looppar_1\looppar_2-1}
\brk!{(\looppar_2-1)^2 \genRLt\otimes\genRPt
-(\looppar_1-1)^2 \genRPt\otimes\genRLt}
\\;\qquad
+\frac{\rnormP}{4\redpar^2}(\looppar_1-1)(\looppar_2-1)(\looppar_1-\looppar_2)
\genRPt\otimes\genRPt
\\;\qquad
-\frac{\rnormL \mref }{32 \redpar^2} (\looppar_1-1)(\looppar_2-1)
(\redrot^a \sleps\slpauli_a)_{jk}\sleps_{lm} \sleps_{rp}
\genQ^{j,lr}\otimes \genQ^{k,mp}
\brk]
\\;
+(\looppar_1-1)^2(\looppar_2-1)^2
\frac{a(\looppar_1)-a(\looppar_2)}{(\looppar_1\minusg\looppar_2)^2}
\\;\quad\cdot
\brk[s]^3[
-\frac{\rnormL}{4\redpar^2}
\frac{1}{\looppar_1\looppar_2-1}
\brk!{(\looppar_2^2-1) \genRLt\otimes\genRPt+(\looppar_1^2-1)
\genRPt\otimes\genRLt}
\\;\qquad
-\frac{\rnormP}{2\redpar^2}
(\looppar_1+\looppar_2)
\genRPt\otimes\genRPt
- \frac{\rnormL \mref}{4 \redpar^2}
\brk{\casQQ - \casJJ_\algleft + \casJJ_\algright}
\brk]
\\;
+\frac{(\looppar-1)^2}{2\redpar}
a'(\looppar)
\brk[s]^2[
\rnormRL\genRDLt[]\otimes\genRCP[\cclabel{+}]
+\rnormRL\genRDPt[]\otimes\genRCL[\cclabel{+}]
+\rnormRP\genRDPt[]\otimes\genRCP[\cclabel{+}]
\\;\qquad
-\rnormRL\genRCP[\cclabel{-}]\otimes\genRDLt[]
-\rnormRL\genRCL[\cclabel{-}]\otimes\genRDPt[]
-\rnormRP\genRCP[\cclabel{-}]\otimes\genRDPt[]
\brk]
\\
\cobra\brk!{a(\looppar)\genRDPt[]}
\&=
-\frac{\rnormL}{2\redpar^2}
(\looppar_1-1)^2 (\looppar_2-1)^2 (\looppar_1+\looppar_2)
\frac{ a(\looppar_1)-a(\looppar_2)}{(\looppar_1\minusg\looppar_2)^2}
\genRPt\otimes\genRPt
\\;
+\rnormRL\frac{(\looppar-1)^2}{2\redpar}a'(\looppar)
\brk[s]!{\genRDPt[]\otimes\genRCP[\cclabel{+}]-
\genRCP[\cclabel{-}]\otimes\genRDPt[]}
\>
Note that for a non-constant prefactor $a(\looppar)$,
these relations thus fail to be strictly anti-\hspace{0pt}symmetric,
while a constant prefactor eliminates all geometric series contributions
leading to no deviations from \eqref{eq:coalgebra1}.
We will comment further on this issue in \secref{sec:Derivations}.

In conclusion, we can accommodate two central charges $\genRCP[\cclabel{\pm}]$
in the bialgebra with loop-level expansion around the corresponding two points
$\looppar_\pm$.
The location of these points can be chosen freely,
with the exception of the points $\looppar=0,\infty$
due to the arguments presented in \secref{sec:r-matrix}.
It makes good sense to assume at least $\looppar_+=+1$ corresponding
to the distinguished point $\shiftpar=\infty$ of the rational bialgebra.
\unskip\footnote{We may as well set $\looppar_-=-1$ corresponding to the point
$\shiftpar=0$.
However, in the rational case, one may shift this point to any other conceivable
point.}
We note that the two charges $\genRCL[\cclabel{\pm}]$ play an analogous role
for the algebra.
We further emphasise that within each pair $\genRCP[\cclabel{\pm}]$ and
$\genRCL[\cclabel{\pm}]$ only one linear combination ties to the rest of the
algebra.
The other linear combination is never generated by contour integrals
and can be viewed as an external $\alg{u}(1)$ charge.

Unfortunately, the above loop-level expansion has two issues
that we cannot resolve simultaneously:
The algebra relations are expressed as Laurent polynomials in $\looppar$,
consequently, they generate poles at $\looppar=0,\infty$.
Formally, this forces us to choose $\looppar_\pm=0,\infty$ as expansion points
such that the poles implied by the algebra relations
can be expressed properly within the chosen basis of functions.
In that case, the arguments presented in \secref{sec:r-matrix}
exclude the placement of central charges at the expansion points.
Central charges serve as duals to the affine derivations,
which therefore also detach from the algebra structures,
see \secref{sec:triangular} for further details.
Effectively, the affine structures reduce to a loop algebra,
which is the situation originally laid out in \cite{Beisert:2022vnc}.
Less formally, we can expand the poles at $\looppar=0,\infty$ as geometric
series.
This approach can be formulated consistently as in \cite{Beisert:2024rke},
but it relies on formal power series and thus misses the goals of the present
work as well.
We will overcome these shortcomings with a more elaborate formulation in the
following \secref{sec:3pt-algebra}.

\subsection{Three-Point Affine Algebra}
\label{sec:3pt-algebra}

We have seen that the points $\looppar=0,\infty$ ($\shiftpar=\pm\redpar$) are
distinguished because the bialgebra relations can be expressed
in terms of Laurent polynomials $\Complex\polyspc{\looppar,\looppar^{-1}}$
which constitute the space of meromorphic functions on $\bar\Complex$ with poles
restricted to $\looppar=0,\infty$.
Likewise, the point $\looppar=1$ ($\shiftpar=\infty$)
plays a distinguished role for the definition of the affine derivation $\genRDLt$
and the central charge $\genRCP[\cclabel{+}]$.
Conversely, the fourth encountered point $\looppar=-1$ ($\shiftpar=0$)
merely serves to counterbalance zeros at $\looppar=1$;
any other point could equivalently take this role,
\unskip\footnote{Note the translational symmetry in $\shiftpar$ generated by
$\genRDLt$.}
and the corresponding central charge $\genRCP[\cclabel{-}]$
is equivalent to $\genRCP[\cclabel{+}]$.
\unskip\footnote{Central charges are defined by their encircling contours which
alternatively encircle the opposite region on the compactified complex plane.}

The original formulation of the algebra in \cite{Beisert:2024rke}
in terms of Laurent polynomials in the variable $\shiftpar$
\[
\Complex\polyspc{\shiftpar,\shiftpar^{-1}}=
\Complex\polyspc!{(\looppar-1)^{-1},(\looppar+1)^{-1}}
\]~
realises the appearance of poles at $\shiftpar=\pm\redpar$ ($\looppar=0,\infty$)
by series expansion and thus by expanding the underlying space
to formal series $\Complex\pwrser{\shiftpar,\shiftpar^{-1}}$.
Here, the central charges $\genRCP[\cclabel{\pm}]$
naturally match with the expansion points $\shiftpar=\infty,0$
($\looppar=\pm 1$).

Conversely, our formulation in terms of Laurent polynomials
in the transformed variable $\looppar$
\[
\Complex\polyspc{\looppar,\looppar^{-1}}
=
\Complex\polyspc!{(\shiftpar+\redpar)^{-1},(\shiftpar-\redpar)^{-1}}
\]~
avoids series expansions in the algebra relations because there are no poles
away from $\looppar=0,\infty$.
Unfortunately, this formulation formally excludes the central charges
$\genRCP[\cclabel{\pm}]$ from appearing in the algebra relations
because all functions are regular at $\looppar=\pm 1$ by construction.
Furthermore, the central charges $\genRCP[\cclabel{0,\infty}]$ at
$\looppar=0,\infty$ are incompatible with the coalgebra relations.

In the following, we formulate a consistent affine Lie bialgebra
with singularities permitted at more than two distinguished points.
\unskip\footnote{We will generally refer to current Lie algebras,
i.e.\ Lie algebras with dependence on exterior coordinates,
as $n$-point affine algebras
if the set of allowable poles in the coordinates is restricted to a set of $n$
points.}
This will enable us to simultaneously consider all algebra relations
without resorting to infinite series and have compatible central charges and
derivations.
For simplicity, we will restrict attention to the minimal set
of the above three distinguished points $\looppar = 0, 1, \infty$.
We thus extend the functional ring for the loop algebra to
\[
\Complex\polyspc!{\looppar,\looppar^{-1},(\looppar-1)^{-1}}
=
\Complex\polyspc!{(\shiftpar+\redpar)^{-1},(\shiftpar-\redpar)^{-1},\shiftpar}
\]~
comprising meromorphic functions with poles
restricted to the relevant points $\looppar=0,1,\infty$.
The resulting algebra is a so-called three-point loop algebra
\cite{Bremner:1994}.
It can be supplemented by derivations and central charges to form
a three-point affine algebra.

A generic meromorphic function $a(\looppar)$ in the three-point algebra can be
expanded in the form
\[@{eq:3ptbasis0}
a(\looppar) =
a_{\const}
+ \sum_{k=1}^{m_0} a_{0,k} \looppar^{-k}
+ \sum_{k=1}^{m_1} a_{1,k} (\looppar-1)^{-k}
+ \sum_{k=1}^{m_\infty} a_{\infty,k} \looppar^{k}
\],
where the $m_{0,1,\infty}$ describe the maximal degrees for the three poles,
and the coefficients $a_{j,k}$, $k\geq 1$, together with $a_{\const}$
describe the higher residues for these poles.
A three-point affine algebra naturally comes with three central charges
located at $\kappa = 0, 1, \infty$, which are thus labelled by
$\cclabel{\kappa}$, but only two of them are independent.

The algebra relations exposed in \secref{sec:reduction}
are perfectly well-defined in the three-point algebra.
In particular, all coefficient functions appearing in the algebra relations
are elements of the three-point ring,
and all relations require only finitely many terms.
As previously discussed, the Jacobi identities hold provided that
the central charges for the points $\looppar=0,\infty$ are excluded.
Similarly, the coalgebra relations in \secref{sec:coalgebra}
are well-defined and consistent in the three-point algebra.

\medskip

It remains to incorporate the r-matrix \eqref{eq:r-rat-red}
in the three-point algebra framework.
Here, we are faced with the common issue that the pole
$1/(\looppar_1-\looppar_2)$ formally does not belong to the three-point algebra.
As in conventional two-point affine algebras, we should expand this
pole into some geometric series.
In order for the expansion of the r-matrix to lead to a valid quasi-triangular
bialgebra, the expansion must involve all basis functions of the three-point
algebra.
Clearly, the basis \eqref{eq:3ptbasis0} is not sufficient for these purposes
because the geometric series expansion would involve only two of the three
relevant points.

To find a suitable basis, we use the framework of Krichever--Novikov algebras.
While the details of the construction can be found in
\appref{app:krichever-novikov}, let us here only state the resulting basis for
the three-point algebra
\[@{eq:basis-kn-3pt}
\KNAbas{n}{0}(\looppar) = \frac{(-\looppar)^n}{(\looppar - 1)^{2n+1}}
\&
\KNAbas{n}{\infty}(\looppar) =  \frac{\looppar^{n+1}}{(\looppar - 1)^{2n+1}}
\]
One can show that this basis gives a straightforward
expansion of the functional dependency in the r-matrix
\[@{eq:expansion}
\mathop{\sum_{n=0}^\infty \sum_{j \in \{0,\infty\}}}
\KNAbas{-n-1}{j}(\looppar_1)\KNAbas{n}{j}(\looppar_2)
=
-\frac{(\looppar_1 - 1)(\looppar_2 - 1)}{\looppar_1 \minush \looppar_2}
\]
Here, the notation $1/(\looppar_1 \minush \looppar_2)$
serves as an alternative to $1/(\looppar_1 \minuslg \looppar_2)$
for the plain geometric series summation of $1/(\looppar_1 -\looppar_2)$
introduced in \eqref{eq:geometric}
\[@{eq:geometric-three}
\frac{1}{\looppar_1 \minush \looppar_2}
\defeq
-(\looppar_1\looppar_2-1)
\mathop{\sum_{n=0}^\infty}
\frac{\looppar_2^n}{\looppar_1^{n+1}}
\frac{(\looppar_1 - 1)^{2n}}{(\looppar_2 - 1)^{2n+2}}
=
\frac{1}{\looppar_1 - \looppar_2}
\]
We can use this expansion to express the r-matrix \eqref{eq:r-rat-repara}
within the basis \eqref{eq:basis-kn-3pt} for the three-point algebra
\<@{eq:3pt-rmat-poincare}
\rmatmsp(\looppar_1, \looppar_2)
\&=
-\frac{\rnormL}{2 \redpar}
\frac{(\looppar_1 - 1)(\looppar_2 - 1)}{\looppar_1 \minush \looppar_2}
\casJJ_{\mspalg}
+ \rtwistL \. \genL^0 \wedge \genP^+
-\rnormL \genZDL\otimes\genRCP
+\ldots
\>
The reduction of the r-matrix in the new basis
is straightforward using the following residues
leading to the central charge $\genRCP\defeq\genRCP[\cclabel{+}]$
\[
\frac{1}{2 \pi \iunit} \oint_{\cclabel{+}} \frac{\der \looppar'}{2 \looppar'}
\KNAbas{n}{0}(\looppar') =
\frac{1}{2 \pi \iunit} \oint_{\cclabel{+}} \frac{\der \looppar'}{2 \looppar'}
\KNAbas{n}{\infty}(\looppar') = \half\delta_{n=0}
\]
In the end, we obtain the following expansion of the reduced r-matrix
to replace \eqref{eq:r-rat-red-series}
\<@{eq:r-rat-red-series-3pt}
\rmatred(\looppar_1, \looppar_2)
\&=
\frac{\rnormL}{2\redpar}
\frac{(\looppar_2-1)^2}{\looppar_1\minush\looppar_2}\genRLt\otimes\genRPt
+ \frac{\rnormL}{2\redpar}
\frac{(\looppar_1-1)^2}{\looppar_1\minush\looppar_2}\genRPt\otimes\genRLt
\\;
+\frac{\rnormP}{\redpar}
\frac{(\looppar_1+\looppar_2)(\looppar_1-1)(\looppar_2-1)}
{\looppar_1\minush\looppar_2}
\genRPt\otimes\genRPt
\\;
+ \frac{\rnormL\mref}{2\redpar}
\frac{(\looppar_1-1)(\looppar_2-1)}{\looppar_1\minush\looppar_2}
 \brk!{\casQQ - \casJJ_{\algleft} + \casJJ_{\algright}}
\\;
-\rnormL \brk[s]!{\genRDLt\otimes\genRCP
  +\genRDPt\otimes\genRCL}
-\rnormP \genRDPt\otimes\genRCP
\>
This form leads to minor adjustments in the cobrackets
$\cobra(a(\looppar)\genRDLt[])$ and $\cobra(a(\looppar)\genRDPt[])$
analogous to \eqref{eq:coalgebra1loop}.

In conclusion, we obtain a fully consistent formulation for the quasi-triangular
bialgebra.
The Krichever--Novikov basis of meromorphic functions spans the three-point
algebra such that all bialgebra relations close without further ado.
In fact, the reduction procedure spelled out in \secref{sec:reduction},
and in particular, the redefinitions of generators $\genRLt, \genRPt, \genRDLt$
in \eqref{eq:L-reduction}, \eqref{eq:P-reduction} and \eqref{eq:D-reduction}
take a regular form within the three-point algebra.
As usual, merely the classical r-matrix requires series expansion.
We note that the multi-point construction may be applied quite generally
to other quasi-triangular affine bialgebras: If there is a need
for additional singular points besides the points distinguished by the classical
r-matrix, usually $\shiftpar=\infty$ (rational case) or $\zpar=0,\infty$
(trigonometric case), these can be accommodated by a Krichever--Novikov basis,
see \appref{app:krichever-novikov} for further details.

\subsection{Comparison of Affine Derivations}
\label{sec:Derivations}

The new parametrisation for the loop algebra
provides a good opportunity to discuss and compare
two plausible choices for the prefactor $a(\looppar)$
of the affine derivation $a(\looppar)\genRDLt[]$.
On the one hand, the trivial prefactor $a(\looppar)=1$
leads to the charge $\genRDLt[]$ which,
by construction,
corresponds to a derivation with respect to
the original loop algebra parameter $\shiftpar$
\[
\genRDLt[]\simeq\frac{\partial}{\partial\shiftpar}
\]
In \cite{Borsato:2017lpf}, the authors constructed a generator
$\genD_{\text{BT}}$ that behaves similarly to an affine derivation.
It was shown in \cite{Beisert:2024rke} that it corresponds to
the prefactor
$f(\shiftpar)=\shiftpar^2(\looppar)-\redpar^2=
\lfrac{4\redpar^2\looppar}{(\looppar-1)^2}$
for the derivation $\genRDLt[]$,
\[
\genD_{\text{BT}}\defeq
\frac{4\redpar^2\looppar}{(\looppar-1)^2}\genRDLt[]
\simeq\brk!{\shiftpar^2-\redpar^2}\frac{\partial}{\partial\shiftpar}
= 2\redpar\frac{\looppar\partial}{\partial\looppar}
= 2\redpar\frac{\partial}{\partial\log\looppar}
\]
Curiously, the generator describes precisely
a logarithmic derivative with respect to the new loop algebra variable
$\looppar$.
In view of this, another curious observation is that
one may introduce the relativistic rapidity variable $\theta$
based on the relations \eqref{eq:Pofu} and \eqref{eq:repar} as
\[
\shiftpar= \redpar \frac{P^0}{\sqrt{P^+P^-}}\stackrel{!}{=}\redpar\coth(\theta)
\&
\looppar=\frac{\coth(\theta)-1}{\coth(\theta)+1}= \exp(-2\theta)
\]
Consequently, the derivation $\genD_{\text{BT}}$ describes a shift in the
rapidity $\theta$.
Let us compare further aspects of the generators.

The generator $\genRDLt[]$ was introduced in \cite{Beisert:2010kk} without
further ado
to describe (the trigonometric counterpart of)
the affine derivation using the original loop algebra variable $\shiftpar$.
This is also the variable in which the classical r-matrix
(almost) takes on the difference form.
Conversely, the generator $\genD_{\text{BT}}$ was constructed in
\cite{Borsato:2017lpf} as the Lorentz boost of a q-deformed two-dimensional
Poincaré algebra.
Consequently, the prefactor of the $\genL$-generator contributions within
$\genD_{\text{BT}}$ is precisely a constant in $\looppar$, see
\eqref{eq:D-reduction}
\[
\genD_{\text{BT}} = \ldots-\redpar\slinv_{ab}\redrot^a \genL^b
\eqnjoin{\text{vs.}}
\genRDLt[] = \ldots
-\frac{(\looppar-1)^2}{4\redpar\looppar}\slinv_{ab}\redrot^a \genL^b
\]
Hence, it makes sense to distinguish the generator $\genRDLt[]$
as an affine derivation $\sim \partial/\partial\shiftpar$
and $\genD_{\text{BT}}$ as a Lorentz boost $\sim \genL$.
The reduction procedure for the algebra relates these two types of generators
via \eqref{eq:D-reduction}
by supplementing them with the generator of the other kind, such that the
resulting algebra relations close.
The resulting two completions $\genRDLt[]$ and $\genD_{\text{BT}}$
are thus proportional but differ in their functional prefactors.

Most importantly, the coalgebra relations \eqref{eq:coalgebra1loop}
for the affine derivations (adjusted to the results of \secref{sec:3pt-algebra})
have qualitatively different structures:
For the affine derivation with prefactor $a(\looppar)=1$, we obtain
\<@{eq:cobraDLplain}
\cobra\brk{\genRDLt[]}
\&=
-\frac{\rnormL}{8\redpar^2}
\frac{\looppar_1\looppar_2+1}{\looppar_1\looppar_2}
\brk!{(\looppar_2-1)^2 \genRLt\otimes\genRPt
-(\looppar_1-1)^2 \genRPt\otimes\genRLt}
\\;
+\frac{\rnormP}{4\redpar^2}
\frac{\looppar_1\looppar_2-1}{\looppar_1\looppar_2}
(\looppar_1-1)(\looppar_2-1)(\looppar_1-\looppar_2)
\genRPt\otimes\genRPt
\\;
-\frac{\rnormL \mref }{32 \redpar^2}
\frac{\looppar_1\looppar_2-1}{\looppar_1\looppar_2}(\looppar_1-1)(\looppar_2-1)
(\redrot^a \sleps\slpauli_a)_{jk}\sleps_{lm} \sleps_{rp}
\genQ^{j,lr}\otimes \genQ^{k,mp}
\\
\cobra\brk{\genRDPt[]}
\&= 0
\>
This expression is a finite Laurent polynomial in the two loop parameters
$\looppar_k$.
Conversely, the Lorentz boost generator $\genD_{\text{BT}}$
has the cobrackets
\<
\cobra\brk{\genD_{\text{BT}}}
\&=
\cobra\brk*{ \frac{4\redpar^2 \looppar}{(\looppar-1)^2} \genRDLt[] }
\\=
- \frac{\rnormL}{\looppar_1\minush\looppar_2}
\brk!{(1-\looppar_2^2) \genRLt \otimes \genRPt
- (\looppar_1^2-1) \genRPt \otimes \genRLt}
\\;
+
\frac{\looppar_1\looppar_2-1}{\looppar_1\minush\looppar_2}
\brk!{
\rnormL \mref \brk{\casQQ - \casJJ_\algleft + \casJJ_\algright}
+ 2\rnormP(\looppar_1+\looppar_2) \genRPt \otimes \genRPt
}
\\;
-2\frac{\looppar_1+1}{\looppar_1-1}
\brk[s]^2{
\rnormRL\genRDLt\otimes\genRCP
+\rnormRL\genRDPt\otimes\genRCL
+\rnormRP\genRDPt\otimes\genRCP
}
\\=
\frac{-2\rnormL}{\shiftpar(\looppar_1)\minush\shiftpar(\looppar_2)}
\brk!{ \shiftpar(\looppar_2) \genRLo \otimes \genRPo
+ \shiftpar(\looppar_1)  \genRPo \otimes \genRLo}
\\;
+
\frac{\shiftpar(\looppar_1)+\shiftpar(\looppar_2)}
{\shiftpar(\looppar_1)\minush\shiftpar(\looppar_2)}
\brk[s]*{
- \rnormL \mref \brk{\casQQ - \casJJ_\algleft + \casJJ_\algright}
- \rnormP\frac{\shiftpar(\looppar_1)\shiftpar(\looppar_2)-\redpar^2 }{\redpar^2}
\genRPo \otimes \genRPo
}
\\;
+\frac{2\shiftpar(\looppar_1)}{\redpar}
\brk[s]^2{
\rnormRL\genRDLt\otimes\genRCP
+\rnormRL\genRDPt\otimes\genRCL
+\rnormRP\genRDPt\otimes\genRCP
}
\>
This expression is in full agreement with \cite{Borsato:2017lpf}.
However, it has a pole at coincident loop algebra parameters
$\looppar_1=\looppar_2$.
This pole is significant because it leads to an infinite geometric series
when the functional dependency is expanded into loop levels
according to \eqref{eq:geometric-three}.
The expression resulting from the geometric series
evidently fails to be anti-symmetric under exchanging the two tensor sites.
\unskip\footnote{Replacing the non-symmetric geometric series
$2/(\looppar\minusg\looppar')$ by its anti-symmetrisation
$1/(\looppar\minusg\looppar')-1/(\looppar\minusl\looppar')$ is not an option:
it leads to unresolvable issues in the dual Jacobi identity
because positive and negative modes are not matched and do not properly cancel
in all cases. This can be observed, e.g.\ by  focusing on terms of the type
$\casJJ_\algleft\otimes\casJJ_\algleft\otimes\casJJ_\algleft$.}
Likewise, the terms pairing the affine derivations with the central charges
are manifestly not anti-symmetric.
Altogether,
the cobracket $\cobra\brk{\genD_{\text{BT}}}$ does not meet the requirements of
a Lie bialgebra (when expanded or in the presence of central charges).

The above discussion shows that the coalgebra structure
can distinguish between different derivations.
There is an interesting continuation on the
compatibility between affine derivations and central charges:
A derivation describes an infinitesimal shift on the complex plane,
and the algebraic relations of central charges depend on closed cycles on the
complex plane.
The derivation is assumed to have trivial algebra relations with the central
charges,
essentially because an infinitesimal shift cannot shift a pole out of an
encircling cycle.
Technically, the derivation turns the single pole into a double pole, which has
no (regular) residue.
However, exponentiating a derivation yields a finite shift.
A finite shift actually does shift a pole away from its original position, and
thus, the algebra relations
between derivations and central charges should be non-trivial in the end.
The exception is where the cycle encircles the fixed points
of the derivation. For the affine derivation $\genRDLt$,
these are at $\looppar=1$, hence central charges defined via cycles
encircling the point $\looppar=1$ can be truly central.
For any other points, in particular $\looppar=0,\infty$,
one should carefully investigate the algebra relations
between the affine derivation and the central charges.

\subsection{Triangular Decomposition}
\label{sec:triangular}

Many quasi-triangular Lie bialgebras admit a triangular decomposition
which establishes them as Manin triples and classical doubles.
A triangular decomposition is a splitting of the algebra into two halves,
$\alg{g}^+$ and $\alg{g}^-$, each of which forms a valid Lie sub(bi)algebra on
its own
\[
\alg{g}=\alg{g}^+ + \alg{g}^-
\&
\liebr{\alg{g}^\pm}{\alg{g}^\pm}\subset \alg{g}^\pm
\&
\cobra(\alg{g}^\pm)\subset\alg{g}^\pm\wedge \alg{g}^\pm
\]
Note that $\alg{g}^+$ and $\alg{g}^-$ can be allowed to overlap on a small
common subalgebra.
The subalgebras must be chosen such that the two tensor sites of the r-matrix
belong to $\alg{g}^+$ and $\alg{g}^-$, respectively
\unskip\footnote{The completion of the tensor product space allows for infinite
series.}
\[
\rmat \in \overline{\alg{g}^-\otimes\alg{g}^+}
\]
Furthermore, the symmetric part of the r-matrix,
which is an invariant bilinear form for $\alg{g}$,
must be invertible.
\unskip\footnote{Altogether, these properties of $\rmat$ ensure that the subalgebras
$\alg{g}^\pm$
also form subcoalgebras of $\alg{g}$.}
Then, the classical double $\algdbl(\alg{g}^\pm)\defeq \alg{g}^+\oplus\alg{g}^-$
of either subalgebra constitutes a Manin triple
$(\algdbl(\alg{g}^\pm),\alg{g}^+,\alg{g}^-)$ which has a canonical
quasi-triangular structure.
The original quasi-triangular Lie bialgebra $\alg{g}$ with r-matrix $\rmat$ is
recovered from the double by dividing out the intersection
$\alg{g}^+\cap\alg{g}^-$ (if non-trivial)
\[
\alg{g}=\algdbl(\alg{g}^\pm)/(\alg{g}^+\cap\alg{g}^-)
\]

In the following, we will demonstrate that the expanded loop algebra
admits a triangular decomposition for all the pole assignments discussed
earlier, including the three-point algebra case.
To streamline our discussion in this section,
we will merely state the triangular decomposition of the reduced algebra.
More details can be found in \appref{app:triang_decomp}.

\medskip

We start with the expansion points $\looppar = \pm 1$ of the new parametrisation
corresponding to $\shiftpar = \infty,0$ in the old formulation.
The triangular decomposition in this case
has already been established in \cite{Beisert:2024rke}
up to the necessity to expand poles at $\looppar = 0,\infty$ corresponding to
$\shiftpar = \pm \redpar$ into power series.
Let us now rephrase the decomposition in the new variables,
where we consider power series from
$\Complex\pwrser{(\looppar-1)^{-1},(\looppar+1)^{-1}}$.
The expanded r-matrix is given in \eqref{eq:r-rat-red-series}. This r-matrix
induces the following decomposition
\<@{eq:g-g+-1}
\alg{g}^- \&=
\frac{\looppar - 1}{\looppar + 1}
\spn!{\brk{\looppar - 1}^{-1}\genRLt,\brk{\looppar - 1}
\genRPt,\genQ,\genJL,\genJR}
\pwrser!{\brk{\looppar + 1}^{-1}}
\oplus \spn!{\genRDLt, \genRDPt, \genRCL[\cclabel{-}], \genRCP[\cclabel{-}]}
\\
\alg{g}^+ \&=
\spn!{\genRLt,\genRPt,\genQ,\genJL,\genJR}
\pwrser!{(\looppar-1)^{-1}}
\oplus \spn!{\genRDLt, \genRDPt, \genRCL[\cclabel{+}], \genRCP[\cclabel{+}]}
\>
Notice that the algebra relations for the affine derivations of the two
subalgebras $\alg{g}^\pm$ lead to two distinct series expansions for the
subalgebras to close.
Hence, they formally cannot be identified in the double bialgebra
$\algdbl(\alg{g}^\pm)$.
Instead, one may drop the affine derivations from one subalgebra
while dropping the corresponding central charges from the other subalgebra.

\medskip

Next, we consider the case of $\looppar = 0,\infty$ corresponding to
$\shiftpar=\pm \redpar$ as expansion points.
While we have already argued in \secref{sec:r-matrix} that the placement of
the central charges at $\looppar = 0,\infty$ is prohibited for the reduced
algebra, it is still possible to expand the loop algebra r-matrix
\eqref{eq:r-rat-red} around these points.
The resulting function space is the standard space of Laurent polynomials
$\Complex\polyspc{\looppar, \looppar^{-1}}$.
We note that the absence of central charges naturally excludes the affine
derivations as their duals from the bialgebra decomposition.
The reduced bialgebra then splits into
\<@{eq:g-g+-2}
\alg{g}^- \&=
\spn!{\genRLt,\genRPt,\genQ,\genJL,\genJR}\polyspc{\looppar}
\\
\alg{g}^+ \&=
\looppar^{-1}\spn!{\genRLt,(\looppar-1)^2\genRPt,
(\looppar-1)\genQ,(\looppar-1)\genJL,(\looppar-1)\genJR
}\polyspc{\looppar^{-1}}
\>
Here, we also observe that the central charges for this pole assignment
are not allowed for the reduced algebra from another point of view:
Let us consider the algebra relations \eqref{eq:deralgebra-reduced}
in the simplified setting of a derivative operator
\[@{eq:problematic_operator}
\genD\defeq(\looppar-1)^2
\brk*{\frac{\partial}{\partial\looppar}-\frac{\alpha}{\looppar}}
\]~
acting on the space $\Complex\polyspc{\looppar,\looppar^{-1}}$ of Laurent
polynomials in $\looppar$.
The relevant question is whether the action of $\genD$ closes
on polynomials in $\looppar$ and/or in $\looppar^{-1}$
(with detailed assignments for the lowest powers to be sorted out).
Clearly,
\[
\genD\looppar^n
= \brk{n-\alpha}(\looppar-1)^2\looppar^{n-1}
= \brk{n-\alpha}\brk!{\looppar^{n+1}-2\looppar^n+\looppar^{n-1}}
\]~
both increases and decreases the powers of $\looppar$.
The only way to separate polynomials in $\looppar$
from polynomials in $\looppar^{-1}$
is via the kernel of $\genD$ given by the monomial $\looppar^\alpha$.
If $\alpha$ were an integer, the action of $\genD$ would close
on polynomials spanned by $\looppar^{\alpha+n}$ with $n\geq0$
and with $n\leq 0$ (the limiting $n=0$ belongs to both subspaces).
However, in our case, $\alpha$ is not an integer
and therefore $\genD$ does not have invariant subspaces.
The algebra of the affine derivation $\genRDLt[]$
therefore does not fit a triangular decomposition
with regard to the points $\looppar=0,\infty$.
Nevertheless, the loop algebra without affine derivations
does admit a triangular decomposition.

\medskip

Finally, let us discuss the triangular decomposition
of the three-point affine bialgebra $\alg{g}$ induced by the r-matrix
\eqref{eq:r-rat-red-series-3pt}.
The series expression \eqref{eq:r-rat-red-series-3pt} for the r-matrix
uses the Krichever--Novikov basis \eqref{eq:basis-kn-3pt}
whose positive and negative modes span the following spaces of meromorphic
functions:
\[
\spn{\KNAbas{n<0}{j}} = (\looppar-1)\Complex\polyspc{\looppar,\looppar^{-1}}
\&
\spn{\KNAbas{n\geq 0}{j}} = \Complex\polyspc!{(\looppar-1)^{-1}}
\]
Hence, the subalgebras $\alg{g}^-$ and $\alg{g}^+$ should be based on
functions with poles at $\looppar=0,\infty$ and $\looppar=1$, respectively.
However, the appropriate distribution of meromorphic functions of low degree,
particularly the constant function, onto the two $\alg{g}^\pm$ subalgebras
depends on the details of the algebra relations and the r-matrix.

We propose the following triangular decomposition
of the three-point affine bialgebra using the basis functions
\eqref{eq:basis-kn-3pt}
\<@{eq:3pt-triangular}
\alg{g}^- &= \spn!{\genRLtm,(\looppar-1)^2\genRPt,(\looppar-1)\genQ,
(\looppar-1)\genJL,(\looppar-1)\genJR} \polyspc{\looppar,\looppar^{-1}}
\oplus \spn!{\genRDLt,\genRDPt}
\\
\alg{g}^+ &=
\spn!{(\looppar-1)^{-1}\genRLt,(\looppar-1)\genRPt,\genQ,\genJL,\genJR}
\polyspc!{(\looppar-1)^{-1}}
\oplus \spn!{\genRCL,\genRCP}
\>
In other words, the elements of $\alg{g}^-$ are given by generators
whose functional prefactors are Laurent polynomials in $\looppar$.
Furthermore, the prefactors for the generators $\genQ,\genJL,\genJR$
must have a zero at $\looppar=1$,
whereas the generator $\genRPt$ must have a double zero.
The algebra relations within $\alg{g}^-$ respect this restriction,
however, terms proportional to $\rnormP$ in the r-matrix
\eqref{eq:r-rat-red-series-3pt} and in the coalgebra relation $\cobra(\genRDLt)$
in \eqref{eq:cobraDLplain} yield $\genRPt$ with just a single zero at
$\looppar=1$.
These terms are eliminated completely by the following redefinition of $\genRLt$
\unskip\footnote{Importantly, the redefinition also uses the prefactor function
$a(\looppar)$ with inverted argument $a(\looppar^{-1})$.}
\[@{eq:Lmix}
a(\looppar)\genRLtm
\defeq
a(\looppar)\genRLt
+\frac{\rnormP}{\rnormL}
\brk!{2\looppar a(\looppar)-(\looppar+\looppar^{-1})a(\looppar^{-1})}\genRPt
\]
Conversely, the elements of $\alg{g}^+$ are given by generators
whose functional prefactors are meromorphic functions with poles at
$\looppar=1$.
Furthermore, the prefactor for the generator $\genRLt$
must have a zero at $\looppar=\infty$
while the prefactor for the generator $\genRPt$ can have a single pole at
$\looppar=\infty$.
\unskip\footnote{We might use $\genRLtm$ in place of $\genRLt$,
but together with $\genRPt$ it spans the same subspace within $\alg{g}^+$.}
The affine derivations and central charges are associated to $\alg{g}^-$ and
$\alg{g}^+$, respectively.
One can verify that the algebra relations in \secref{sec:reduction},
the coalgebra relations in \secref{sec:coalgebra}
as well as the r-matrix \eqref{eq:r-rat-red-series-3pt}
satisfy the above triangular decomposition properties.
The three-point affine bialgebra, therefore, is the classical double of
$\alg{g}^\pm$
\[
\alg{g} = \algdbl(\alg{g}^\pm)
\]

\medskip

In conclusion, we point out that the three-point algebra formulation
(for $\looppar = 1,0,\infty$)
yields the only proper expansion of the bialgebra relations
with a consistent and complete triangular decomposition.
The two-point formulations only work with power series expansions
(for $\looppar = \pm 1$)
or without affine extensions
(for $\looppar = 0,\infty$).
We further note that the reduction from a three-point Poincaré superalgebra
is compatible with the triangular decomposition
as is shown in \appref{sec:3ptreduction}.

\subsection{Three-Point Algebra in Original Coordinates}

For convenience, we express the three-point algebra formulation in
\secref{sec:3pt-algebra}
and the corresponding triangular decomposition in \secref{sec:triangular}
in terms of the original variables and generators introduced in
\cite{Beisert:2024rke}.
The translation spelled out in \secref{sec:reparametrisation} and
\secref{sec:reduction} can be summarised as
\[
\looppar=\frac{\shiftpar-\redpar}{\shiftpar+\redpar}
\&
\genRLt=-(\looppar-1)\genRLo
\&
\genRPt=-(\looppar-1)^{-1}\genRPo
\]
The algebraic relations are the same as in \cite{Beisert:2024rke},
but the r-matrix \eqref{eq:r-rat-red-series-3pt} now takes the form
\<
\rmatred(\looppar_1, \looppar_2)
\&=
\frac{\rnormL}{\shiftpar_1\minush\shiftpar_2}
\brk[s]!{\genRLo\otimes\genRPo+\genRPo\otimes\genRLo}
+ \frac{\rnormP}{\redpar^2}
\frac{\shiftpar_1\shiftpar_2+\redpar^2}{\shiftpar_1\minush\shiftpar_2}
\genRPo\otimes\genRPo
\\;
+ \frac{\rnormL\mref}{\shiftpar_1\minush\shiftpar_2}
 \brk!{\casQQ - \casJJ_{\algleft} + \casJJ_{\algright}}
-\rnormL \brk[s]!{\genRDLt\otimes\genRCP
  +\genRDPt\otimes\genRCL}
-\rnormP \genRDPt\otimes\genRCP
\>~
with the expansion \eqref{eq:geometric-three} expressed in the original
coordinates
\[
\frac{1}{\shiftpar_1 \minush \shiftpar_2}
\defeq
(\shiftpar_1+\shiftpar_2)
\sum_{n \ge 0}
\frac{(\shiftpar_2^2-\redpar^2)^n}{(\shiftpar_1^2-\redpar^2)^{n+1}}
=
\frac{1}{\shiftpar_1-\shiftpar_2}
\]
The r-matrix $\rmatred\in\overline{\alg{g}^-\otimes\alg{g}^+}$ describes a
duality relation
between the two subalgebras $\alg{g}^\pm$ for a triangular decomposition.
The elements of $\alg{g}^-$ are given by generators
whose functional prefactors are meromorphic functions with poles at
$\shiftpar=\pm\redpar$.
Furthermore, the prefactors for the generators
$\genRLom,\genRPo,\genQ,\genJL,\genJR$
must have a zero at $\shiftpar=\infty$,
where we defined a mixed generator $\genRLom$ corresponding to \eqref{eq:Lmix}
in order to remove contributions proportional to $\rnormP\genRPo$
without a zero at $\shiftpar=\infty$
\[
a(\shiftpar)\genRLom
\defeq
a(\shiftpar)\genRLo
+\frac{\rnormP}{2\redpar^2\rnormL}\brk!{(\shiftpar^2-\redpar^2)a(\shiftpar)
+(\shiftpar^2+\redpar^2)a(-\shiftpar)}\genRPo
\]
Conversely, the elements of $\alg{g}^+$ are given by generators
$\genRLo,\genRPo,\genQ,\genJL,\genJR$ whose functional prefactors are
polynomials in $\shiftpar$.
The triangular decomposition in the original variables thus takes the form
\<
\alg{g}^- &=
(\shiftpar+\redpar)^{-1}\spn!{\genRLom,\genRPo,\genQ,\genJL,\genJR}
\polyspc{(\shiftpar+\redpar)^{-1}}
\oplus (\shiftpar-\redpar)^{-1}\spn!{\genRLom,\genRPo,\genQ,\genJL,\genJR}
\polyspc{(\shiftpar-\redpar)^{-1}}
\\;
\oplus \spn{\genRDLt,\genRDPt}
\\
\alg{g}^+ &= \spn!{\genRLo,\genRPo,\genQ,\genJL,\genJR}
\polyspc{\shiftpar}\oplus \spn{\genRCL,\genRCP}
\>
%

\section{Trigonometric Case}
\label{sec:trigonometric}

In this final section, we discuss the derivation
of the trigonometric AdS/CFT r-matrix,
which is relevant to q-deformations.
Similarly to the rational case, we first introduce the reparametrisation and
reduction and subsequently compute the r-matrix and cobrackets.
One beneficial side effect of the new parametrisation
will be that trigonometric expressions are very close
to their rational counterparts discussed in \secref{sec:rational}.

\subsection{Reparametrisation and Reduction}

The trigonometric version of the r-matrix \eqref{eq:r-rational} reads
\cite{Beisert:2022vnc}
\<@{eq:r-trig}
\rmatmsp(\looppar_1,\looppar_2)
\&=
-\frac{1}{2}
\frac{\zpar(\looppar_1) + \zpar(\looppar_2)}{\zpar(\looppar_1)-\zpar(\looppar_2)}
\brk!{\rnormL \. \casJJ_{\mspalg}+\rnormP \. \casPP}
+ \rtwistL \. \genL^0 \wedge \genP^+
+ \rtwistP \. \genP^0 \wedge \genP^+
\\;
+ \quarter\rnormL\brk[s]!{2 \genL^0 \wedge \genP^0
- \genL^+ \wedge \genP^-
+ \genL^- \wedge \genP^+}
- \quarter\rnormP \. \genP^+ \wedge \genP^-
\\;
+ \quarter\mref\rnormL\brk[s]!{\sleps_{lm}\sleps_{rp}
\genQ^{\spinup,lr}\wedge\genQ^{\spindown,mp}
- \genJL^+ \wedge \genJL^-
+ \genJR^+ \wedge \genJR^-}
\>
Again, the quadratic invariants $\casJJ_{\mspalg}$ and $\casPP$
are given in \eqref{eq:superpoincareinvariant} and \eqref{eq:poincareinvariant}.
Here, $\rtwistP$ is an additional null twist parameter
required for the trigonometric reduction.

The analogue of the Lorentz subalgebra \eqref{eq:Lofu}
in the trigonometric case is given by the direction
\[
\genRLo = \frac{\trigq}{\trigq-\trigq^{-1}} (\zpar -1) \genL^0
- \half \econst^{+\iunit\redang} \genL^-
- \half \trigq^2 \econst^{-\iunit\redang} \zpar \genL^+
\],
and the analogue of the momentum reduction \eqref{eq:Pofu} reads
(up to central charge terms)
\[
\genP^0 = \frac{\trigq}{\trigq-\trigq^{-1}} (\zpar -1) \genRPo+\ldots
\&
\genP^+ = \econst^{+\iunit\redang} \genRPo+\ldots
\&
\genP^- = \trigq^2 \econst^{-\iunit\redang} \zpar \genRPo+\ldots
\]
Here, $\trigq$ is the parameter of the reduction,
it is related to the combinations $h,h',\zpar_*^\pm$ introduced in
\cite{Beisert:2010kk} as
\[
h = -\ihalf (\trigq-\trigq^{-1})
\&
h'  = \sqrt{1-h^2}= \half (\trigq+\trigq^{-1})
\&
\zpar_*^\pm = (\iunit h \pm h')^2 = \trigq^{\pm2}
\]

We want to map the multiplicative evaluation parameter $\zpar$
for the trigonometric loop algebra to a new evaluation parameter $\looppar$
such that the singular points $\zpar=\zpar_*^\pm=\trigq^{\pm 2}$
are mapped to $\looppar=0,\infty$.
A suitable Möbius transformation reads
\[
\zpar(\looppar)
=\trigq^{-2} \frac{\looppar-\trigq^2}{\looppar-\trigq^{-2}}
\&
\looppar(\zpar) = \trigq^{-2}\frac{\zpar - \trigq^2}{\zpar-\trigq^{-2}}
\]
Applying the coordinate transformation to $\genRLo$ and rescaling
$\genRLt \defeq -(\looppar-\trigq^{-2})\genRLo$ yields
the analogue of \eqref{eq:L-reduction}
\[
\genRLt
=
\slinv_{ab}\reddir^a(\looppar) \genL^b
\]
Once again, the restriction to the subalgebra activates two ideals among the
momentum generators.
Dividing out these ideals, the momentum generators can be expressed in terms
of a single base generator $\genRPt\defeq-\genRPo/(\looppar-\trigq^{-2})$,
cf.\ \cite{Beisert:2024rke},
which serves as the analogue of \eqref{eq:P-reduction}
\[
a(\looppar)\genP^a
=
-a(\looppar)\reddir^a(\looppar)\genRPt
+
\frac{\redrot^a}{2\pi \iunit}\oint_{\cclabel{\kappa}}
\frac{\diff{\looppar}}{2\looppar}a(\looppar)
\genOCP[\cclabel{\kappa}]
\]
The form of the above relations remains unchanged with respect to the rational
case, merely the vectors $\reddir^a(\looppar)$ and $\redrot^a$ are now given by
\[
\<*
\reddir^0(\looppar)\\
\reddir^+(\looppar)\\
\reddir^-(\looppar)
\>
=
\<*
-(\looppar+1)\\
(\looppar-\trigq^{-2}) \econst^{+\iunit\redang}\\
(\looppar-\trigq^2) \econst^{-\iunit\redang}
\>
\&
\<* \redrot^0\\\redrot^+\\\redrot^-\>
=
\frac{2}{\trigq+\trigq^{-1}}
\<*
\half(\trigq-\trigq^{-1})\\
\trigq^{-1}\econst^{+\iunit\redang}\\
-\trigq\econst^{-\iunit\redang}
\>
\]
As in the rational case,
the vectors have fixed $\alg{sl}(2)$-invariant products
\[
\slinv_{ab}\reddir^a(\looppar)\reddir^b(\looppar)=(\trigq-\trigq^{-1})^2\looppar
\&
\slinv_{ab}\reddir^a(\looppar)\redrot^b=0
\&
\slinv_{ab}\redrot^a\redrot^b=-1
\],
and $\reddir^a(\looppar)$ satisfies a differential equation
\[
2\looppar \frac{\partial}{\partial \looppar}\reddir^a(\looppar)
=
\reddir^a(\looppar)
+ \iunit \slstruct^{a}{}_{bc} \redrot^b \reddir^c(\looppar)
\]
We can then adjust the affine derivations
to make them close on the subalgebra
in analogy to \eqref{eq:D-reduction}
\<
\genRDLt[] \&=
\frac{(\looppar-\trigq^{-2})(\looppar-\trigq^2)}{\looppar}
\brk!{ \genODL  - \half \slinv_{ab}\redrot^a \genL^b}
\\
\genRDPt[]
\&=
\frac{4}{(\trigq + \trigq^{-1})^2}
\frac{(\looppar-\trigq^{-2})(\looppar-\trigq^2)}{\looppar} \genODP
\>
The formal expressions deviate from the rational case by their prefactors only.
We have normalised the base generator
$\genRDLt[]$ such that it corresponds to a logarithmic derivative
in the variable $\zpar(\looppar)$, i.e.
\[
\frac{\zpar(\looppar)\partial}{\partial \zpar(\looppar)}
=
\frac{(\looppar-\trigq^{-2})(\looppar-\trigq^2)}{\trigq^2 - \trigq^{-2}}
\frac{\partial}{\partial \looppar}
\]
An additional prefactor for $\genRDPt[]$ is included to conveniently
absorb some common deformation factor from its relations.
Finally, we introduce an appropriate rescaling of the central charges before
and after the reduction
\[
\genOCP = \genRCP
\&
\genOCL = \frac{4}{(\trigq+\trigq^{-1})^{2}}\genRCL
\]
In the new parametrisation, the algebra relations
take exactly the same form as their rational counterparts in
\eqref{eq:algebra-reduced},
where only the vectors $\reddir^a(\looppar)$ and $\redrot^a$ take different
values.
Also, the form of the derivation relations \eqref{eq:deralgebra-reduced}
is the same up to a replacement of the common prefactor
according to the replacement rule
\[
\frac{(\looppar-1)^2}{2\redpar}
\eqnjoin*\longrightarrow
(\looppar-\trigq^{-2})(\looppar-\trigq^2)
\]

Once again, the pattern of $\genRDLt[]$-charges arising from the above algebra
relations hints at a twist of the loop algebra structure
for $\genRLt$ and $\genRPt$ vs.\ $\genJL^a$ and $\genJR^a$.
In analogy to the discussion in \secref{sec:reduction}, the matrix
$\redrot^a\slpauli_a$ in the trigonometric case is given by
\[
\redrot^a\slpauli_a
=
\frac{2}{\trigq+\trigq^{-1}}
\<* \half(\trigq-\trigq^{-1}) & \iunit \trigq^{-1}\econst^{+\iunit\redang} \\
-\iunit\trigq\econst^{-\iunit\redang} &-\half(\trigq-\trigq^{-1}) \>
\]
Its eigenvectors are
\[
\<
\redrot^a\slpauli_a\psi_\wedge \&= +\psi_\wedge \\
\redrot^a\slpauli_a\psi_\vee \&= -\psi_\vee
\>
\&~
\psi_\wedge=
\<*
1\\
-\iunit \econst^{-\iunit\redang}
\>
\&
\psi_\vee=
\<*
-\iunit\econst^{\iunit\redang}\\
\trigq^2
\>
\]
We can thus expand the odd generators in this basis as
\[
\genQ^{j,lr} = \psi_\vee^j \genRQt^{\vee,lr} + \psi_\wedge^j \genRQt^{\wedge,lr}
\],
and find the following algebra relations with the derivation $\genRDLt[]$
\<
\comm!{a(\looppar)\genRDLt[]}{b(\looppar)\genRQt^{\wedge,lr}}
\&=
(\looppar-\trigq^{-2})(\looppar-\trigq^2) a(\looppar) \brk[s]*{b'(\looppar)
+ \frac{b(\looppar)}{4 \looppar}}
\genRQt^{\wedge,lr}
\\
\comm!{a(\looppar)\genRDLt[]}{b(\looppar)\genRQt^{\vee,lr}}
\&=
(\looppar-\trigq^{-2})(\looppar-\trigq^2) a(\looppar) \brk[s]*{b'(\looppar)
- \frac{b(\looppar)}{4 \looppar}}
\genRQt^{\vee,lr}
\>.
Thus, again, we see that the loop algebra structure is twisted.
For the algebra relations of $\genRQt$ with $\genRLt$,
we express the matrix $\reddir^a(\looppar)\slpauli_a$ in the eigenvector basis
\[
\reddir^a(\looppar)\slpauli_a\psi_\wedge =-\iunit\econst^{-\iunit\redang}
\frac{\trigq+\trigq^{-1}}{\trigq} \psi_\vee
\&
\reddir^a(\looppar)\slpauli_a\psi_\vee =\iunit\econst^{\iunit\redang}
\looppar\frac{\trigq+\trigq^{-1}}{\trigq^{-1}} \psi_\wedge
\]~
and find
\<
\comm!{a(\looppar)\genRLt}{b(\looppar)\genRQt^{\wedge,lr}}
\&=
-\iunit\econst^{\iunit\redang} \frac{\trigq + \trigq^{-1}}{2\trigq}
a(\looppar) b(\looppar) \genRQt^{\vee,lr}
\\
\comm!{a(\looppar)\genRLt}{b(\looppar)\genRQt^{\vee,lr}}
\&=
\iunit\econst^{-\iunit\redang}\frac{\trigq + \trigq^{-1}}{2\trigq^{-1}}
a(\looppar) b(\looppar) \looppar \genRQt^{\wedge,lr}
\>
%

\subsection{r-Matrix and Coalgebra}

We compute the reduced r-matrix from \eqref{eq:r-trig}.
For this, we expand $\casJJ_{\mspalg}$ and $\casPP$.
Moreover, we have to tune the twist phase to
\[
\rtwistL=\frac{\econst^{-\iunit\redang}\rnormL\trigq}{\trigq-\trigq^{-1}}
\]~
and remove the central charges at $\looppar=0$ and $\looppar=\infty$,
see the discussion in \secref{sec:r-matrix}.
Furthermore, we define
\[
\rtwistP=-\rfrac{1}{4}\econst^{-\iunit\redang}\rnormP\trigq(\trigq-\trigq^{-1})
\]~
for later convenience, and introduce the rescaling
\[
\rnormRL=\frac{\rnormL}{\trigq^2-\trigq^{-2}}
\&
\rnormRP=\frac{\rnormP}{4}\frac{\trigq+\trigq^{-1}}{\trigq-\trigq^{-1}}
\]
Subsequently, we arrive at
\<@{eq:r-trig-red}
\rmatred(\looppar_1,\looppar_2)
\&=
\rnormRL\frac{(\looppar_2-\trigq^2)(\looppar_2-\trigq^{-2})}
{\looppar_1\minusg\looppar_2}
\genRLt \otimes\genRPt
+\rnormRL\frac{(\looppar_1-\trigq^2)(\looppar_1-\trigq^{-2})}
{\looppar_1\minusg\looppar_2}
\genRPt \otimes\genRLt
\\;
+2\rnormRP
\frac{\looppar_1\looppar_2-\half(\trigq^2+\trigq^{-2})(\looppar_1+\looppar_2)+1}
{\looppar_1\minusg\looppar_2}
(\looppar_1+\looppar_2)\genRPt\otimes \genRPt
\\;
- \mref\. \rnormRL\frac{\looppar_1\looppar_2
-\half(\trigq^2+\trigq^{-2})(\looppar_1+\looppar_2)+1}
{\looppar_1\minusg\looppar_2}
\brk!{-\genJL^0 \otimes \genJL^0 + \genJR^0 \otimes \genJR^0 }
\\;
+\frac{\mref \. \rnormRL}{2}
\frac{(\looppar_1-\trigq^2)(\looppar_2-\trigq^{-2})}
{\looppar_1\minusg\looppar_2}
\brk!{
-\genJL^+ \otimes \genJL^- + \genJR^+ \otimes \genJR^-
+\sleps_{lm} \sleps_{rp} \genQ^{\spinup, lr}\otimes \genQ^{\spindown, mp}
}
\\;
+ \frac{\mref \. \rnormRL}{2}
\frac{(\looppar_1-\trigq^{-2})(\looppar_2-\trigq^2)}
{\looppar_1\minusg\looppar_2}
\brk!{
- \genJL^- \otimes \genJL^+  + \genJR^- \otimes \genJR^+
- \sleps_{lm} \sleps_{rp} \genQ^{\spindown, lr}\otimes \genQ^{\spinup, mp}
}
\\;
-
\rnormRL
\brk!{
\genRDLt\otimes \genRCP[\cclabel{+}]
+\genRDPt\otimes \genRCL[\cclabel{+}]
-\genRCP[\cclabel{-}]\otimes\genRDLt
-\genRCL[\cclabel{-}]\otimes\genRDPt
}
\\;
-
\rnormRP
\brk!{
\genRDPt\otimes \genRCP[\cclabel{+}]
-\genRCP[\cclabel{-}]\otimes\genRDPt
}
\>
This r-matrix agrees with the result of \cite{Beisert:2024rke}.
Here, we have implemented the loop-level expansion
discussed in \secref{sec:rational-extra}
by means of the two fixed points
$\looppar_\pm=\trigq^{\mp 2}$
of the affine derivation and the
associated notation \eqref{eq:geometric}
for the geometric series
\[
\frac{1}{\looppar_1\minusg\looppar_2}
\&?\defeq
(\trigq^2-\trigq^{-2})\sum_{n=0}^\infty
\frac{(\looppar_1-\trigq^{-2})^n}{(\looppar_1-\trigq^2)^{n+1}}
\frac{(\looppar_2-\trigq^2)^n}{(\looppar_2-\trigq^{-2})^{n+1}}
\]

The respective coalgebra relations (without the terms due to the supersymmetric
extension) are
as follows.
For the affine derivation, we find
\<
\cobra\brk!{a(\looppar)\genRDLt[]}
\&=
\frac{\looppar_1^{-1}(\looppar_1-\trigq^{-2})(\looppar_1-\trigq^2)a(\looppar_1)
  -\looppar_2^{-1}(\looppar_2-\trigq^{-2})(\looppar_2-\trigq^2)a(\looppar_2)}
  {\looppar_1-\looppar_2}
\\;\quad\cdot
\brk[s]^3[
-\frac{\rnormRL}{2}\frac{\looppar_1\looppar_2+1}{\looppar_1\looppar_2-1}
\brk!{
(\looppar_2-\trigq^{-2})(\looppar_2-\trigq^2)\genRLt \otimes\genRPt
-(\looppar_1-\trigq^{-2})(\looppar_1-\trigq^2)\genRPt \otimes\genRLt
}
\\;\qquad\quad
+
\rnormRP\brk!{\looppar_1\looppar_2-\half(\trigq^2+\trigq^{-2})
(\looppar_1+\looppar_2)+1}
(\looppar_1-\looppar_2)\genRPt\otimes \genRPt
\brk]
\\;
+\frac{\rnormRL\mref}{8}\frac{1}{\looppar_1-\looppar_2}
E_{jk}(\looppar_1,\looppar_2)\sleps_{lm} \sleps_{rp}
\\;\quad\cdot
\brk[s]^3[
\frac{(\looppar_1-\trigq^{-2})(\looppar_1-\trigq^2)a(\looppar_1)}{\looppar_1}
(\redrot^a\slpauli_a)^j{}_i \genQ^{i,lr}\otimes \genQ^{k,mp}
\\;\qquad\quad
+\frac{(\looppar_2-\trigq^{-2})(\looppar_2-\trigq^2)a(\looppar_2)}{\looppar_2}
(\redrot^a\slpauli_a)^k{}_i \genQ^{j,lr}\otimes \genQ^{i,mp}
\brk]
\\;
+
(\looppar_1-\trigq^{-2})(\looppar_1-\trigq^2)(\looppar_2-\trigq^{-2})
(\looppar_2-\trigq^2)
\frac{a(\looppar_1)-a(\looppar_2)}{(\looppar_1\minusg\looppar_2)^2}
\\;\quad\cdot
\brk[s]^3[
-\rnormRL\frac{1}{\looppar_1\looppar_2-1}
\brk!{(\looppar_2^2-1)\genRLt
\otimes\genRPt+(\looppar_1^2-1)\genRPt \otimes\genRLt}
\\;\qquad\quad
-2\rnormRP(\looppar_1+\looppar_2)\genRPt\otimes \genRPt
-\rnormRL\mref(\casQQ - \casJJ_{\algleft} + \casJJ_{\algright})
\brk]
\\;
+(\looppar-\trigq^{-2})(\looppar-\trigq^2)
a'(\looppar)
\brk[s]^2[
\rnormRL\genRDLt[]\otimes\genRCP[\cclabel{+}]
+\rnormRL\genRDPt[]\otimes\genRCL[\cclabel{+}]
+\rnormRP\genRDPt[]\otimes\genRCP[\cclabel{+}]
\\;\qquad
-\rnormRL\genRCP[\cclabel{-}]\otimes\genRDLt[]
-\rnormRL\genRCL[\cclabel{-}]\otimes\genRDPt[]
-\rnormRP\genRCP[\cclabel{-}]\otimes\genRDPt[]
\brk]
\\
\cobra\brk!{a(\looppar)\genRDPt[]}
\&=
-2\rnormRL
(\looppar_1-\trigq^{-2})(\looppar_1-\trigq^2)(\looppar_2-\trigq^2)
(\looppar_2-\trigq^{-2})
(\looppar_1+\looppar_2)
\frac{a(\looppar_1)-a(\looppar_2)}{(\looppar_1\minusg\looppar_2)^2}
\genRPt\otimes\genRPt
\\;
+\rnormRL(\looppar-\trigq^{-2})(\looppar-\trigq^2)a'(\looppar)
\brk[s]!{\genRDPt[]\otimes\genRCP[\cclabel{+}]
-\genRCP[\cclabel{-}]\otimes\genRDPt[]}
\>,
where we have defined the matrix
\[
E(\looppar_1,\looppar_2) \defeq
\<* 0 & +(\looppar_1-\trigq^{2})(\looppar_2-\trigq^{-2}) \\
-(\looppar_1-\trigq^{-2})(\looppar_2-\trigq^{2}) & 0 \>
\]
Note that these expressions are properly anti-symmetric only
for the plain affine derivations $\genRDLt[],\genRDPt[]$
without a functional prefactor, $a(\looppar)=1$,
where they simplify to
\<
\cobra\brk{\genRDLt[]}
\&=
-\frac{\rnormRL}{2}
\frac{\looppar_1\looppar_2+1}{\looppar_1\looppar_2}
\brk!{
(\looppar_2-\trigq^{-2})(\looppar_2-\trigq^2)\genRLt \otimes\genRPt
-(\looppar_1-\trigq^{-2})(\looppar_1-\trigq^2)\genRPt \otimes\genRLt
}
\\;
+\rnormRP\frac{\looppar_1\looppar_2-1}{\looppar_1\looppar_2}
\brk!{\looppar_1\looppar_2-\half(\trigq^2+\trigq^{-2})(\looppar_1+\looppar_2)+1}
(\looppar_1-\looppar_2)\genRPt\otimes \genRPt
\\;
+\frac{\rnormRL\mref}{8}\frac{1}{\looppar_1-\looppar_2}
E_{jk}(\looppar_1,\looppar_2)\sleps_{lm} \sleps_{rp}
\\;\quad\cdot
\brk[s]^3[
\frac{(\looppar_1-\trigq^{-2})(\looppar_1-\trigq^2)}{\looppar_1}
(\redrot^a\slpauli_a)^j{}_i \genQ^{i,lr}\otimes \genQ^{k,mp}
\\;\qquad\quad
+\frac{(\looppar_2-\trigq^{-2})(\looppar_2-\trigq^2)}{\looppar_2}
(\redrot^a\slpauli_a)^k{}_i \genQ^{j,lr}\otimes \genQ^{i,mp}
\brk]
\\
\cobra\brk{\genRDPt[]}\&=0
\>
The cobrackets for the reduced Poincaré generators read
\<
\cobra\brk!{a(\looppar)\genRLt}
\&=
- \frac{\rnormRL\mref}{4}
\frac{1}{\looppar_1-\looppar_2}
E_{jk}(\looppar_1,\looppar_2) \sleps_{lm} \sleps_{rp}
\\;\quad\cdot
\brk[s]!{
a(\looppar_1)\brk!{\reddir^a(\looppar_1)\slpauli_a}{}^j{}_i
\genQ^{i,lr}\otimes\genQ^{k,mp}
+ a(\looppar_2)\brk!{\reddir^a(\looppar_2)\slpauli_a}{}^k{}_i
\genQ^{j,lr}\otimes\genQ^{i,mp}
}
\\;
+
\frac{1}{2\pi \iunit} \oint_{\cclabel{\kappa}}
\frac{\diff{\looppar'}}{\looppar-\looppar'}
\brk[s]*{ a'(\looppar') + \frac{a(\looppar')}{2\looppar'}  }
\\;\quad\cdot
\brk[s]^3[
\rnormRL(\looppar'-\trigq^2)(\looppar'-\trigq^{-2})
\genRLt\wedge\genRCP[\cclabel{\kappa}]
 \\;\qquad\quad
+
4 \rnormRL \looppar'(\looppar-\trigq^2)(\looppar-\trigq^{-2})
\genRPt\wedge\genRCL[\cclabel{\kappa}]
\\;\qquad\quad
+
2\rnormRP
\brk!{\looppar\looppar'-\half(\trigq^2+\trigq^{-2})(\looppar+\looppar')+1}
(\looppar+\looppar') \genRPt\wedge \genRCP[\cclabel{\kappa}]
\brk]
\\
\cobra\brk!{a(\looppar)\genRPt}
\&=
\rnormRL
\frac{1}{2\pi \iunit} \oint_{\cclabel{\kappa}}
\frac{\diff{\looppar'}}{\looppar-\looppar'}
\brk[s]*{ a'(\looppar') - \frac{a(\looppar')}{2\looppar'}  }
(\looppar-\trigq^2)(\looppar-\trigq^{-2})
 \genRPt\wedge\genRCP[\cclabel{\kappa}]
\>
%

\subsection{Four-Point Algebra}
\label{sec:4p-trig}

Similarly to the rational case, the formulation of the trigonometric bialgebra
singles out several points for $\looppar$:
On the one hand, the algebraic relations are expressed
as Laurent polynomials in $\looppar$,
hence the points $\looppar=\infty,0$ are required to accommodate poles.
On the other hand, the r-matrix and the affine derivation
have zeros at the points $\looppar=\trigq^{\pm2}$,
which are commonly used as expansion points for the r-matrix.
Central charges cannot be hosted at $\looppar=\infty,0$,
and it makes sense to use $\looppar=\trigq^{\pm2}$ instead.
In the following, we consider all four distinguished points
as expansion points for the formulation of
the affine extension of the Poincaré superalgebra.
The underlying associative algebra for meromorphic functions
will be the four-point algebra
\[
\Complex\polyspc{\looppar, \looppar^{-1}, (\looppar - \trigq^2)^{-1},
(\looppar - \trigq^{-2})^{-1}}
\]

Towards establishing a Krichever--Novikov basis for the algebra,
see \appref{app:krichever-novikov} for more details,
we have to choose a separating contour which partitions
the singular points into two groups
and which serves as a contour for defining the central charge contributions.
There are various conceivable choices:
The absence of central charges at $\looppar=\infty,0$ suggests that
we associate this pair of points to the same grouping, such that the separating
contour cannot produce a residue integral that distinguishes between
$\looppar=\infty$ and $\looppar=0$.
We assume that the two fixed points $\looppar=\trigq^{\pm 2}$ of the affine
derivation form the other grouping.
Alternatively, one could use the partitions
$\looppar=\infty,0,\trigq^{\mp 2}$ vs.\ $\looppar=\trigq^{\pm 2}$
or consider a three-point algebra $\looppar=\infty,0$ vs.\
$\looppar=\trigq^{\pm 2}$.
\unskip\footnote{It is also conceivable to add any number of singular points to
the former grouping;
the minimum requirement is that the points $\looppar=\infty,0$
are separated from at least one of the points $\looppar=\trigq^{\pm 2}$.}

In the rational case, the Krichever--Novikov basis directly provides a suitable
basis to expand the r-matrix in a multi-point algebra setting
and to induce a natural triangular decomposition of the basis.
The situation in the trigonometric case is complicated by the fact that
the r-matrix involves the various Lie generators with different functional
prefactors.
We can address this issue by introducing two alternative Krichever--Novikov bases
for the four-point algebra at hand.
One such basis is given by
\[@{eq:2+2pt-KN-bas-minus}
\KNAbas{n}{0}^- = \frac{(-\looppar)^n}{(\looppar - \trigq^2)^{n+1}
(\looppar - \trigq^{-2})^{n}}
\&
\KNAbas{n}{\infty}^- = \frac{\looppar^{n+1}}{(\looppar - \trigq^2)^{n+1}
(\looppar - \trigq^{-2})^{n}}
\]
The alternative basis takes a very similar form
\[@{eq:2+2pt-KN-bas-plus}
\KNAbas{n}{0}^+ = \frac{(-\looppar)^n}{(\looppar - \trigq^2)^{n}
(\looppar - \trigq^{-2})^{n+1}}
\&
\KNAbas{n}{\infty}^+ = \frac{\looppar^{n+1}}{(\looppar - \trigq^2)^{n}
(\looppar - \trigq^{-2})^{n+1}}
\]
It differs from \eqref{eq:2+2pt-KN-bas-minus}
by exchanging the roles of $\trigq^{2}$ and $\trigq^{-2}$.
Alternatively, the two sets of basis functions are related by multiplication
with $(\looppar - \trigq^2)/(\looppar - \trigq^{-2})$.
As this function is among the four-point functions,
both spaces equivalently span this space.

Let us now consider the elementary trigonometric r-matrix for $\alg{sl}(2)$
on which the Poincaré algebra and superalgebra rely heavily
\<@{eq:trig-r}
\rmat(\looppar_1, \looppar_2)
\&=
\frac{1}{2}
\frac{(\looppar_1-\trigq^2)(\looppar_2 - \trigq^{-2})}{\looppar_1 - \looppar_2}
\brk!{\genJ^+ \otimes \genJ^- - \genJ^0 \otimes \genJ^0 }
\\;
+\frac{1}{2}
\frac{(\looppar_1-\trigq^{-2})(\looppar_2-\trigq^{2})}{\looppar_1 - \looppar_2}
\brk!{\genJ^- \otimes \genJ^+ - \genJ^0 \otimes \genJ^0}
\>
At $\trigq\to 1$, it readily reduces to the rational r-matrix in
\eqref{eq:r-rat-repara-gen-DC}.
The above two Krichever--Novikov bases
provide an expansion of the two relevant prefactors in the trigonometric
r-matrix in terms of meromorphic functions with four singular points
\<@{eq:trig_KN_expansion}
\sum_{n=0}^\infty \sum_{j \in\{0,\infty\}} \KNAbas{-n-1}{j}^+
(\looppar_1) \KNAbas{n}{j}^-(\looppar_2)
\&=
-\frac{(\looppar_1 - \trigq^2)(\looppar_2 - \trigq^{-2})}
{\looppar_1 \minush \looppar_2}
\\
\sum_{n=0}^\infty \sum_{j \in\{0,\infty\}} \KNAbas{-n-1}{j}^-(\looppar_1)
\KNAbas{n}{j}^+(\looppar_2)
\&=
-\frac{(\looppar_1 - \trigq^{-2})(\looppar_2 - \trigq^{2})}
{\looppar_1 \minush \looppar_2}
\>,
where we introduce a four-point generalisation $1/(\looppar_1\minush\looppar_2)$
of the three-point geometric series \eqref{eq:geometric-three}
\[@{eq:geometric-four}
\frac{1}{\looppar_1 \minush \looppar_2}
\defeq
-(\looppar_1\looppar_2-1)\sum_{n=0}^\infty
\frac{\looppar_2^n(\looppar_1-\trigq^2)^n (\looppar_1-\trigq^{-2})^n }
{\looppar_1^{n+1}(\looppar_2-\trigq^2)^{n+1} (\looppar_2-\trigq^{-2})^{n+1}}
=
\frac{1}{\looppar_1 - \looppar_2}
\]

Towards establishing a triangular decomposition associated to the trigonometric
r-matrix,
we point out that the Krichever--Novikov basis functions with non-negative vs.\
negative mode numbers span the following functional spaces
\<
\spn!{\KNAbas{n<0}{0,\infty}^\pm} \&=
(\looppar-\trigq^{\pm 2})\Complex\polyspc{\looppar,\looppar^{-1}}
\\
\spn!{\KNAbas{n\geq0}{0,\infty}^\pm} \&=
\Complex\polyspc!{(\looppar-\trigq^{2})^{-1},(\looppar-\trigq^{-2})^{-1}}
\>
It is evident that non-negative and negative modes, respectively,
close under functional multiplication.
Therefore, the r-matrix splits the $\alg{sl}(2)$ four-point algebra into the
following two subalgebras
\<
\alg{g}^-
\&=
\spn!{(\looppar-\trigq^{2})\genJ^+ , \genJ^0, (\looppar-\trigq^{-2})\genJ^- }
\polyspc{\looppar,\looppar^{-1}}
\\
\alg{g}^+
\&=
\spn!{\genJ^+ , \genJ^0, \genJ^- }
\polyspc!{(\looppar-\trigq^{2})^{-1},(\looppar-\trigq^{-2})^{-1}}
\>
Notice that the functional dependency of $\genJ^0$ in $\alg{g}^-$ is slightly
relaxed
due to the fact that $\genJ^0$ appears with both
functional prefactors \eqref{eq:trig_KN_expansion} in \eqref{eq:trig-r}:
The Laurent polynomials with zeros at either $\looppar=\trigq^{2}$ or
$\looppar=\trigq^{-2}$ together span all Laurent polynomials.

\medskip

Let us now consider the reduced $\mspalg$ algebra.
The r-matrix singles out the pair $\genRDLt,\genRDPt$ as the appropriate pair of
affine derivations.
\unskip\footnote{All singular points which are not fixed points of
$\genRDLt,\genRDPt$ must belong to the same subalgebra, say $\alg{g}^-$.
In the multi-point case, there are non-fixed points,
hence $\alg{g}^+$  may not contain a copy of the affine derivation.}
The corresponding pair of central charges $\genRCL,\genRCP$
is the sum of central charges at $\looppar=\trigq^{\pm2}$,
which are enclosed by the separating cycle in the Krichever--Novikov algebra
(see \appref{app:krichever-novikov}).
\unskip\footnote{A central charge at individual points $\looppar=\trigq^{\pm2}$
could be supported by the bialgebra,
but it cannot be dual to admissible affine derivations.}
Namely, the affine extension in the r-matrix must take the form
\[
\rmatred
= \ldots
-\rnormL \brk[s]!{ \genRDLt \otimes \genRCP + \genRDPt \otimes \genRCL }
-\rnormP \genRDPt \otimes \genRCP
\]
The final reduced bialgebra can also be expressed in the language of a
four-point algebra, the corresponding trigonometric r-matrix is given by
\eqref{eq:r-trig-red} with appropriate replacements of the meromorphic functions
by \eqref{eq:geometric-four}.
Moreover, one has to combine the central charges at $\looppar = \trigq^2$
and $\looppar = \trigq^{-2}$ and modify the affine terms in the r-matrix
accordingly.
A suitable set of adjusted Krichever--Novikov basis functions
for the geometric series expansion of the $\genRLt,\genRPt$ contributions in the
r-matrix is thus given by
\<
\KNAbas{n}{0}^{\genL} \&=
\frac{(-\looppar)^n}{(\looppar - \trigq^2)^{n+1} (\looppar - \trigq^{-2})^{n+1}}
\&
\KNAbas{n}{\infty}^{\genL} \&=
\frac{\looppar^{n+1}}{(\looppar - \trigq^2)^{n+1}(\looppar - \trigq^{-2})^{n+1}}
\\
\KNAbas{n}{0}^{\genP} \&=
\frac{(-\looppar)^n}{(\looppar - \trigq^2)^{n} (\looppar - \trigq^{-2})^{n}}
\&
\KNAbas{n}{\infty}^{\genP} \&=
\frac{\looppar^{n+1}}{(\looppar - \trigq^2)^{n} (\looppar - \trigq^{-2})^{n}}
\>.
These have geometric series expressions, which nicely
reproduce the corresponding functional factors within the r-matrix
\<
\sum_{n=0}^\infty \sum_{j \in\{0,\infty\}}
\KNAbas{-n-1}{j}^{\genL}(\looppar_1) \KNAbas{n}{j}^{\genP}(\looppar_2)
\&=
-\frac{(\looppar_2 - \trigq^2)(\looppar_2 - \trigq^{-2})}
{\looppar_1 \minush \looppar_2}
\\
\sum_{n=0}^\infty \sum_{j \in\{0,\infty\}}
\KNAbas{-n-1}{j}^{\genP}(\looppar_1) \KNAbas{n}{j}^{\genL}(\looppar_2)
\&=
-\frac{(\looppar_1 - \trigq^{-2})(\looppar_1 - \trigq^{2})}
{\looppar_1 \minush \looppar_2}
\>

Finally, we propose a triangular decomposition analogous to
\eqref{eq:3pt-triangular}
\<@{eq:4pt-triangular}
\alg{g}^- &=
\spn!{\genRLtm,(\looppar-\trigq^{-2})(\looppar-\trigq^2)\genRPt}
\polyspc{\looppar,\looppar^{-1}}
\oplus \spn!{\genRDLt,\genRDPt}
\\;
\oplus
\brk!{
(\looppar-\trigq^{2})\spn!{\genQ^{\spinup,\spinupdown\spinupdown},
\genJL^+,\genJR^+}
\oplus\spn!{\genJL^0,\genJR^0}
\oplus(\looppar-\trigq^{-2})
\spn!{\genQ^{\spindown,\spinupdown\spinupdown},\genJL^-,\genJR^-}
}
\polyspc{\looppar,\looppar^{-1}}
\\
\alg{g}^+ &=
(\looppar-\trigq^{2})^{-1}\genRLt\polyspc!{(\looppar-\trigq^{2})^{-1}}
\oplus(\looppar-\trigq^{-2})^{-1}\genRLt\polyspc!{(\looppar-\trigq^{-2})^{-1}}
\oplus \spn!{\genRCL[\cclabel{+}],\genRCP[\cclabel{+}]}
\\;
\oplus
\spn!{(\looppar-\trigq^{-2})\genRPt,\genQ,\genJL,\genJR}
\polyspc!{(\looppar-\trigq^{2})^{-1},(\looppar-\trigq^{-2})^{-1}}
\>
In other words, the negative subalgebra $\alg{g}^-$
contains all generators $\genRLtm,\genRPt,\genQ,\genJL,\genJR$
whose prefactors are Laurent polynomials in $\looppar$.
To accommodate the case $\rnormP\neq 0$, we must introduce
a mixing between $\genRLt$ and $\genRPt$
deforming the mixing relation \eqref{eq:Lmix}
\[
a(\looppar)\genRLtm
\defeq
a(\looppar)\genRLt
+\frac{2\rnormRP}{\rnormRL}\brk!{2\looppar a(\looppar)-(\looppar+\looppar^{-1})
a(\looppar^{-1})}\genRPt
\]
Furthermore, the prefactor for the generators
$\genRPt,\genQ^{\spinup},\genJL^+,\genJR^+$ must have a zero at
$\looppar=\trigq^{2}$ while the prefactor for the generators
$\genRPt,\genQ^{\spindown},\genJL^-,\genJR^-$ must have a zero at
$\looppar=\trigq^{-2}$.
Consequently, the prefactor for $\genRPt$ must have
zeros at both $\looppar=\trigq^{\pm 2}$.
The positive subalgebra $\alg{g}^+$
contains all generators $\genRLt,\genRPt,\genQ,\genJL,\genJR$
whose prefactors can have poles at $\looppar=\trigq^{\pm 2}$.
In addition, $\genRLt$ must have a zero at $\looppar=\infty$
and $\genRPt$ can have a pole at $\looppar=\infty$.
The four-point affine bialgebra is the classical double
modulo the identification of elements $\genJL^0$ and $\genJR^0$
appearing in both subalgebras
\[
\alg{g}=\lfrac!{\algdbl(\alg{g}^\pm)}{\spn{\genJL^{0,+}
-\genJL^{0,-},\genJR^{0,+}-\genJR^{0,-}}}
\]
%

\section{Conclusions and Outlook}
\label{sec:conclusions}

In this work, we have discussed and reformulated
the complex analytical structure for
the classical affine Lie bialgebra relevant
to the one-dimensional Hubbard model and to the AdS/CFT worldsheet scattering
problem \cite{Beisert:2007ty,Beisert:2010kk,Beisert:2022vnc,Beisert:2024rke}.
This affine algebra is special in that its analytical structure
displays additional singularity points next to the two common expansion points.

To this end, we have applied a Möbius transformation $\shiftpar\to\looppar$ to
the loop variable which relocates the additional poles at
$\shiftpar=\pm \redpar$,
generated by the reduction introduced in \cite{Beisert:2024rke},
to the standard locations $\looppar=0,\infty$ for the expansion of the affine
algebra.
In the transformed form, the puzzling feature of additional singularities
naturally disappears, and the resulting algebra takes on a more standard form.
Moreover, the algebra and coalgebra relations,
in particular those involving the affine derivation and central charge,
become more transparent and more homogeneous.
For instance, the new presentation for the rational and trigonometric cases
of the affine algebra derived in \secref{sec:rational} and
\secref{sec:trigonometric},
respectively, now deviate only in details from each other.
This helps in exposing the relation between the affine structure and
the non-difference form of the r-matrix more clearly.

In \secref{sec:analysis} we have discussed various aspects and details of the
new formulation
related to the affine derivation and central charge.
While the new formulation provides more rigorous foundations for the algebra,
the conventional basis of generators for the affine algebra
still does not accommodate the affine derivation and central charge
in a way that is compatible with the quasi-triangular structure.
Hence, we develop a multi-point formulation for the algebra with a
Krichever--Novikov basis that takes into account both the original expansion
point(s) and the additional singularities.
In this new basis, the quasi-triangular Lie bialgebra
properly splits in a triangular way into positive and negative modes
on which the algebra and coalgebra relations can now close.

\medskip

The main motivation for this work is to improve the starting point towards
quantisation of the bialgebra
into a quasi-triangular Hopf algebra.
The additional poles of the old parametrisation
were previously addressed by expanding them in power series around
$\shiftpar=\infty,0$ thus complicating the description of the reduced
generators.
A common feature in quantum affine algebras is that the evaluation parameter
gets shifted by discrete multiples of $\hbar$ in various expressions.
This effect clearly relates to the affine derivation, which naturally generates
such shifts.
Consequently, the additional singularities will come along with their shifted
images in the quantum algebra, and new branch cuts can form between these.
In particular, the superalgebra representation theory is known to
turn the additional singularities into square-root branch points.
Altogether, the connectivity structure of the underlying quantum Riemann surface
becomes very involved, see \cite{Dorey:2007xn} for a discussion.
Given that the expressions for Yangian and quantum affine algebras generators
at generic levels are very cumbersome,
the new parametrisation is attractive from the practical standpoint.
In particular, the present formulation seems ideally suited for understanding
how the affine derivation should be incorporated at the quantum level.
Here, the Krichever--Novikov basis of generators can come in as an essential
tool which may even incorporate the extensions of the Riemann surfaces imposed
by quantisation.
It will therefore be very interesting to lift the present
construction to the Yangian and quantum affine algebras and to clarify whether
this alternative affine structure can be implemented directly
on the universal R-matrix.
For instance, the Krichever--Novikov basis might
lead to a suitable prescription for writing the universal R-matrix
as an ordered product of (q-)exponentiated pairs of basis elements.
We hope that the results of this work
will help make these questions more tractable.

The application of multi-point affine algebras may also provide
a new perspective on more general integrable systems.
For instance, reflection matrices may depend on the evaluation parameter
\cite{Cherednik:1984vvp,Sklyanin:1988yz}
through a factor $(\shiftpar\pm w)/(\shiftpar\mp w)$,
in which case they introduce new poles at $\shiftpar=\pm w$
to the analytical structure of the algebra;
a formal treatment of these poles requires an extension of the basis of
meromorphic functions.
Furthermore, one may twist a given classical r-matrix by generators
depending on the evaluation parameters.
The simplest example of such a twist of the rational $\alg{sl}(2)$ r-matrix
uses terms of the form $\genJ^+\wedge (\shiftpar^{-1}\genJ^-)$.
This term effectively exchanges
the pair of elements $\genJ^+$ and $\shiftpar^{-1}\genJ^-$
between the positive and negative subalgebras $\alg{g}^\pm$
in the ordinary two-point expansion of the algebra around $\shiftpar=\infty,0$.
Considering the three-point algebra expansion instead
suggests that an analogous exchange of basis elements \eqref{eq:basis-kn-3pt}
between $\alg{g}^\pm$ may work as well.
This leads to twists with meromorphic functional dependencies
\[
\genJ^+\wedge \frac{\genJ^-}{\shiftpar-w}
\eqnjoin{\text{and}}
\genJ^+\wedge \frac{\genJ^-}{\shiftpar-w}
+
\brk!{(\shiftpar-w)\genJ^+}\wedge \frac{\genJ^-}{(\shiftpar-w)(\shiftpar-w')}
\]
Indeed, these twists satisfy the classical Yang--Baxter equation,
and it may be interesting to investigate further such twists
and to discuss their connection with quasi-triangular multi-point affine
bialgebras.

\subsection*{Acknowledgements}

The work of NB and EI was supported by the Swiss National Science Foundation
through the NCCR SwissMAP.

\appendix

\section{Dual Jacobi Identity}
\label{app:dual-jacobi}

Since the reduced rational r-matrix \eqref{eq:r-rat-red} is anti-symmetric and
satisfies the cYBE, the cobrackets \eqref{eq:coalgebra1}--\eqref{eq:coalgebra4}
must satisfy the dual Jacobi identity.
For illustration, we discuss how one sample part of the dual Jacobi identity
is satisfied.
This calculation shows explicitly
that central charges sourced at $\looppar=0,\infty$
lead to issues with the dual Lie algebra structure.

The cobracket was defined in \eqref{eq:cobra} as a map
$\cobra \colon \alg{g} \to \alg{g} \otimes \alg{g}$.
Its dual, the Lie bracket, can be defined similarly as a map
$\cobra^\ast \colon \alg{g} \otimes \alg{g} \to \alg{g}$ with
$\cobra^\ast(X \otimes Y) = \liebr{X}{Y}$.
Then, the Jacobi identity can be written as
\[
\cobra^\ast \brk[s]!{ (\cobra^\ast \otimes 1)
\brk!{ \operatorname{Alt} (X \otimes Y \otimes Z)}}
= 0
\],
where `$\operatorname{Alt}$' is the projector onto the alternating part
of a (three-site) tensor product.
Subsequently, the dual Jacobi identity is
\[@{eq:dual-jacobi}
\operatorname{Alt} \brk[s]!{(\cobra \otimes 1)\brk!{\cobra(X)}}
= 0
\]

For the dual Jacobi identity of $a(\looppar)\genRDLt[]$ we find
(focusing only on the term proportional to
$\genQ \otimes \genQ \otimes \genRPt$):
\<
\&:
\operatorname{Alt} \brk[s]^{2}{
    (\cobra \otimes 1)
    \brk!{
        \cobra\brk!{a(\looppar)\genRDLt[]}
        }}
\\=
\operatorname{Alt} \brk[s]^4[
    (\cobra \otimes 1)
     \brk^4[
\frac{1}{\looppar_1-\looppar_2}
\brk[s]*{\frac{(\looppar_1-1)^2}{\looppar_1}a(\looppar_1)
-\frac{(\looppar_2-1)^2}{\looppar_2}a(\looppar_2)}
\\;\qquad\cdot
\brk[s]^3[
-\frac{\rnormL}{8\redpar^2}
\frac{\looppar_1\looppar_2+1}{\looppar_1\looppar_2-1}
\brk!{(\looppar_2-1)^2 \genRLt\otimes\genRPt
-(\looppar_1-1)^2 \genRPt\otimes\genRLt}
\\;\qquad\qquad
+\frac{\rnormP}{4\redpar^2}(\looppar_1-1)(\looppar_2-1)(\looppar_1-\looppar_2)
\genRPt\otimes\genRPt
\\;\qquad\qquad
-\frac{\rnormL \mref }{32 \redpar^2} (\looppar_1-1)(\looppar_2-1)
(\redrot^a  \sleps \slpauli_a )_{jk}\sleps_{lm}
\sleps_{rp}  \genQ^{j,lr}\otimes \genQ^{k,mp}
\brk]
\brk]\brk]
\\;
+
\operatorname{Alt} \brk[s]^4[
    (\cobra \otimes 1)
     \brk^4[
(\looppar_1-1)^2(\looppar_2-1)^2
\frac{a(\looppar_1)-a(\looppar_2)}{(\looppar_1-\looppar_2)^2}
\\;\qquad\cdot
\brk[s]^3[
-\frac{\rnormL}{4\redpar^2}
\frac{1}{\looppar_1\looppar_2-1}
\brk!{(\looppar_2^2-1) \genRLt\otimes\genRPt
+(\looppar_1^2-1) \genRPt\otimes\genRLt}
\\;\qquad\qquad
-\frac{\rnormP}{2\redpar^2}
(\looppar_1+\looppar_2)
\genRPt\otimes\genRPt
- \frac{\rnormL \mref}{4 \redpar^2}
\brk{\casQQ - \casJJ_\algleft + \casJJ_\algright}
\brk]
\brk]\brk]
\\=
-\frac{\rnormL^2 \mref}{64\redpar^3}
\operatorname{Alt} \brk[s]^4[
\frac{(\looppar_1-1)(\looppar_2-1)(\looppar_3-1)^2}
{(\looppar_1-\looppar_2)\looppar_3}
\\;\qquad\cdot
\brk^4[
   - \frac{(\looppar_1+1)(\looppar_1-1)^2
   (\looppar_1+\looppar_3)\looppar_3 a(\looppar_1)}
{\looppar_1(\looppar_1-\looppar_3)^2}
\\;\qquad\qquad
    + \frac{(\looppar_1+1)(\looppar_3-1)(\looppar_1-3\looppar_3(1-\looppar_1)
    - \looppar_3^2)a(\looppar_3)}{(\looppar_1-\looppar_3)^2}
\\;\qquad\qquad
    + \frac{(\looppar_2+1)(\looppar_2-1)^2(\looppar_2+\looppar_3)
    \looppar_3 a(\looppar_2)}
{\looppar_2(\looppar_2-\looppar_3)^2}
\\;\qquad\qquad
    - \frac{(\looppar_2+1)(\looppar_3-1)(\looppar_2-3\looppar_3(1-\looppar_2)
    -\looppar_3^2) a(\looppar_3)}
{(\looppar_2-\looppar_3)^2}
    \brk]
\\;\qquad\cdot
    (\sleps \slpauli_0)_{jk} \sleps_{lm} \sleps_{rp}
\genQ^{j,lr}\otimes \genQ^{k,mp} \otimes \genRPt
\brk]
\\;
-
\frac{\rnormL^2 \mref }{128 \redpar^3}
\operatorname{Alt} \brk[s]^4[\frac{ (\looppar_1-1)(\looppar_2-1)(1-\looppar_3)}
{\looppar_1-\looppar_2}
\\;\qquad\cdot
\brk^4[
\frac{\looppar_1-1}{\looppar_1-\looppar_3}
\brk[s]*{\frac{(\looppar_1-1)^2 a(\looppar_1)}{\looppar_1}
-\frac{(\looppar_3-1)^2 a(\looppar_3)}{\looppar_3}}
\\;\qquad\qquad
-
\frac{\looppar_2-1}{\looppar_2-\looppar_3}
\brk[s]*{\frac{(\looppar_2-1)^2a(\looppar_2)}{\looppar_2}
-\frac{(\looppar_3-1)^2a(\looppar_3)}{\looppar_3}}
\brk]
\\;\qquad\cdot
(\sleps \slpauli_0 )_{jk} \sleps_{lm} \sleps_{rp}
\brk[s]*{
(\looppar_2-1)\genQ^{j,lr}\otimes \genRPt \otimes \genQ^{k,mp}
-
(\looppar_1-1)\genRPt \otimes \genQ^{j,lr}\otimes  \genQ^{k,mp}
}
\brk]
\\;
 -
\frac{\rnormL^2 \mref }{32 \redpar^3}
\operatorname{Alt} \brk[s]^4[\frac{1}{\looppar_1-\looppar_2}
\\;\qquad\cdot
\brk^4[
\frac{(\looppar_1+1)(\looppar_1-1)^2(\looppar_2-1)(\looppar_3-1)^2
(a(\looppar_1)-a(\looppar_3))}
{(\looppar_1-\looppar_3)^3}
\\;\qquad\qquad
-
\frac{(\looppar_2+1)(\looppar_1-1)(\looppar_2-1)^2(\looppar_3-1)^2
(a(\looppar_2)-a(\looppar_3))}
{(\looppar_2-\looppar_3)^3}
\brk]
\\;\qquad\cdot
(\sleps \slpauli_0)_{jk} \sleps_{lm} \sleps_{rp}
\brk*{
(\looppar_2-1) \genQ^{j,lr}\otimes \genRPt \otimes \genQ^{k,mp}
-(\looppar_1-1) \genRPt \otimes \genQ^{j,lr}\otimes  \genQ^{k,mp}
}
\brk]
\\;
+ \ldots
\\=
0
\>
The cancellations happen upon working out the anti-symmetrisations.
The remaining terms are similar to those spelled out here.
We note that the dual Jacobi identity apparently holds
even for non-constant prefactor functions $a(\looppar)$;
however, the (multiple) poles in $1/(\looppar_j-\looppar_k)$
lead to residual (distributional) terms,
see \secref{sec:rational-extra} and \secref{sec:Derivations}.

Let us now consider the terms in the dual Jacobi identity of $\genRDLt[]$
proportional to $\genRLt\otimes\genRPt \otimes \genRCP[\cclabel{\kappa}]$:
\<
\&:
\operatorname{Alt}\brk[s]!{
(\cobra\otimes1)\brk!{\cobra\brk{\genRDLt[]}}
}
\\=
\operatorname{Alt}\brk[s]^3[
\frac{\rnormL^2}{16\redpar^3}
 \frac{1}{2\pi\iunit} \oint_{\cclabel{\kappa}}  \diff{\looppar'}
\brk*{
\frac{\looppar'-1}{\looppar_2-\looppar'}
\frac{3\looppar_1\looppar'^2+\looppar_1\looppar'+\looppar'+3}{\looppar_1\looppar'^2}
+
\frac{(\looppar'-1)^2}{\looppar_1-\looppar'}
\frac{\looppar_2\looppar'-1}{\looppar_2\looppar'^2}
}
\\;\qquad\cdot
(\looppar_2-1)^2\genRLt\otimes\genRPt \otimes \genRCP[\cclabel{\kappa}]
\brk]
+\ldots
\\=
\operatorname{Alt}\brk[s]*{
\frac{\rnormL^2}{16\redpar^3}
\brk!{-4+1/\looppar_1+\looppar_1-1/\looppar_2+3\looppar_2}
(\looppar_2-1)^2\genRLt\otimes\genRPt \otimes \genRCP[\cclabel{\infty}]}
\\;
+\operatorname{Alt}\brk[s]*{
\frac{\rnormL^2}{16\redpar^3}
\frac{4-1/\looppar_1-\looppar_1-3/\looppar_2+\looppar_2}{\looppar_1\looppar_2}
(\looppar_2-1)^2\genRLt\otimes\genRPt \otimes \genRCP[\cclabel{0}]}+\ldots
\>
Here we observe non-vanishing contributions involving the central charges
located at $\looppar=\infty,0$ due to residues at the corresponding locations.
These terms can only be avoided by dropping the central charges
$\genRCP[\cclabel{\infty}]$ and $\genRCP[\cclabel{0}]$
from the bialgebra.

\section{Krichever--Novikov Algebras}
\label{app:krichever-novikov}

In \secref{sec:3pt-algebra} we employed the Krichever--Novikov basis for
a three-point algebra to establish a triangular decomposition for
a current or affine Lie algebra admitting more than two punctures on a sphere.
In this appendix, we briefly review the construction of the Krichever--Novikov
algebras and discuss their implications for the bialgebra constructions.

In general, Krichever--Novikov type algebras broadly reference various
algebraic structures that are related to meromorphic forms and vector fields
on a Riemann surface \cite{Schlichenmaier+2014}. One starts by considering a
space of global differential forms
$\KNalg = \bigoplus_\lambda \KNalg^\lambda$
on a Riemann surface $\RSurf$ of genus $g$ with $N$ punctures,
where $\lambda$ is the degree of the forms.
\unskip\footnote{We will effectively only consider functions with $\lambda=0$
and their derivatives with $\lambda=1$.}
There is a natural associative product `$\cdot$' on this space
that maps $\KNalg^\lambda \times \KNalg^\mu \to \KNalg^{\lambda + \mu}$.
In the local coordinates $\looppar$, the product reads
\[
f(\looppar) \diff{\looppar^\lambda} \cdot g(\looppar) \diff{\looppar^\mu}
=
f(\looppar) g(\looppar) \diff{\looppar^{\lambda + \mu}}
\]
For our purposes, the most relevant part of this space is the genus $g=0$
associative algebra
$\KNalg^0 \eqdef \KNfuns$, which is the algebra of all meromorphic functions
with poles at the $N$ punctures on a sphere.
The Krichever--Novikov structure on this space identifies with a particular
basis that makes some features transparent. The choice of the basis relies on
a splitting of the
distinguished $N$ points into “in” points $I = \set{p_1, \ldots, p_K}$
and “out” points $O = \set{q_1, \ldots, q_M}$ with $K + M = N$.
Once the splitting is fixed, one can accordingly split the spaces
$\KNalg^\lambda$ into homogeneous subspaces
$\KNalg^\lambda = \bigoplus_{n \in \Integer} \KNalg^\lambda_n$ each of
dimension $K$, such that the space of meromorphic forms admits an
almost-grading.
There exists a constant $R$ such that
\[@{eq:KN-almost-grading}
\KNalg^\lambda_n \cdot \KNalg^\mu_m \subseteq
\bigoplus_{k=n+m}^{n+m+R} \KNalg^{\lambda + \mu}_{k}
\]
This also induces a triangular decomposition into positive modes given by
$\KNalg^\lambda_{n > 0}$, negative modes $\KNalg^\lambda_{n < -R}$ and
zero modes $\KNalg^\lambda_{-R \le n \le 0}$. Notice that in the
conventional loop algebra case (which corresponds to a two-point algebra of
zero-forms $\KNfuns$ on a sphere) $R=0$. Additionally, this splitting of points
identifies a special
closed curve on the surface, the separating cycle $C_{\text{S}}$,
which encircles all the “in” points and excludes all the “out” points.
This cycle is then used to define the Krichever--Novikov pairing
as the non-degenerate bilinear form
\[@{eq:KN-pairing}
\knpair{f}{g} \defeq \frac{1}{2\pi\iunit}\oint_{C_{\text{S}}} f \cdot g
\]

There exists a prescription to construct a basis $\KNbas{\lambda}{n}{p}$ for
$\KNalg^\lambda_n$, where $p \in I$. It is determined up to a
multiplicative factor by fixing the order of the pole (or zero)
at each of the distinguished points. Namely, we require
\[
\ord_{p_i} \KNbas{\lambda}{n}{p_j}= n + 1 - \lambda - \delta_{i,j}
\],
and some more complicated conditions at the “out” points,
which ultimately make the Riemann--Roch theorem satisfied
(see \cite{Schlichenmaier+2014} for more details).
In the case of a single “out” point, we require that
$\ord_{q_1} \KNbas{\lambda}{n}{p_j} = - K(n+1-\lambda) + (2 \lambda -1)(g - 1)$.
If we additionally require that in local coordinates $\looppar$ in the
vicinity of $p_i$, the basis forms behave as
\[@{eq:kn-bas-normalisation}
\KNbas{\lambda}{n}{p_i}(\looppar) \sim (\looppar - p_i)^{n-\lambda}
\],
then \eqref{eq:KN-pairing} establishes the duality relation
\[
\knpair{\KNbas{\lambda}{n}{p}}{\KNbas{1-\lambda}{m}{q}}
= \delta_{n,-m} \delta_{p,q}
\]

In the context of Lie algebras, the Krichever--Novikov algebras allow us to
promote a given finite-dimensional Lie algebra $\alg{g}$ to an affine
algebra $\alg{g} \otimes \KNfuns$ that inherits the useful structures discussed
above.
In particular, the pairing $\knpair{\cdot}{\cdot}$ induces a non-degenerate
bilinear form on $\KNfuns$
\[@{eq:knbiform}
\biform{f}{g} = \knpair{f}{g \cdot \omega}
\],
for $f, g \in \KNfuns$ and some $\omega \in \KNalg^{1}$, with the divisor
of $\omega$ completely supported by the $N$ punctures. The latter is required in
order to ensure that the map
$\KNfuns \mapsto \KNalg^{1}: f \mapsto f \cdot \omega$  is an isomorphism.
We can further
extend to a bilinear form on the affine Lie algebra as
\[
\biform{x\otimes f}{y \otimes g} =
\kappa(x,y) \biform{f}{g}
\],
where $x, y \in \alg{g}$ and $\kappa$ is the Killing form on $\alg{g}$
(we assume that it exists, e.g.\ $\alg{g}$ is simple).
If we now restrict to a Riemann sphere $g=0$, the resulting affine algebra
is an $N$-point algebra discussed in \secref{sec:3pt-algebra} and
\secref{sec:4p-trig}.
The precise choice of $\omega$ will be important in what follows in view of the
classical r-matrix discussed next.

In this work, we are motivated to use the Krichever--Novikov structures
because they provide a means to express the quasi-triangular bialgebra within
an $N$-point algebra. To this end, let us consider the following
Krichever--Novikov basis for three- and four-point algebras.
For a three-point algebra
with $I=\set{p_1,p_2}$ and $O = \set{q_1}$, we fix a basis on $\KNfuns$ as
\unskip\footnote{A puncture at $\looppar = \infty$ is realised by dropping
the factors involving it. An adjustment of the numerical prefactor may be
required to make the following relations work out literally.}
\footnote{Notice that the normalisation of these basis functions differs from
the standard one \eqref{eq:kn-bas-normalisation}.}
\[
\KNAbas{n}{p_1} = \frac{(p_2 - p_1)^n}{(p_1 - q_1)^{2n}}
\frac{(\looppar -p_1)^n (\looppar - p_2)^{n+1}}{(\looppar - q_1)^{2n+1}}
\&
\KNAbas{n}{p_2} = \frac{(p_1 - p_2)^n}{(p_2 - q_1)^{2n}}
\frac{(\looppar -p_2)^n (\looppar - p_1)^{n+1}}{(\looppar - q_1)^{2n+1}}
\]
For a four-point algebra with the splitting $I=\set{p_1,p_2}$ and
$O = \set{q_1, q_2}$ we choose
\unskip\footnote{Identifying $q_1=q_2=q$ yields the above three-point algebra.}
\<
\KNAbas{n}{p_1}
\&=
\frac{(p_2 - p_1)^n}{(p_1 - q_1)^n(p_1-q_2)^n}
\frac{(\looppar -p_1)^n (\looppar - p_2)^{n+1}}
{(\looppar - q_1)^{n} (\looppar - q_2)^{n+1}}
\\
\KNAbas{n}{p_2} \&=
\frac{(p_1 - p_2)^n}{(p_2 - q_1)^n(p_2 - q_2)^n}
\frac{(\looppar -p_2)^n (\looppar - p_1)^{n+1}}
{(\looppar - q_1)^{n} (\looppar - q_2)^{n+1}}
\>,
and with the splitting $I=\set{p_1,p_2,p_3}$ and $O = \set{q_1}$ we write
\<
\KNAbas{n}{p_1}
\&=
\frac{(p_2 - p_1)^n(p_1 - p_3)^n}{(p_1 - q_1)^{3n}}
\frac{(\looppar -p_1)^n (\looppar - p_2)^{n+1} (\looppar - p_3)^{n+1}}
{(\looppar - q_1)^{3n + 2}}
\\
\KNAbas{n}{p_2} \&=
\frac{(p_3 - p_2)^n(p_2 - p_1)^n}{(p_2 - q_1)^{3n}}
\frac{(\looppar -p_2)^n (\looppar - p_3)^{n+1} (\looppar - p_1)^{n+1}}
{(\looppar - q_1)^{3n + 2}}
\\
\KNAbas{n}{p_3} \&=
\frac{(p_1 - p_3)^n(p_3 - p_2)^n}{(p_3 - q_1)^{3n}}
\frac{(\looppar -p_3)^n (\looppar - p_1)^{n+1} (\looppar - p_2)^{n+1}}
{(\looppar - q_1)^{3n + 2}}
\>.
In all of these cases, the functional prefactor of the rational r-matrix
in \eqref{eq:r-rat-red} can be written as the (geometric) series
\[
\frac{(\looppar_1 - q_1)(\looppar_2 - q_1)}{\looppar_1 - \looppar_2}
=
\sum_{n = 0}^\infty \sum_{p \in I} \KNAbas{n}{p}(\looppar_1)
\KNAbas{-n-1}{p}(\looppar_2)
\]
Therefore, the rational r-matrix for an $N$-point algebra on a
simple Lie algebra $\alg{g}$ is naturally expressed within the $N$-point algebra
as
\[@{eq:kn_r_mat_rat}
\rmat_{\alg{g}}(\looppar_1, \looppar_2)
=
\sum_{n = 0}^\infty \sum_{p \in I} \KNAbas{n}{p}(\looppar_1)
\KNAbas{-n-1}{p}(\looppar_2) \. \casJJ_{\alg{g}}
\],
where $\casJJ_{\alg{g}}$ denotes the quadratic invariant for $\alg{g}$,
i.e.\ the inverse of the Killing form $\kappa$.
Importantly, the almost-graded structure \eqref{eq:KN-almost-grading} of
the Krichever--Novikov algebra together with the symmetric bilinear form
\eqref{eq:knbiform} ensure that the classical Yang--Baxter equation holds via the
Manin triple: For a genus-zero algebra of meromorphic functions the constant
$R$ in the almost-grading relation \eqref{eq:KN-almost-grading}
is guaranteed to be $\le 1$. Therefore, the subspaces
\[
\KNfuns_+ = \bigoplus_{n \ge 0} \KNalg^0_{n}
\&
\KNfuns_- = \bigoplus_{n < 0} \KNalg^0_{n}
\]~
are subalgebras. Moreover, in each of these cases
we can fix the one-form $\omega$ in \eqref{eq:knbiform} to be
\[
\omega = - \frac{\der \looppar}{\brk{\looppar - q_1}^2}
\],
such that the subalgebras $\KNfuns_\pm$ are isotropic w.r.t.\ this
non-degenerate form. Thus, the triple
\[
\brk{\alg{g} \otimes \KNfuns, \alg{g} \otimes \KNfuns_+, \alg{g}
\otimes \KNfuns_-}
\]~
is a Manin triple and \eqref{eq:kn_r_mat_rat} satisfies the cYBE.
As we demonstrate in \secref{sec:4p-trig}, the Krichever--Novikov algebra can
also be used to
express a trigonometric r-matrix, although the expression is not as
concise and the choice of the symmetric bilinear form is different.

The Krichever--Novikov algebras can also be used to extend the affine algebra
with derivations.
While in the conventional case of affine algebras, the derivations form the Witt
algebra,
for Krichever--Novikov algebras, the derivations are generalised by
the vector fields $\KNders \defeq \KNalg^{-1}$. However, the invariant bilinear
form \eqref{eq:knbiform} imposes a constraint on the form of the derivation:
the invariance of the form on the affine algebra translates to
\[
\biform!{\genD\brk{f}}{g} + \biform!{f}{\genD\brk{g}} = 0
\eqnjoin*{\implies}
\frac{1}{2\pi\iunit}\oint_{C_{\text{S}}} \genD\brk{h} \. \omega = 0
\],
where $f, g \in \KNfuns$ and $\genD \in \KNders$ for all $h \in \KNfuns$. Since
multiplication by $\omega$ gives an isomorphism between $\KNfuns$ and
$\KNalg^{1}$, we can write $\der h = k \. \omega$ for some $k \in \KNfuns$,
therefore
\[
\genD\brk{h} = k \. \omega(\genD) \eqdef k \. \phi
\eqnjoin*{\implies}
\genD\brk{h} \. \omega = \phi \. \der h
\]
Using integration by parts, we arrive at the conclusion
\[
\oint_{C_{\text{S}}} h \diff{\phi} \propto \knpair{h}{\der \phi} = 0
\],
which must hold for all $h \in \KNfuns$. Due to the non-degeneracy of the
Krichever--Novikov pairing, we conclude that
\[
\der \phi= 0
\eqnjoin*{\implies}
\genD = c\. \omega^{-1}
\]~
for some constant $c$. Therefore, the derivation (up to a
numerical multiplier) is fixed by the choice of the bilinear form, which stems
from the expansion of the classical r-matrix.

Another important feature of Krichever--Novikov algebras is that they admit
a central extension that respects the almost-grading. This can be used to
generate a central extension on the corresponding affine algebra
\[
\liebr{x \otimes f}{y \otimes g}
=
\liebr{x}{y} \otimes f  \cdot g + \kappa(x,y) \frac{1}{2 \pi \iunit}
\oint_{C_{\text{S}}} f \diff{g} \genC
\],
where $f, g \in \KNfuns$ and $x, y\in \alg{g}$. In principle,
one could replace the separating cycle by any non-singular closed contour,
placing a central charge $\genC_{\cclabel{p}}$ at each distinguished point $p$.
However, only  for the separating contour $C_{\text{S}}$, the centrally extended
algebra remains  almost-graded.  Moreover, other central charges spoil the
invariance of the bilinear  form \eqref{eq:knbiform}.
If we place an independent central charge at each puncture
\[
\liebr{x \otimes f}{y \otimes g}
=
\liebr{x}{y} \otimes f  \cdot g + \kappa(x,y) \frac{1}{2 \pi \iunit} \sum_{p}
\oint_{p} f \diff{g} \genC_{\cclabel{p}}
\],
the invariance of the bilinear form\footnote{
Here, we make an implicit assumption that the underlying Lie
algebra $\alg{g}$ is perfect $\liebr{\alg{g}}{\alg{g}} = \alg{g}$,
which is true for the algebras considered in this paper. Otherwise,
non-trivial pairings of $\genD$ with other generators than $\genC_{\cclabel{p}}$
may occur.
}
\[
\biform!{\liebr{x \otimes f}{y\otimes g}}{\genD}
= \biform!{x \otimes f}{\liebr{y \otimes g}{\genD}}
\],
then implies
\[
\sum_p \biform{\genC_{\cclabel{p}}}{\genD} \oint_{p} f \diff{g}
= - c \oint_{C_{\text{S}}} f \diff{g}
\]
Therefore, the central charges are only compatible with the bilinear form if
they come in a combination forming the separating curve, and their
normalisations are fixed by the same pairing with the derivation
\[
\biform{\genC_{\cclabel{p}}}{\genD} = -c
\]
%

\section{Triangular Decomposition}
\label{app:triang_decomp}

This appendix is devoted to an extended examination of the triangular
decomposition discussed in \secref{sec:triangular}. It serves two purposes:
On the one hand, we would like to provide the reader with a broader context in
which the triangular decomposition appears.
On the other hand, we also use it as an opportunity to demonstrate that the
reduction is compatible with the triangular decomposition in the sense that we
can relate the Poincaré superalgebra before and after the reduction at the level
of the triangular decomposition as well.

In the context of \secref{sec:triangular}, we understand the triangular
decomposition to be the construction of the bialgebra of interest as a
classical double.
That is, given a Lie bialgebra $\alg{b}$ and its double
$\algdbl\brk{\alg{b}} = \alg{b} \oplus \alg{b}^*$,
where $\alg{b}^*$ is the vector space dual to $\alg{b}$,
we have a quasi-triangular bialgebra structure
on $\algdbl\brk{\alg{b}}$.
The Lie bracket on $\algdbl\brk{\alg{b}}$ is induced by the conjugation of the
cobracket, i.e.\ for all $x \in \alg{b}$ and $\xi, \eta \in \alg{b}^*$
\[
\biform!{\cobra\brk{x}}{\xi \otimes \eta} = \biform!{x}{\liebr{\xi}{\eta}}
\],
where $\biform{\cdot}{\cdot}$ is the canonical pairing of the dual spaces.
The Lie bracket between the generators from the two subalgebras is given by
\[
\liebr{x}{\xi} = \ad_x^* \brk{\xi} - \ad_\xi^*\brk{x}
\],
where $\ad_x^*$ is the coadjoint action of $\alg{b}$
on $\alg{b}^*$, respectively $\ad_\xi^*$ is the coadjoint action of $\alg{b}^*$
on $\alg{b}$, defined by
\[
\biform!{\ad_\xi^*\brk{x}}{\eta} = - \biform!{x}{\liebr{\xi}{\eta}}
\&
\biform!{y}{\ad_x^*\brk{\xi}} = - \biform!{\liebr{x}{y}}{\xi}
\],
for all $x, y \in \alg{b}$ and $\xi, \eta \in \alg{b}^*$.
Notice that for this definition to be consistent, the one-cocycle condition for
the cobracket $\cobra$
\[@{eq:1-cocycle}
\cobra\brk!{\liebr{x}{y}} = \liebr!{x}{\cobra\brk{y}}
- \liebr!{y}{\cobra\brk{x}}
\]~
needs to be satisfied. For the quasi-triangular cobracket \eqref{eq:cobra} this
automatically holds.

The coalgebra structure on the double is induced by the classical r-matrix
$\rmat \in \alg{b} \otimes \alg{b}^*$  given by the identity operator on
$\alg{b}$.
The canonical pairing $\biform{\cdot}{\cdot}$ then determines an invariant
bilinear form on the double $D\brk{\alg{b}}$, w.r.t.\ which the subalgebras
$\alg{b}$ and $\alg{b}^*$ are isotropic.

\subsection{\texorpdfstring{$\looppar = \pm 1$}{zeta = pm 1} Expansion}

To begin with, we consider the triangular decomposition of the Poincaré
(super)algebra $\mspalg$. Subsequently, we will apply the reduction
to the decomposition to obtain the classical double construction for the
reduced algebra. For the expansion around $\looppar = \pm 1$,
we have to consider the space of meromorphic functions that only have
poles at $\pm 1$, i.e.\ the corresponding loop algebra is
$\alg{g} = \mspalg \otimes
\Complex\polyspc{(\looppar - 1)^{-1},(\looppar + 1)^{-1}}$.
The decomposition is then given by the splitting of the Laurent polynomials
into those that have a pole at $\looppar = +1$ vs.\ $\looppar = -1$.
Additionally, the level-0 elements (i.e.\ the constants) need to be attributed
to one of these two subspaces. Following the expansion \eqref{eq:geometric}, we
decompose the loop algebra as
\[@{eq:pm1_decomp}
\alg{g}^-
=
\frac{\looppar - 1}{\looppar + 1}
\mspalg\polyspc!{(\looppar + 1)^{-1}}
\&
\alg{g}^+
=
\mspalg\polyspc!{(\looppar - 1)^{-1}}
\]

We fix the basis for $\alg{g}^+$ to be
\unskip\footnote{The rest of generators in $\alg{h}$
split similarly between the positive and negative modes.}
\[
\genL_n^a \defeq \brk*{\frac{\looppar + 1}{\looppar - 1}}^n \genL^a
\&
\genP_n^a \defeq \brk*{\frac{\looppar + 1}{\looppar - 1}}^n \genP^a
\],
for $n \ge 0$.
The dual algebra $\alg{g}^- = (\alg{g}^+)^*$ is then spanned by the generators
\[@{eq:pm1-dual-untwisted}
\genL^*_{n,a} \defeq
\frac{\rnormL}{\redpar} \brk*{\frac{\looppar - 1}{\looppar + 1}}^{n+1}
\slinv_{ab}\genP^b
\&
\genP^*_{n,a} \defeq
\frac{1}{\redpar}\brk*{\frac{\looppar - 1}{\looppar + 1}}^{n+1}
\brk!{\rnormL \slinv_{ab}\genL^b + \rnormP \slinv_{ab}\genP^b}
\]

In the presence of the twist, $\rtwistL\neq 0$, the twist term
has constant prefactors in both factors and thus does not belong to
$\alg{g}^-\otimes\alg{g}^+$.
A proper triangular decomposition is achieved in this case by
deforming the generators for all prefactor functions
$a(\looppar) \in \Complex\polyspc{(\looppar - 1)^{-1},(\looppar + 1)^{-1}}$
\footnote{The map $a(\looppar)\mapsto a(\looppar^{-1})$ is the involution
of the loop algebra which fixes both punctures $\looppar=\pm 1$.
It therefore preserves both subspaces in \eqref{eq:pm1_decomp}.}
\<@{eq:pm1-twisted-generators}
a(\looppar)\genLtw^a \&=
a(\looppar)\genL^a
+ \frac{\rnormP}{2\rnormL} \brk!{a(\looppar)-a(\looppar^{-1})}\genP^a
\\;
+ \delta^{a,-} \frac{2\redpar\rtwistL}{\rnormL}\frac{\looppar+1}{\looppar-1}
\brk[s]*{
 a(\looppar)\genL^0
+ \frac{\rnormP}{2\rnormL} \brk!{a(\looppar)+a(\looppar^{-1})}\genP^0
}
\\
a(\looppar)\genPtw^a \&=
a(\looppar)\genP^a
+\delta^{a,0}\frac{\redpar\rtwistL}{\rnormL} \frac{\looppar+1}{\looppar-1}
a(\looppar)\genP^+
\>
In these variables, we write the basis elements
\[
\genLtw_n^a \defeq \brk*{\frac{\looppar + 1}{\looppar - 1}}^n \genLtw^a
\&
\genPtw_n^a \defeq \brk*{\frac{\looppar + 1}{\looppar - 1}}^n \genPtw^a
\],
for $n \ge 0$, which clearly span $\alg{g}^+$ as before.
The dual basis in the presence of the twist is given by
\[
\genLtw^*_{n,a} = \frac{\rnormL}{\redpar}
\brk*{\frac{\looppar - 1}{\looppar + 1}}^{n+1}\slinv_{ab}\genPtw^b
\&
\genPtw^*_{n,a} = \frac{\rnormL}{\redpar}
\brk*{\frac{\looppar - 1}{\looppar + 1}}^{n+1}\slinv_{ab}\genLtw^b
\]
The decomposition then reads
\[@{eq:pm1-decomp-twisted}
\alg{g}^-
=
\frac{\looppar - 1}{\looppar + 1}
\spn{\genLtw,\genPtw,\genQ,\genJL,\genJR}\polyspc!{(\looppar + 1)^{-1}}
\&
\alg{g}^+
=
\mspalg\polyspc!{(\looppar - 1)^{-1}}
\]
Although \eqref{eq:pm1-twisted-generators} applies to all levels alike,
the deformation moves elements between the two subspaces only at level $n=0$:
At all higher levels, it amounts to a change of basis within $\alg{g}^-$.
At $n=0$, the deformed generators coincide with the dual basis
\eqref{eq:pm1-dual-untwisted} shifted by the twist,
\[@{eq:pm1-decomp-twisted-n=0}
\genLtw^*_{0,a} \&= \genL^\ast_{0,a} - \delta_{a,0}\rtwistL \genP^+
\&
\genPtw^*_{0,a} \&= \genP^\ast_{0,a} + \delta_{a,+} \rtwistL \genL^0
\]
Notice that the deformation affects only terms with a simple pole at
$\looppar=-1$ in the $\genL^0$ and $\genP^+$ directions,
which does not spoil the closure of $\alg{g}^-$: the
Lie bracket between supercharges produces at least a second-order pole
within $\alg{g}^-$.

One can define an invariant non-degenerate bilinear form on $\alg{g}$
\[@{eq:biform_poincare}
\biform!{f(\looppar)\gen{X}}{g(\looppar)\gen{Y}}
=
2\redpar\.\kappa\brk{\gen{X},\gen{Y}}
\frac{1}{2\pi\iunit}\oint_{\cclabel{+}}
\frac{f(\looppar)\.g(\looppar)}{\brk{\looppar-1}^2}\der\looppar
\],
where $\gen{X},\gen{Y}\in\mspalg$, the contour $\cclabel{+}$ encircles
$\looppar=+1$ counter-clockwise, and $\kappa$ is the invariant form on $\mspalg$
which is inverse to the quadratic invariant appearing in the symmetric part
of the r-matrix \eqref{eq:r-rational}.
Explicitly, on the Lorentz and momentum generators, we have
\[
\kappa\brk{\genL^a,\genP^b} = \frac{1}{\rnormL}\slinv^{ab}
\&
\kappa\brk{\genL^a,\genL^b} = -\frac{\rnormP}{\rnormL^2}\slinv^{ab}
\&
\kappa\brk{\genP^a,\genP^b} = 0
\]
The form \eqref{eq:biform_poincare} depends only on the symmetric part
of the r-matrix. In particular, it does not depend on the twist parameter
$\rtwistL$.
Both $\alg{g}^+$ and $\alg{g}^-$ are isotropic with respect to it,
with or without the twist. Therefore, we obtain a proper triangular
decomposition of the bialgebra $\alg{g} = \alg{g}^+ \oplus \alg{g}^-$.

We now extend this decomposition of the loop algebra to the full affine algebra.
To this end, we allow for central charges $\genRCL[\cclabel{\kappa}]$ and
$\genRCP[\cclabel{\kappa}]$
at both poles $\looppar = \pm 1$. We assign the poles to the subalgebras
$\alg{g}^\pm$ as follows
\unskip\footnote{Different assignments (including mixing) of the central charges
are permitted.
This will only modify a prefactor of the dual derivations. The assignment
chosen here is best suited for the reduction.}
\<
\alg{g}^- &= \frac{\looppar - 1}{\looppar + 1}
\mspalg\polyspc!{(\looppar + 1)^{-1}}
 \oplus \spn!{\genRCL[\cclabel{-}],\genRCP[\cclabel{-}]}
\\
\alg{g}^+ &= \mspalg\polyspc!{(\looppar - 1)^{-1}} \oplus
\spn!{\genRCL[\cclabel{+}], \genRCP[\cclabel{+}]}
\>
Since $\genRCL[\cclabel{\pm}]$ and $\genRCP[\cclabel{\pm}]$ are central, the
bilinear form must vanish whenever a central element is paired with any other
generator. Therefore, in order
to keep the bilinear form non-degenerate, we need to complement the central
charges with four new generators: $\genRDL[\cclabel{+}]$ and
$\genRDP[\cclabel{+}]$, which we assign to $\alg{g}^-$, and
$\genRDL[\cclabel{-}]$ and $\genRDP[\cclabel{-}]$, which we assign to $
\alg{g}^+$.
These are precisely the derivations of the affine
algebra, which we can generally write as
\[
\genRDL[\cclabel{\pm}]
=
\Psi^\pm_{\genL}(\looppar) \genODL + \Phi^\pm_{\genL}(\looppar) \genODP
\&
\genRDP[\cclabel{\pm}]
=
\Psi^\pm_{\genP}(\looppar) \genODL + \Phi^\pm_{\genP}(\looppar) \genODP
\],
where $\Psi^\pm_{\genL, \genP}$ and $\Phi^\pm_{\genL, \genP}$
are some functions of $\looppar$. They can be fixed up to a multiplicative
factor by requiring that the action of the derivations closes on $\alg{g}^\pm$
and the bilinear form remains invariant in the presence of the derivations.
The normalisation follows from the duality constraint
\[
\biform!{\genRDL[\cclabel{\pm}]}{\genRCL[\cclabel{\pm}]} = 1
\&
\biform!{\genRDP[\cclabel{\pm}]}{\genRCP[\cclabel{\pm}]} = 1
\],
giving
\unskip\footnote{The resulting two derivations in $\alg{g}^-$ span the same
subspace within the full affine algebra as the other two derivations from
$\alg{g}^+$.
Therefore, the double algebra $D(\alg{g}^-)$ is slightly bigger than $\alg{g}$,
with the derivations being duplicated.
In order to fix this issue, we have to divide out the centre
spanned by $\genRDL[+] + \genRDP[+]$ and $\genRDL[-] + \genRDP[-]$.}
\[
\genRDL[\cclabel{\pm}] = \mp \frac{\brk{\looppar - 1}^2}{2 \redpar \looppar}
\rnormL \genODP
\&
\genRDP[\cclabel{\pm}] = \mp \frac{\brk{\looppar - 1}^2}{2 \redpar \looppar}
\brk!{\rnormL \genODL + \rnormP \genODP}
\],
which are precisely the derivations that we had to add to the classical
r-matrix in \secref{sec:rational-extra}. In the end, we obtain a triangular
decomposition  of the full affine bialgebra on the Poincaré Lie superalgebra.
In fact, this is the same decomposition as was given in \cite{Beisert:2024rke}.

Now, we can apply the reduction at the level of the triangular decomposition.
Let us start with the $\alg{g}^+$ subalgebra without the affine extension.
Considering $\alg{g}^+$ as a Lie algebra
only, the reduction consists in taking the subalgebra $\alg{s}_-$ spanned by
\[
\lambda\.\genLtw_{n+1}^0
+ \frac{\econst^{-\iunit\redang}}{2}\genLtw_n^+
+ \frac{\econst^{+\iunit\redang}}{2}\genLtw_n^-
\&
\genPtw_n^a
\],
where $\lambda \defeq 1-\frac{\econst^{\iunit\redang}\redpar\rtwistL}{\rnormL}$
for $n \ge 0$, and modding out the ideal $\alg{i}_-$ generated by
\[
\econst^{\iunit\redang}\genPtw_n^- - \econst^{-\iunit\redang}\genPtw_n^+
\&
\econst^{\iunit\redang}\genPtw_n^0 + \lambda\.\genPtw_{n+1}^+
\],
for $n \ge 0$. Notice that so far, all the operations are well-defined and
the quotient $\alg{s}_- / \alg{i}_-$ is a Lie algebra.

In order to lift this construction to the full bialgebra, notice that the
reduction produces the Lie bialgebra
$\brk{\alg{s}_+ / \alg{i}_+, \alg{s}_- / \alg{i}_-}$,
where $\alg{s}_+ = \alg{i}_-^\perp$ and $\alg{i}_+ = \alg{s}_-^\perp$.
The orthogonal
subspace is defined through the invariant bilinear form as
\[
\alg{i}_-^\perp = \setcond{ x \in \alg{g}^- }{\biform{x}{\alg{i}_-} = 0}
\&
\alg{s}_-^\perp = \setcond{ x \in \alg{g}^- }{\biform!{x}{\alg{s}_-} = 0}
\]
The resulting bialgebra is well-defined if $\alg{s}_+$ is a Lie subalgebra of
$\alg{g}^-$ and $\alg{i}_+$ is an algebra ideal in $\alg{s}_+$.
In order to ensure that these conditions are satisfied, we actually have to tune
the twist term: $\alg{s}_+ = \alg{i}_-^\perp$ is spanned by the generators
\[
\econst^{\iunit\redang}\genPtw^*_{n,+} + \econst^{-\iunit\redang}\genPtw^*_{n,-}
- \brk{1-\delta_{n,0}}\lambda\.\genPtw^*_{n-1,0}
\&
\genLtw^*_{n,a}
\],
for $n \ge 0$.
This is where setting
$\rtwistL=\econst^{-\iunit \redang}\lfrac{\rnormL}{\redpar}$ is necessary:
It turns $\alg{s}_+$ into a proper Lie subalgebra.

Finally, it is straightforward to verify that $\alg{i}_+ = \alg{s}_-^\perp$
forms a Lie algebra ideal of $\alg{s}_+$ spanned by
\[
\econst^{\iunit\redang}\genLtw^*_{n,+} - \econst^{-\iunit\redang}\genLtw^*_{n,-}
\&
\genLtw^*_{n,0}
\],
for $n \ge 0$. The resulting bialgebra is well-defined, and the
triangular decomposition schematically takes the form
\unskip\footnote{The reduction does not affect the other $\alg{sl}(2)$
subalgebras spanned by $\genJL^a$ and $\genJR^a$.
The presence of the $\looppar$-dependent matrix $\reddir(\looppar)$
in the supersymmetry algebra can be resolved by going to the eigenstate basis,
and the resulting twisted loop algebra certainly has a triangular structure.}
\<
\alg{g}^- \&=
\frac{\looppar - 1}{\looppar + 1}
\spn!{\brk{\looppar - 1}^{-1}\genRLt,\brk{\looppar - 1}
\genRPt,\genQ,\genJL,\genJR}
\polyspc!{\brk{\looppar + 1}^{-1}}
\\
\alg{g}^+ \&=
\spn!{\genRLt,\genRPt,\genQ,\genJL,\genJR}
\polyspc!{\brk{\looppar-1}^{-1}}
\>
To obtain the full form presented in \eqref{eq:g-g+-1}, we must
add central charges and derivations, which introduces a small complication.
The action of the derivation $\genODL$ does not close on the
subalgebras $\alg{g}^\pm$. In \eqref{eq:D-reduction}, this problem was solved by
adding some generators to the derivation. However, these terms are not parts
of the subalgebras $\alg{g}^\pm$: they have a pole at $\looppar = 0$.
This is exactly the problem encountered in \cite{Beisert:2024rke}.
There, it was resolved by promoting the space of polynomials in the loop
counting variable $\shiftpar$ to formal power series and expanding away the
undesired pole.
The same procedure can be applied in the case of the pole at $\looppar = 0$ by
expanding it around $\looppar = \pm 1$.
The resulting double bialgebra is exactly the one discussed in
\secref{sec:rational-extra}.
However, it would be beneficial to avoid resorting to infinite series due to
the pole at $\looppar = 0$. Next, we will thus consider the alternative
expansion which takes into account this pole.

\subsection{\texorpdfstring{$\looppar = 0, \infty$}{zeta = 0, infty} Expansion}

Alternatively, we can consider the decomposition for the expansion
around $\looppar = 0, \infty$. The construction is qualitatively the same as
in the previous case. The loop algebra splits into
\unskip\footnote{Notice that this is not a unique splitting for the same
bialgebra $\alg{g}$.
Alternatively, one could equivalently choose
$\alg{g}^- = \mspalg\polyspc{\looppar}$ and
$\alg{g}^+ = \mspalg\polyspc{\looppar^{-1}}$.}
\[
\alg{g}^-
=
\mspalg\polyspc{\looppar}
\&
\alg{g}^+ =
(1 - \looppar^{-1})\mspalg\polyspc{\looppar^{-1}}
\],
up to minor modifications at low levels similar to
\eqref{eq:pm1-decomp-twisted-n=0} to accommodate the twist.
The corresponding symmetric invariant bilinear form on $\alg{g}$ is
given by the same formula \eqref{eq:biform_poincare}
with the contour to encircle $\looppar = 0$ instead.
The central charges and derivations can be added in a similar fashion as
before. The central charges $\genRCL[\cclabel{0, \infty}]$ and
$\genRCP[\cclabel{0, \infty}]$ are now placed at $\looppar = 0, \infty$. The
derivations can be constructed out of $\genODL$ and $\genODP$ analogously.
Moreover, the reduction works exactly the same within the new triangular
decomposition.
The final reduced loop bialgebra splits as \eqref{eq:g-g+-2}
\<
\alg{g}^- \&=
\spn!{\genRLt,\genRPt,\genQ,\genJL,\genJR}\polyspc{\looppar}
\\
\alg{g}^+ \&=
\looppar^{-1}\spn!{\genRLt,(\looppar-1)^2\genRPt,
(\looppar-1)\genQ,(\looppar-1)\genJL,(\looppar-1)\genJR
}\polyspc{\looppar^{-1}}
\>

As discussed in the main text, adding the central extensions and
derivations turns out to be problematic.
Nevertheless, we concluded that the reduced Lie bialgebra in the novel
parametrisation still  admits the triangular decomposition, albeit with an
important caveat.
When we consider $\looppar = 0$ as
an expansion point, the affine derivation spoils the decomposition.
This problem could be resolved by resorting to the functions
$\sqrt{\looppar}\Complex\polyspc{\looppar, \looppar^{-1}}$.
While the square root allows the action of the derivation to close on
the subalgebras $\alg{g}^\pm$,
the new set of functions introduces
significant issues with the r-matrix and the cobrackets,
and we will not follow it further.
Instead, let us consider the option of unifying the
relevant singular points within a single multi-point algebra.

\subsection{Three-Point Algebra}

The problem of incompatibility of the affine derivation with the triangular
decomposition can be resolved by considering a three-point algebra
instead of the two-point algebras we have discussed so far.
In \secref{sec:3pt-algebra} we saw that it is indeed
possible and more natural to formulate the reduced quasi-triangular Lie
bialgebra in the setting of the three-point algebra
$\Complex\polyspc{\looppar, \looppar^{-1}, (\looppar-1)^{-1}}$.
In the following, let us demonstrate how the triangular decomposition works in
this case.

In \appref{app:krichever-novikov} we have seen that the Krichever--Novikov
algebras give a natural basis \eqref{eq:basis-kn-3pt} to construct the r-matrix
in the presence of multiple punctures as well as the associated Manin triple.
This provides us with the triangular decomposition for free:
The loop algebra $\alg{g} = \alg{h}[\looppar, \looppar^{-1}, (\looppar-1)^{-1}]$
splits into the subalgebras
\[
\alg{g}^- = \mspalg \otimes \KNfuns_- \simeq
(\looppar-1)\mspalg\polyspc{\looppar,\looppar^{-1}}
\&
\alg{g}^+ = \mspalg \otimes \KNfuns_+ \simeq
\mspalg\polyspc!{(\looppar-1)^{-1}}
\]

The Krichever--Novikov pairing \eqref{eq:KN-pairing}
induces the non-degenerate symmetric bilinear form \eqref{eq:knbiform}, with
respect to which the subalgebras are isotropic. Explicitly, the bilinear form
reads
\[@{eq:biform-3pt}
\biform!{\KNAbas{n}{i}}{\KNAbas{m}{j}} = \frac{1}{2 \pi \iunit} \oint_{\cclabel{1}}
\frac{\KNAbas{n}{i}(\looppar)\KNAbas{m}{j}(\looppar)}{(\looppar-1)^2}
\diff{\looppar} = \delta_{n+m=-1} \delta_{i=j}
\]
This bilinear form can be viewed as a duality relation between dual
algebras $\alg{g}^\pm$. We can verify that the duality provides a proper
cobracket on $\alg{g}^+$.
Therefore, this gives a valid triangular decomposition.

Moreover, as pointed out in \appref{app:krichever-novikov}, the affine extension
and the derivations naturally fit within this bialgebra. The bilinear
form \eqref{eq:biform-3pt}
fixes the functional form of the allowed derivations
\[
\genD \propto \omega^{-1} = (\looppar - 1)^2 \partial_\looppar
\]
The actual derivations $\genZDL$ and $\genZDP$ from \eqref{eq:der-repara}
exactly align with this form.

In conclusion, we obtain a valid Lie bialgebra $\alg{g}^+$ within the
three-point algebra framework, and its double $\algdbl(\alg{g}^+) = \alg{g}$ is
a three-point affine bialgebra with a quasi-triangular structure, whose r-matrix
is given by \eqref{eq:3pt-rmat-poincare}.
This bialgebra admits the pole at $\looppar = 0$, which makes the
reduction of the affine derivation well-defined within the space of Laurent
polynomials,
while it also accommodates derivations and central charges placed at
$\looppar = +1$.

\subsection{Reduction for Three-Point Algebra}
\label{sec:3ptreduction}

Finally, we spell out triangular decomposition
for the three-point Poincaré (super)algebra
and show that the reduction is compatible with it.
First, the triangular decomposition for the untwisted Poincaré superalgebra
with $\rnormP=0$ reads
\<
\alg{g}^- &= (\looppar-1)\mspalg\polyspc{\looppar,\looppar^{-1}}
\oplus \spn{\genZDL,\genZDP}
\\
\alg{g}^+ &= \mspalg\polyspc!{(\looppar-1)^{-1}}
\oplus \spn{\genZCL,\genZCP}
\>
In order to implement a non-trivial parameter $\rnormP\neq 0$
and the off-diagonal twist $\genL^0\wedge\genP^+$,
the basic generators must be deformed slightly
\unskip\footnote{The non-trivial terms involving $a(\looppar^{-1})$
essentially correspond to
(the absence of) a twist $\sim\genP^0\wedge\genP^+$,
which we discarded from the start, even though it is admissible.
The application of $a(\looppar^{-1})$ to the functional form of the r-matrix
acts similarly to the reflection equation.
It essentially projects out the bare rational r-matrix contribution,
but it leaves behind a residue from the twist contribution.}
\<
a(\looppar)\genLtw^a \&=
a(\looppar)\genL^a
+ \frac{\rnormP}{2\rnormL} \brk!{a(\looppar)-a(\looppar^{-1})}\genP^a
\\;
+ \delta^{a,-} \frac{2\redpar\rtwistL}{\rnormL}\frac{\looppar+1}{\looppar-1}
\brk[s]*{
 a(\looppar)\genL^0
+ \frac{\rnormP}{2\rnormL} \brk!{a(\looppar)+a(\looppar^{-1})}\genP^0
}
\\
a(\looppar)\genPtw^a \&=
a(\looppar)\genP^a
+\delta^{a,0}\frac{\redpar\rtwistL}{\rnormL} \frac{\looppar+1}{\looppar-1}
a(\looppar)\genP^+
\\
\genZDLtw \&=
\genZDL + \frac{\rnormP}{\rnormL} \genZDP
\>
A proper triangular decomposition is then achieved by
replacing the bare generators $\genL,\genP,\genZDL$
by the deformed ones $\genLtw,\genPtw,\genZDLtw$
\<
\alg{g}^- &= (\looppar-1)\spn{\genLtw,\genPtw,\genQ,\genJL,\genJR}
\polyspc{\looppar,\looppar^{-1}}
\oplus \spn{\genZDLtw,\genZDP}
\\
\alg{g}^+ &= \mspalg\polyspc!{(\looppar-1)^{-1}}
\oplus \spn{\genZCL,\genZCP}
\>
These two subalgebras close.
There is no concern for the subalgebra $\alg{g}^+$
because new poles appear only at $\looppar=1$ and they are naturally
accommodated.
The closure for $\alg{g}^-$ comes about by having two zeros at $\looppar=1$
from the argument prefactors which compensate at most one pole
to yield at least one zero in the result.
The only Lie bracket involving two poles is $\liebr{\genLtw^-}{\genPtw^0}$
(with suitable prefactor functions),
which, however, does close within $\alg{g}^-$ due to cancellations.
Lie brackets involving the affine derivation and central charges also close
without further ado.
Altogether, the r-matrix \eqref{eq:r-rat-repara}
with its affine extension
simplifies to
\[
\rmatmsp(\looppar_1,\looppar_2)
=
-\frac{\rnormL}{2\redpar}\frac{(\looppar_1-1)(\looppar_2-1)}
{\looppar_1-\looppar_2}\casJJtw_{\mspalg}
-\rnormL \brk[s]!{\genZDLtw\otimes\genZCP+\genZDP\otimes\genZCL}
\]
This shows that the two subalgebras are canonically dual to each other.

Let us now consider the reduction step.
The essential property of the reduction is that the two subalgebras
are never mixed in the relations.
For instance, the reduced generator $\genRLt$ in \eqref{eq:L-reduction}
with the adjustment $\genRLtm$ in \eqref{eq:Lmix} concerning $\rnormP\neq 0$
now takes the form
\<
a(\looppar)\genRLtm
&=
a(\looppar)\genRLt
+\frac{\rnormP}{\rnormL} \brk!{2\looppar a(\looppar)-(\looppar+\looppar^{-1})
a(\looppar^{-1})}\genRPt
\\
&=
\half (\looppar-1)a(\looppar)
\brk!{\econst^{-\iunit \redang} \genLtw^+ + \econst^{+\iunit \redang} \genLtw^-}
\>
Importantly, the combination $\genLtw^+$ and $\genLtw^-$ never mixes the
subalgebras.
Next, the reduction imposes two constraints \eqref{eq:P-reduction} by dividing
out algebra ideals.
One ideal is projected out by setting
\[
\econst^{+\iunit \redang}a(\looppar)\genPtw^-
- \econst^{-\iunit \redang}a(\looppar)\genPtw^+
\&=
\econst^{+\iunit \redang}a(\looppar)\genP^-
- \econst^{-\iunit \redang}a(\looppar)\genP^+
=
0
\]
The other ideal involves the modified generator $\genPtw^0$ and the central
charge
\[
a(\looppar)(\looppar-1)\genPtw^0
\&=
a(\looppar)(\looppar+1)\genP^+ + a(\looppar)(\looppar-1)\genP^0
=
\frac{\econst^{+\iunit \redang}}{2\pi \iunit}\oint_{\cclabel{+}}
\frac{\diff{\looppar}}{2\looppar}(\looppar+1)a(\looppar) \genZCP
\]
Again, the mixing is manifestly only between generators of the same subalgebra.
In particular, the central charge, which is associated to $\alg{g}^+$
can arise only for functions $a(\looppar)$ with poles at $\looppar=1$
which also belong to $\alg{g}^+$.
Finally, the reduction for the affine derivation \eqref{eq:D-reduction}
takes the form
\<
\genRDLt
\&=
\genZDL
+ \frac{(\looppar-1)^2}{8\redpar\looppar}
\brk!{\econst^{-\iunit \redang} \genL^+ - \econst^{+\iunit \redang} \genL^- }
\\=
\genZDLtw - \frac{\rnormP}{\rnormL} \genZDP
+ \frac{(\looppar-1)^2}{8\redpar\looppar}
\brk[s]*{ \econst^{-\iunit \redang} \genLtw^+  -\econst^{+\iunit \redang}
\genLtw^-  +2\frac{\looppar+1}{\looppar-1}\genLtw^0}
\>
Here, $\genZDLtw$ must belong to the subalgebra $\alg{g}^-$
which accommodates $a(\looppar)\genLtw$ with poles at $\looppar=0,\infty$.
In particular, this combination implies that
$a(\looppar)\genLtw$ with poles at $\looppar=0$
and $a(\looppar)\genLtw$ with poles $\looppar=\infty$
cannot be separated but must belong to the same subalgebra.

Altogether, this shows that the reduction procedure acts within
the two subalgebras $\alg{g}^\pm$ of the three-point algebra.
As a consequence, the triangular decomposition structure is manifestly
maintained.
Implementing the prefactors of the modified generators, it reads
\eqref{eq:3pt-triangular}
\<
\alg{g}^- &= \spn!{\genRLtm,(\looppar-1)^2\genRPt,(\looppar-1)\genQ,
(\looppar-1)\genJL,(\looppar-1)\genJR} \polyspc{\looppar,\looppar^{-1}}
\oplus \spn!{\genRDLt,\genRDPt}
\\
\alg{g}^+ &=
\spn!{(\looppar-1)^{-1}\genRLt,(\looppar-1)\genRPt,\genQ,\genJL,\genJR}
\polyspc!{(\looppar-1)^{-1}}
\oplus \spn!{\genRCL,\genRCP}
\>

\ifnum\version<\verarxiv\relax
\begin{bibtex}[\jobname]

  @article{Beisert:2007ty,
  author = "Beisert, Niklas and Spill, Fabian",
  title = "The Classical r-matrix of AdS/CFT and its Lie Bialgebra Structure",
  eprint = "0708.1762",
  archivePrefix = "arXiv",
  primaryClass = "hep-th",
  reportNumber = "AEI-2007-116, HU-EP-07-31",
  doi = "10.1007/s00220-008-0578-2",
  journal = "Commun. Math. Phys.",
  volume = "285",
  pages = "537--565",
  year = "2009"
}

@article{Beisert:2010kk,
  author = "Beisert, Niklas",
  title = "The Classical Trigonometric r-Matrix for the Quantum-Deformed Hubbard Chain",
  eprint = "1002.1097",
  archivePrefix = "arXiv",
  primaryClass = "math-ph",
  reportNumber = "AEI-2010-016",
  doi = "10.1088/1751-8113/44/26/265202",
  journal = "J. Phys. A",
  volume = "44",
  pages = "265202",
  year = "2011"
}

@article{Beisert:2024rke,
    author = "Beisert, Niklas and Im, Egor",
    title = "{Affine Classical Lie Bialgebras for AdS/CFT Integrability}",
    eprint = "2401.10327",
    archivePrefix = "arXiv",
    primaryClass = "hep-th",
    month = "1",
    year = "2024"
}

@article{Borsato:2017lpf,
  author = "Borsato, Riccardo and Torrielli, Alessandro",
  title = "{$q$-Poincaré supersymmetry in AdS$_5$ / CFT$_4$}",
  eprint = "1706.10265",
  archivePrefix = "arXiv",
  primaryClass = "hep-th",
  reportNumber = "DMUS-MP-17-07, NORDITA-2017-066",
  doi = "10.1016/j.nuclphysb.2018.01.017",
  journal = "Nucl. Phys. B",
  volume = "928",
  pages = "321--355",
  year = "2018"
}

@article{Bremner:1994,
  title={Generalized Affine Kac-Moody Lie Algebras Over Localizations of the Polynomial
  Ring in One Variable},
  volume={37},
  DOI={10.4153/CMB-1994-004-8},
  number={1},
  journal={Can. Math. Bull.},
  publisher={Cambridge University Press},
  author={Bremner, Murray},
  year={1994},
  pages={21–28}
}

@article{Bremner:1995,
  ISSN = {00029939, 10886826},
  doi = {10.2307/2160931},
  author = {Murray Bremner},
  journal = {Proc. Am. Math. Soc.},
  number = {7},
  pages = {1981--1989},
  publisher = {American Mathematical Society},
  title = {Four-Point Affine Lie Algebras},
  urldate = {2023-01-25},
  volume = {123},
  year = {1995}
}

@article{Hartwig:2007,
  title = {The Tetrahedron algebra, the Onsager algebra, and the sl$_2$ loop algebra},
  journal = {J. Alg.},
  volume = {308},
  number = {2},
  pages = {840-863},
  year = {2007},
  issn = {0021-8693},
  doi = {https://doi.org/10.1016/j.jalgebra.2006.09.011},
  author = {Brian Hartwig and Paul Terwilliger},
  eprint = "0511004",
  archivePrefix = "arXiv",
  primaryClass = "math-ph",
}

@article{Beisert:2022vnc,
    author = "Beisert, Niklas and Im, Egor",
    title = "{Classical Lie Bialgebras for AdS/CFT Integrability by Contraction and Reduction}",
    eprint = "2210.11150",
    archivePrefix = "arXiv",
    primaryClass = "hep-th",
    doi = "10.21468/SciPostPhys.14.6.157",
    journal = "SciPost Phys.",
    volume = "14",
    pages = "157",
    year = "2023"
}

@article{Stolin:1990,
 ISSN = {00255521, 19031807},
 URL = {http://www.jstor.org/stable/24492600},
 author = {A. Stolin},
 journal = {Math. Scand.},
 number = {1},
 pages = {57--80},
 publisher = {Mathematica Scandinavica},
 title = {On Rational Solutions of Yang-Baxter Equation for $\alg{sl}(n)$},
 urldate = {2023-05-12},
 volume = {69},
 year = {1991}
}

@Book{Chari:1994pz,
     author    = "Chari, V. and Pressley, A.",
     title     = "A guide to quantum groups",
     address   = "Cambridge, UK",
     year      = "1994",
     pages     = "651",
     publisher = "Cambridge University Press"
}

@book{Klimyk:1997eb,
    author = "Klimyk, A. and Schmudgen, K.",
    title = "{Quantum groups and their representations}",
    year = "1997",
    doi = "10.1007/978-3-642-60896-4",
    pages     = "552",
    publisher = "Springer Berlin, Heidelberg"
}

@article{Young:2007wd,
    author = "Young, C. A. S.",
    title = "{q-deformed supersymmetry and dynamic magnon representations}",
    eprint = "0704.2069",
    archivePrefix = "arXiv",
    primaryClass = "hep-th",
    reportNumber = "DCPT-07-09",
    doi = "10.1088/1751-8113/40/30/033",
    journal = "J. Phys. A",
    volume = "40",
    pages = "9165--9176",
    year = "2007"
}

@article{Belavin:1982,
    author = "Drinfel'd, Vladimir Gershonovich and Belavin, Aleksandr Abramovich",
    title = "Solutions of the classical Yang-Baxter equation for simple Lie algebras",
    journal = "Func. Anal. Appl.",
    volume = "16",
    issue = "3",
    pages = "159--180",
    year = "1982",
    doi = "10.1007/BF01081585"
}

@misc{Abedin:2021,
      title={Geometrization of solutions of the generalized classical Yang-Baxter
        equation and a new proof of the Belavin-Drinfeld trichotomy},
      author={Raschid Abedin},
      year={2022},
      eprint={2107.10722},
      archivePrefix={arXiv},
      primaryClass={math.AG}
}

@article{Drinfel'd:1985,
    author = "Drinfel'd, Vladimir Gershonovich",
    title = "Hopf algebras and the quantum Yang–Baxter equation",
    journal   = "Sov. Math. Dokl.",
    volume    = "32",
    pages     = "254-258",
    year      = "1985"
}

@article{Drinfel'd:1988,
    author = "Drinfel'd, Vladimir Gershonovich",
    title = "Quantum groups",
    doi = "10.1007/BF01247086",
    journal = "J. Sov. Math.",
    volume = "41",
    issue = "2",
    pages = "898--915",
    year = "1988"
}

@article{Hubbard:1963,
    ISSN = {00804630},
    URL = {http://www.jstor.org/stable/2414761},
    author = {J. Hubbard},
    journal = {Proc. R. Soc. London A},
    number = {1365},
    pages = {238--257},
    publisher = {The Royal Society},
    title = {Electron Correlations in Narrow Energy Bands},
    urldate = {2022-09-21},
    volume = {276},
    year = {1963}
}

@Book{Essler:2005aa,
     author    = "Essler, F. H. L. and Frahm, H. and Göhmann, F. and Klümper, A. and Korepin, V. E.",
     title     = "The one-dimensional Hubbard model",
     address   = "Cambridge, UK",
     year      = "2005",
     pages     = "690",
     publisher = "Cambridge University Press"
}

@Article{Shastry:1986bb,
    author = "Shastry, B. Sriram",
    title     = "Exact Integrability of the One-Dimensional Hubbard Model",
    journal   = "Phys. Rev. Lett.",
    volume    = "56",
    year      = "1986",
    pages     = "2453-2455",
    doi = "10.1103/PhysRevLett.56.2453",
}

@article{Beisert:2010jr,
    author = "Beisert, Niklas and others",
    title = "{Review of AdS/CFT Integrability: An Overview}",
    eprint = "1012.3982",
    archivePrefix = "arXiv",
    primaryClass = "hep-th",
    reportNumber = "AEI-2010-175, CERN-PH-TH-2010-306, HU-EP-10-87, HU-MATH-2010-22, KCL-MTH-10-10, UMTG-270, UUITP-41-10",
    doi = "10.1007/s11005-011-0529-2",
    journal = "Lett. Math. Phys.",
    volume = "99",
    pages = "3--32",
    year = "2012"
}

@article{Bombardelli:2016rwb,
    author = "Bombardelli, Diego and others",
    title = "{An integrability primer for the gauge-gravity correspondence: An introduction}",
    eprint = "1606.02945",
    archivePrefix = "arXiv",
    primaryClass = "hep-th",
    reportNumber = "CNRS-16-03, DCPT-16-19, DESY-16-083, DMUS-MP-16-09, HU-EP-16-13, NORDITA-2016-33, HU-MATH-16-08",
    doi = "10.1088/1751-8113/49/32/320301",
    journal = "J. Phys. A",
    volume = "49",
    number = "32",
    pages = "320301",
    year = "2016"
}

@article{Maldacena:1997re,
    author = "Maldacena, Juan Martin",
    title = "{The Large N limit of superconformal field theories and supergravity}",
    eprint = "hep-th/9711200",
    archivePrefix = "arXiv",
    reportNumber = "HUTP-97-A097, HUTP-98-A097",
    doi = "10.4310/ATMP.1998.v2.n2.a1",
    journal = "Adv. Theor. Math. Phys.",
    volume = "2",
    pages = "231--252",
    year = "1998"
}

@article{Minahan:2002ve,
    author = "Minahan, J. A. and Zarembo, K.",
    title = "{The Bethe ansatz for N = 4 superYang-Mills}",
    eprint = "hep-th/0212208",
    archivePrefix = "arXiv",
    reportNumber = "UUITP-17-02, ITEP-TH-73-02",
    doi = "10.1088/1126-6708/2003/03/013",
    journal = "JHEP",
    volume = "03",
    pages = "013",
    year = "2003"
}

@article{Beisert:2004ry,
    author = "Beisert, Niklas",
    title = "{The Dilatation operator of N = 4 super Yang-Mills theory and integrability}",
    eprint = "hep-th/0407277",
    archivePrefix = "arXiv",
    reportNumber = "AEI-2004-057",
    doi = "10.1016/j.physrep.2004.09.007",
    journal = "Phys. Rept.",
    volume = "405",
    pages = "1--202",
    year = "2004"
}

@article{Beisert:2003ys,
    author = "Beisert, Niklas",
    title = "{The su(2$/$3) dynamic spin chain}",
    eprint = "hep-th/0310252",
    archivePrefix = "arXiv",
    reportNumber = "AEI-2003-087",
    doi = "10.1016/j.nuclphysb.2003.12.032",
    journal = "Nucl. Phys. B",
    volume = "682",
    pages = "487--520",
    year = "2004"
}

@article{Beisert:2005fw,
    author = "Beisert, Niklas and Staudacher, Matthias",
    title = "{Long-range psu(2,2$/$4) Bethe Ansatze for gauge theory and strings}",
    eprint = "hep-th/0504190",
    archivePrefix = "arXiv",
    reportNumber = "AEI-2005-092, PUTP-2159",
    doi = "10.1016/j.nuclphysb.2005.06.038",
    journal = "Nucl. Phys. B",
    volume = "727",
    pages = "1--62",
    year = "2005"
}

@article{Beisert:2006qh,
    author = "Beisert, Niklas",
    title = "{The Analytic Bethe Ansatz for a Chain with Centrally Extended su(2$/$2) Symmetry}",
    eprint = "nlin/0610017",
    archivePrefix = "arXiv",
    reportNumber = "AEI-2006-074, PUTP-2211",
    doi = "10.1088/1742-5468/2007/01/P01017",
    journal = "J. Stat. Mech.",
    volume = "0701",
    pages = "P01017",
    year = "2007"
}

@article{Arutyunov:2007tc,
    author = "Arutyunov, Gleb and Frolov, Sergey",
    title = "{On String S-matrix, Bound States and TBA}",
    eprint = "0710.1568",
    archivePrefix = "arXiv",
    primaryClass = "hep-th",
    reportNumber = "ITP-UU-07-50, SPIN-07-37, TCDMATH-07-15",
    doi = "10.1088/1126-6708/2007/12/024",
    journal = "JHEP",
    volume = "12",
    pages = "024",
    year = "2007"
}

@article{Bombardelli:2009ns,
    author = "Bombardelli, Diego and Fioravanti, Davide and Tateo, Roberto",
    title = "{Thermodynamic Bethe Ansatz for planar AdS/CFT: A Proposal}",
    eprint = "0902.3930",
    archivePrefix = "arXiv",
    primaryClass = "hep-th",
    doi = "10.1088/1751-8113/42/37/375401",
    journal = "J. Phys. A",
    volume = "42",
    pages = "375401",
    year = "2009"
}

@article{Arutyunov:2009ur,
    author = "Arutyunov, Gleb and Frolov, Sergey",
    title = "{Thermodynamic Bethe Ansatz for the AdS$_5$ $\times$ S$_5$ Mirror Model}",
    eprint = "0903.0141",
    archivePrefix = "arXiv",
    primaryClass = "hep-th",
    reportNumber = "ITP-UU-09-09, SPIN-09-09, TCDMATH-09-09, HMI-09-05",
    doi = "10.1088/1126-6708/2009/05/068",
    journal = "JHEP",
    volume = "05",
    pages = "068",
    year = "2009"
}

@article{Gromov:2009bc,
    author = "Gromov, Nikolay and Kazakov, Vladimir and Kozak, Andrii and Vieira, Pedro",
    title = "{Exact Spectrum of Anomalous Dimensions of Planar N = 4 Supersymmetric Yang-Mills Theory: TBA and excited states}",
    eprint = "0902.4458",
    archivePrefix = "arXiv",
    primaryClass = "hep-th",
    reportNumber = "DESY-09-041",
    doi = "10.1007/s11005-010-0374-8",
    journal = "Lett. Math. Phys.",
    volume = "91",
    pages = "265--287",
    year = "2010"
}

@article{Gromov:2013pga,
    author = "Gromov, Nikolay and Kazakov, Vladimir and Leurent, Sebastien and Volin, Dmytro",
    title = "{Quantum Spectral Curve for Planar N = 4 Super-Yang-Mills Theory}",
    eprint = "1305.1939",
    archivePrefix = "arXiv",
    primaryClass = "hep-th",
    reportNumber = "IMPERIAL-TP-13-SL-02",
    doi = "10.1103/PhysRevLett.112.011602",
    journal = "Phys. Rev. Lett.",
    volume = "112",
    number = "1",
    pages = "011602",
    year = "2014"
}

@article{Gromov:2014caa,
    author = "Gromov, Nikolay and Kazakov, Vladimir and Leurent, Sébastien and Volin, Dmytro",
    title = "{Quantum spectral curve for arbitrary state/operator in AdS$_{5}$/CFT$_{4}$}",
    eprint = "1405.4857",
    archivePrefix = "arXiv",
    primaryClass = "hep-th",
    reportNumber = "NORDITA-2014-60-TCDMATH-14-05",
    doi = "10.1007/JHEP09(2015)187",
    journal = "JHEP",
    volume = "09",
    pages = "187",
    year = "2015"
}

@article{Gromov:2017blm,
    author = "Gromov, Nikolay",
    title = "{Introduction to the Spectrum of N = 4 SYM and the Quantum Spectral Curve}",
    eprint = "1708.03648",
    archivePrefix = "arXiv",
    primaryClass = "hep-th",
    month = "8",
    year = "2017"
}

@article{Levkovich-Maslyuk:2019awk,
    author = "Levkovich-Maslyuk, Fedor",
    title = "{A review of the AdS/CFT Quantum Spectral Curve}",
    eprint = "1911.13065",
    archivePrefix = "arXiv",
    primaryClass = "hep-th",
    doi = "10.1088/1751-8121/ab7137",
    journal = "J. Phys. A",
    volume = "53",
    number = "28",
    pages = "283004",
    year = "2020"
}

@article{Beisert:2005tm,
    author = "Beisert, Niklas",
    title = "{The SU(2$/$2) dynamic S-matrix}",
    eprint = "hep-th/0511082",
    archivePrefix = "arXiv",
    reportNumber = "PUTP-2181, NSF-KITP-05-92",
    doi = "10.4310/ATMP.2008.v12.n5.a1",
    journal = "Adv. Theor. Math. Phys.",
    volume = "12",
    pages = "945--979",
    year = "2008"
}

@article{Dorey:2006dq,
    author = "Dorey, Nick",
    title = "{Magnon Bound States and the AdS/CFT Correspondence}",
    eprint = "hep-th/0604175",
    archivePrefix = "arXiv",
    doi = "10.1088/0305-4470/39/41/S18",
    journal = "J. Phys. A",
    volume = "39",
    pages = "13119--13128",
    year = "2006"
}

@article{Chen:2006gq,
    author = "Chen, Heng-Yu and Dorey, Nick and Okamura, Keisuke",
    title = "{On the scattering of magnon boundstates}",
    eprint = "hep-th/0608047",
    archivePrefix = "arXiv",
    reportNumber = "DAMTP-06-62, UT-06-16",
    doi = "10.1088/1126-6708/2006/11/035",
    journal = "JHEP",
    volume = "11",
    pages = "035",
    year = "2006"
}

@article{deLeeuw:2008dp,
    author = "de Leeuw, Marius",
    title = "{Bound States, Yangian Symmetry and Classical r-matrix for the AdS$_5$ $\times$ S$^5$ Superstring}",
    eprint = "0804.1047",
    archivePrefix = "arXiv",
    primaryClass = "hep-th",
    reportNumber = "ITP-UU-08-18, SPIN-08-17",
    doi = "10.1088/1126-6708/2008/06/085",
    journal = "JHEP",
    volume = "06",
    pages = "085",
    year = "2008"
}

@article{Arutyunov:2008zt,
    author = "Arutyunov, Gleb and Frolov, Sergey",
    title = "{The S-matrix of String Bound States}",
    eprint = "0803.4323",
    archivePrefix = "arXiv",
    primaryClass = "hep-th",
    reportNumber = "ITP-UU-08-15, SPIN-08-14, TCDMATH-08-03",
    doi = "10.1016/j.nuclphysb.2008.06.005",
    journal = "Nucl. Phys. B",
    volume = "804",
    pages = "90--143",
    year = "2008"
}

@article{Arutyunov:2009mi,
    author = "Arutyunov, Gleb and de Leeuw, Marius and Torrielli, Alessandro",
    title = "{The Bound State S-Matrix for AdS$_5$ $\times$ S$^5$ Superstring}",
    eprint = "0902.0183",
    archivePrefix = "arXiv",
    primaryClass = "hep-th",
    reportNumber = "ITP-UU-09-06, SPIN-09-06",
    doi = "10.1016/j.nuclphysb.2009.03.024",
    journal = "Nucl. Phys. B",
    volume = "819",
    pages = "319--350",
    year = "2009"
}

@article{Chen:2006gp,
    author = "Chen, Heng-Yu and Dorey, Nick and Okamura, Keisuke",
    title = "{The Asymptotic spectrum of the N = 4 super Yang-Mills spin chain}",
    eprint = "hep-th/0610295",
    archivePrefix = "arXiv",
    reportNumber = "DAMTP-06-64, UY-06-17",
    doi = "10.1088/1126-6708/2007/03/005",
    journal = "JHEP",
    volume = "03",
    pages = "005",
    year = "2007"
}

@article{Matsumoto:2014cka,
    author = "Matsumoto, Takuya and Molev, Alexander",
    title = "{Representations of centrally extended Lie superalgebra psl(2$/$2)}",
    eprint = "1405.3420",
    archivePrefix = "arXiv",
    primaryClass = "math.RT",
    doi = "10.1063/1.4896396",
    journal = "J. Math. Phys.",
    volume = "55",
    pages = "091704",
    year = "2014"
}

@article{Berenstein:2002jq,
    author = "Berenstein, David Eliecer and Maldacena, Juan Martin and Nastase, Horatiu Stefan",
    title = "{Strings in flat space and pp waves from N = 4 superYang-Mills}",
    eprint = "hep-th/0202021",
    archivePrefix = "arXiv",
    doi = "10.1088/1126-6708/2002/04/013",
    journal = "JHEP",
    volume = "04",
    pages = "013",
    year = "2002"
}

@article{Beisert:2003yb,
    author = "Beisert, Niklas and Staudacher, Matthias",
    title = "{The N = 4 SYM integrable super spin chain}",
    eprint = "hep-th/0307042",
    archivePrefix = "arXiv",
    reportNumber = "AEI-2003-057",
    doi = "10.1016/j.nuclphysb.2003.08.015",
    journal = "Nucl. Phys. B",
    volume = "670",
    pages = "439--463",
    year = "2003"
}

@article{Beisert:2003tq,
    author = "Beisert, N. and Kristjansen, C. and Staudacher, M.",
    title = "{The Dilatation operator of conformal N = 4 superYang-Mills theory}",
    eprint = "hep-th/0303060",
    archivePrefix = "arXiv",
    reportNumber = "AEI-2003-028",
    doi = "10.1016/S0550-3213(03)00406-1",
    journal = "Nucl. Phys. B",
    volume = "664",
    pages = "131--184",
    year = "2003"
}

@article{Metsaev:1998it,
    author = "Metsaev, R. R. and Tseytlin, Arkady A.",
    title = "{Type IIB superstring action in AdS$_5$ $\times$ S$^5$ background}",
    eprint = "hep-th/9805028",
    archivePrefix = "arXiv",
    reportNumber = "FIAN-TD-98-21, IMPERIAL-TP-97-98-44, NSF-ITP-98-055",
    doi = "10.1016/S0550-3213(98)00570-7",
    journal = "Nucl. Phys. B",
    volume = "533",
    pages = "109--126",
    year = "1998"
}

@article{Brink:1976bc,
    author = "Brink, Lars and Schwarz, John H. and Scherk, Joel",
    title = "{Supersymmetric Yang-Mills Theories}",
    reportNumber = "CALT-68-574",
    doi = "10.1016/0550-3213(77)90328-5",
    journal = "Nucl. Phys. B",
    volume = "121",
    pages = "77--92",
    year = "1977"
}

@article{Arutyunov:2009ga,
    author = "Arutyunov, Gleb and Frolov, Sergey",
    title = "{Foundations of the AdS$_{5} \times S^{5}$ Superstring. Part I}",
    eprint = "0901.4937",
    archivePrefix = "arXiv",
    primaryClass = "hep-th",
    reportNumber = "ITP-UU-09-05, SPIN-09-05, TCD-MATH-09-06, HMI-09-03",
    doi = "10.1088/1751-8113/42/25/254003",
    journal = "J. Phys. A",
    volume = "42",
    pages = "254003",
    year = "2009"
}

@article{Bena:2003wd,
    author = "Bena, Iosif and Polchinski, Joseph and Roiban, Radu",
    title = "{Hidden symmetries of the AdS$_5$ $\times$ S$^5$ superstring}",
    eprint = "hep-th/0305116",
    archivePrefix = "arXiv",
    reportNumber = "NSF-KITP-03-34, UCLA-03-TEP-14",
    doi = "10.1103/PhysRevD.69.046002",
    journal = "Phys. Rev. D",
    volume = "69",
    pages = "046002",
    year = "2004"
}

@article{Frolov:2006cc,
    author = "Frolov, Sergey and Plefka, Jan and Zamaklar, Marija",
    title = "{The AdS$_5$ $\times$ S$^5$ superstring in light-cone gauge and its Bethe equations}",
    eprint = "hep-th/0603008",
    archivePrefix = "arXiv",
    reportNumber = "AEI-2006-011, HU-EP-06-08",
    doi = "10.1088/0305-4470/39/41/S15",
    journal = "J. Phys. A",
    volume = "39",
    pages = "13037--13082",
    year = "2006"
}

@article{Arutyunov:2004yx,
    author = "Arutyunov, Gleb and Frolov, Sergey",
    title = "{Integrable Hamiltonian for classical strings on AdS$_5$ $\times$ S$^5$}",
    eprint = "hep-th/0411089",
    archivePrefix = "arXiv",
    reportNumber = "AEI-2004-105",
    doi = "10.1088/1126-6708/2005/02/059",
    journal = "JHEP",
    volume = "02",
    pages = "059",
    year = "2005"
}

@article{Arutyunov:2006ak,
    author = "Arutyunov, Gleb and Frolov, Sergey and Plefka, Jan and Zamaklar, Marija",
    title = "{The Off-shell Symmetry Algebra of the Light-cone AdS$_5$ $\times$ S$^5$ Superstring}",
    eprint = "hep-th/0609157",
    archivePrefix = "arXiv",
    reportNumber = "AEI-2006-071, HU-EP-06-31, ITP-UU-06-39, SPIN-06-33, TCDMATH-06-13",
    doi = "10.1088/1751-8113/40/13/018",
    journal = "J. Phys. A",
    volume = "40",
    pages = "3583--3606",
    year = "2007"
}

@article{Arutyunov:2004vx,
    author = "Arutyunov, Gleb and Frolov, Sergey and Staudacher, Matthias",
    title = "{Bethe ansatz for quantum strings}",
    eprint = "hep-th/0406256",
    archivePrefix = "arXiv",
    reportNumber = "AEI-2004-046",
    doi = "10.1088/1126-6708/2004/10/016",
    journal = "JHEP",
    volume = "10",
    pages = "016",
    year = "2004"
}

@article{Klose:2006zd,
    author = "Klose, Thomas and McLoughlin, Tristan and Roiban, Radu and Zarembo, Konstantin",
    title = "{Worldsheet scattering in AdS$_5$ $\times$ S$^5$}",
    eprint = "hep-th/0611169",
    archivePrefix = "arXiv",
    reportNumber = "ITEP-TH-61-06, UUITP-15-06",
    doi = "10.1088/1126-6708/2007/03/094",
    journal = "JHEP",
    volume = "03",
    pages = "094",
    year = "2007"
}

@article{Roiban:2007jf,
    author = "Roiban, R. and Tirziu, A. and Tseytlin, Arkady A.",
    title = "{Two-loop world-sheet corrections in AdS$_5$ $\times$ S$^5$ superstring}",
    eprint = "0704.3638",
    archivePrefix = "arXiv",
    primaryClass = "hep-th",
    reportNumber = "IMPERIAL-TP-AT-2007-1",
    doi = "10.1088/1126-6708/2007/07/056",
    journal = "JHEP",
    volume = "07",
    pages = "056",
    year = "2007"
}

@article{Klose:2007rz,
    author = "Klose, T. and McLoughlin, T. and Minahan, J. A. and Zarembo, K.",
    title = "{World-sheet scattering in AdS$_5$ $\times$ S$^5$ at two loops}",
    eprint = "0704.3891",
    archivePrefix = "arXiv",
    primaryClass = "hep-th",
    reportNumber = "ITEP-TH-17-7, UUITP-07-07",
    doi = "10.1088/1126-6708/2007/08/051",
    journal = "JHEP",
    volume = "08",
    pages = "051",
    year = "2007"
}

@article{Janik:2006dc,
    author = "Janik, Romuald A.",
    title = "{The AdS$_5$ $\times$ S$^5$ superstring worldsheet S-matrix and crossing symmetry}",
    eprint = "hep-th/0603038",
    archivePrefix = "arXiv",
    doi = "10.1103/PhysRevD.73.086006",
    journal = "Phys. Rev. D",
    volume = "73",
    pages = "086006",
    year = "2006"
}

@article{Hernandez:2006tk,
    author = "Hernández, Rafael and López, Esperanza",
    title = "{Quantum corrections to the string Bethe ansatz}",
    eprint = "hep-th/0603204",
    archivePrefix = "arXiv",
    reportNumber = "CERN-PH-TH-2006-048, IFT-UAM-CSIC-06-14",
    doi = "10.1088/1126-6708/2006/07/004",
    journal = "JHEP",
    volume = "07",
    pages = "004",
    year = "2006"
}

@article{Arutyunov:2006iu,
    author = "Arutyunov, G. and Frolov, S.",
    title = "{On AdS$_5$ $\times$ S$^5$ String S-matrix}",
    eprint = "hep-th/0604043",
    archivePrefix = "arXiv",
    reportNumber = "ITP-UU-06-15, SPIN-06-13",
    doi = "10.1016/j.physletb.2006.06.064",
    journal = "Phys. Lett. B",
    volume = "639",
    pages = "378--382",
    year = "2006"
}

@article{Beisert:2006ib,
    author = "Beisert, Niklas and Hernández, Rafael and López, Esperanza",
    title = "{A Crossing-symmetric phase for AdS$_5$ $\times$ S$^5$ strings}",
    eprint = "hep-th/0609044",
    archivePrefix = "arXiv",
    reportNumber = "AEI-2006-068, CERN-PH-TH-2006-176, IFT-UAM-CSIC-06-44, PUTP-2208",
    doi = "10.1088/1126-6708/2006/11/070",
    journal = "JHEP",
    volume = "11",
    pages = "070",
    year = "2006"
}

@article{Beisert:2006ez,
    author = "Beisert, Niklas and Eden, Burkhard and Staudacher, Matthias",
    title = "{Transcendentality and Crossing}",
    eprint = "hep-th/0610251",
    archivePrefix = "arXiv",
    reportNumber = "AEI-2006-079, ITP-UU-06-44, SPIN-06-34",
    doi = "10.1088/1742-5468/2007/01/P01021",
    journal = "J. Stat. Mech.",
    volume = "0701",
    pages = "P01021",
    year = "2007"
}

@article{Spill:2008tp,
    author = "Spill, Fabian and Torrielli, Alessandro",
    title = "{On Drinfeld's second realization of the AdS/CFT su(2$/$2) Yangian}",
    eprint = "0803.3194",
    archivePrefix = "arXiv",
    primaryClass = "hep-th",
    reportNumber = "MIT-CTP-3935, IMPERIAL-TP-08-FS-01, HU-EP-08-05",
    doi = "10.1016/j.geomphys.2009.01.001",
    journal = "J. Geom. Phys.",
    volume = "59",
    pages = "489--502",
    year = "2009"
}

@article{Beisert:2014hya,
    author = "Beisert, Niklas and de Leeuw, Marius",
    title = "{The RTT realization for the deformed gl(2$/$2) Yangian}",
    eprint = "1401.7691",
    archivePrefix = "arXiv",
    primaryClass = "math-ph",
    doi = "10.1088/1751-8113/47/30/305201",
    journal = "J. Phys. A",
    volume = "47",
    pages = "305201",
    year = "2014"
}

@article{Beisert:2016qei,
    author = "Beisert, Niklas and de Leeuw, Marius and Hecht, Reimar",
    title = "{Maximally extended sl(2$/$2) as a quantum double}",
    eprint = "1602.04988",
    archivePrefix = "arXiv",
    primaryClass = "math-ph",
    doi = "10.1088/1751-8113/49/43/434005",
    journal = "J. Phys. A",
    volume = "49",
    number = "43",
    pages = "434005",
    year = "2016"
}

@article{Matsumoto:2022nrk,
    author = "Matsumoto, Takuya",
    title = "{Drinfeld realization of the centrally extended psl(2$/$2) Yangian algebra with the manifest coproducts}",
    eprint = "2208.11889",
    archivePrefix = "arXiv",
    primaryClass = "math.QA",
    doi = "10.1063/5.0124333",
    journal = "J. Math. Phys.",
    volume = "64",
    number = "4",
    pages = "041704",
    year = "2023"
}

@article{Gomez:2006va,
    author = "Gómez, César and Hernández, Rafael",
    title = "{The Magnon kinematics of the AdS/CFT correspondence}",
    eprint = "hep-th/0608029",
    archivePrefix = "arXiv",
    reportNumber = "CERN-PH-TH-2006-140, IFT-UAM-CSIC-06-37",
    doi = "10.1088/1126-6708/2006/11/021",
    journal = "JHEP",
    volume = "11",
    pages = "021",
    year = "2006"
}

@article{Plefka:2006ze,
    author = "Plefka, Jan and Spill, Fabian and Torrielli, Alessandro",
    title = "{On the Hopf algebra structure of the AdS/CFT S-matrix}",
    eprint = "hep-th/0608038",
    archivePrefix = "arXiv",
    reportNumber = "HU-EP-06-22",
    doi = "10.1103/PhysRevD.74.066008",
    journal = "Phys. Rev. D",
    volume = "74",
    pages = "066008",
    year = "2006"
}

@article{Beisert:2006fmy,
    author = "Beisert, Niklas",
    editor = "Faddeev, L. and Henneaux, M. and Kashaev, R. and Volkov, A. and Lambert, F.",
    title = "{The S-matrix of AdS/CFT and Yangian symmetry}",
    eprint = "0704.0400",
    archivePrefix = "arXiv",
    primaryClass = "nlin.SI",
    reportNumber = "AEI-2007-019",
    doi = "10.22323/1.038.0002",
    journal = "PoS",
    volume = "SOLVAY",
    pages = "002",
    year = "2006"
}

@article{Drummond:2009fd,
    author = "Drummond, James M. and Henn, Johannes M. and Plefka, Jan",
    editor = "Liu, Feng and Xiao, Zhigang and Zhuang, Pengfei",
    title = "{Yangian symmetry of scattering amplitudes in N = 4 super Yang-Mills theory}",
    eprint = "0902.2987",
    archivePrefix = "arXiv",
    primaryClass = "hep-th",
    reportNumber = "HU-EP-09-06, LAPTH-1308-09",
    doi = "10.1088/1126-6708/2009/05/046",
    journal = "JHEP",
    volume = "05",
    pages = "046",
    year = "2009"
}

@article{Beisert:2017pnr,
    author = "Beisert, Niklas and Garus, Aleksander and Rosso, Matteo",
    title = "{Yangian Symmetry and Integrability of Planar N = 4 Supersymmetric Yang-Mills Theory}",
    eprint = "1701.09162",
    archivePrefix = "arXiv",
    primaryClass = "hep-th",
    reportNumber = "HU-EP-17-03",
    doi = "10.1103/PhysRevLett.118.141603",
    journal = "Phys. Rev. Lett.",
    volume = "118",
    number = "14",
    pages = "141603",
    year = "2017"
}

@article{Matsumoto:2007rh,
    author = "Matsumoto, Takuya and Moriyama, Sanefumi and Torrielli, Alessandro",
    title = "{A Secret Symmetry of the AdS/CFT S-matrix}",
    eprint = "0708.1285",
    archivePrefix = "arXiv",
    primaryClass = "hep-th",
    reportNumber = "MIT-CTP-3853",
    doi = "10.1088/1126-6708/2007/09/099",
    journal = "JHEP",
    volume = "09",
    pages = "099",
    year = "2007"
}

@article{deLeeuw:2012jf,
    author = "de Leeuw, Marius and Matsumoto, Takuya and Moriyama, Sanefumi and Regelskis, Vidas and Torrielli, Alessandro",
    title = "{Secret Symmetries in AdS/CFT}",
    eprint = "1204.2366",
    archivePrefix = "arXiv",
    primaryClass = "hep-th",
    reportNumber = "DMUS-MP-12-04, NORDITA-2012-26",
    doi = "10.1088/0031-8949/86/02/028502",
    journal = "Phys. Scripta",
    volume = "02",
    pages = "028502",
    year = "2012"
}

@article{Moriyama:2007jt,
    author = "Moriyama, Sanefumi and Torrielli, Alessandro",
    title = "{A Yangian double for the AdS/CFT classical r-matrix}",
    eprint = "0706.0884",
    archivePrefix = "arXiv",
    primaryClass = "hep-th",
    reportNumber = "MIT-CTP-3843",
    doi = "10.1088/1126-6708/2007/06/083",
    journal = "JHEP",
    volume = "06",
    pages = "083",
    year = "2007"
}

@article{Torrielli:2007mc,
    author = "Torrielli, Alessandro",
    title = "{Classical r-matrix of the su(2$/$2) SYM spin-chain}",
    eprint = "hep-th/0701281",
    archivePrefix = "arXiv",
    reportNumber = "MIT-CTP-3809",
    doi = "10.1103/PhysRevD.75.105020",
    journal = "Phys. Rev. D",
    volume = "75",
    pages = "105020",
    year = "2007"
}

@article{Beisert:2017xqx,
    author = "Beisert, Niklas and Hecht, Reimar and Hoare, Ben",
    title = "{Maximally extended sl(2$/$2), q-deformed d(2,1;$\epsilon$) and 3D kappa-Poincaré}",
    eprint = "1704.05093",
    archivePrefix = "arXiv",
    primaryClass = "math-ph",
    doi = "10.1088/1751-8121/aa7a2f",
    journal = "J. Phys. A",
    volume = "50",
    number = "31",
    pages = "314003",
    year = "2017"
}

@article{Matsumoto:2008ww,
    author = "Matsumoto, Takuya and Moriyama, Sanefumi",
    title = "{An Exceptional Algebraic Origin of the AdS/CFT Yangian Symmetry}",
    eprint = "0803.1212",
    archivePrefix = "arXiv",
    primaryClass = "hep-th",
    doi = "10.1088/1126-6708/2008/04/022",
    journal = "JHEP",
    volume = "04",
    pages = "022",
    year = "2008"
}

@article{Delduc:2013qra,
    author = "Delduc, Francois and Magro, Marc and Vicedo, Benoit",
    title = "{An integrable deformation of the AdS$_5$ $\times$ S$^5$ superstring action}",
    eprint = "1309.5850",
    archivePrefix = "arXiv",
    primaryClass = "hep-th",
    doi = "10.1103/PhysRevLett.112.051601",
    journal = "Phys. Rev. Lett.",
    volume = "112",
    number = "5",
    pages = "051601",
    year = "2014"
}

@article{Arutyunov:2013ega,
    author = "Arutyunov, Gleb and Borsato, Riccardo and Frolov, Sergey",
    title = "{S-matrix for strings on $\eta$-deformed AdS$_5$ $\times$ S$^5$}",
    eprint = "1312.3542",
    archivePrefix = "arXiv",
    primaryClass = "hep-th",
    reportNumber = "ITP-UU-13-31, SPIN-13-23, HU-MATHEMATIK-2013-24, TCD-MATH-13-16",
    doi = "10.1007/JHEP04(2014)002",
    journal = "JHEP",
    volume = "04",
    pages = "002",
    year = "2014"
}

@article{Delduc:2014kha,
    author = "Delduc, Francois and Magro, Marc and Vicedo, Benoit",
    title = "{Derivation of the action and symmetries of the q-deformed AdS$_5$ $\times$ S$^5$ superstring}",
    eprint = "1406.6286",
    archivePrefix = "arXiv",
    primaryClass = "hep-th",
    doi = "10.1007/JHEP10(2014)132",
    journal = "JHEP",
    volume = "10",
    pages = "132",
    year = "2014"
}

@article{Nahm:1977tg,
    author = "Nahm, W.",
    title = "{Supersymmetries and their Representations}",
    reportNumber = "CERN-TH-2341",
    doi = "10.1016/0550-3213(78)90218-3",
    journal = "Nucl. Phys. B",
    volume = "135",
    pages = "149",
    year = "1978"
}

@article{VanderJeugt:1985,
    author = {Van der Jeugt,J. },
    title = {Irreducible representations of the exceptional Lie superalgebras D(2,1;$\alpha$)},
    journal = {J. Math. Phys.},
    volume = {26},
    number = {5},
    pages = {913-924},
    year = {1985},
    doi = {10.1063/1.526547}
}

@article{deLeeuw:2011fr,
    author = "de Leeuw, Marius and Regelskis, Vidas and Torrielli, Alessandro",
    title = "{The Quantum Affine Origin of the AdS/CFT Secret Symmetry}",
    eprint = "1112.4989",
    archivePrefix = "arXiv",
    primaryClass = "hep-th",
    reportNumber = "DMUS-MP-11-02",
    doi = "10.1088/1751-8113/45/17/175202",
    journal = "J. Phys. A",
    volume = "45",
    pages = "175202",
    year = "2012"
}

@article{Beisert:2008tw,
    author = "Beisert, Niklas and Koroteev, Peter",
    title = "{Quantum Deformations of the One-Dimensional Hubbard Model}",
    eprint = "0802.0777",
    archivePrefix = "arXiv",
    primaryClass = "hep-th",
    reportNumber = "AEI-2008-003, ITEP-TH-06-08",
    doi = "10.1088/1751-8113/41/25/255204",
    journal = "J. Phys. A",
    volume = "41",
    pages = "255204",
    year = "2008"
}

@article{Beisert:2011wq,
    author = "Beisert, Niklas and Galleas, Wellington and Matsumoto, Takuya",
    title = "{A Quantum Affine Algebra for the Deformed Hubbard Chain}",
    eprint = "1102.5700",
    archivePrefix = "arXiv",
    primaryClass = "math-ph",
    reportNumber = "AEI-2011-005",
    doi = "10.1088/1751-8113/45/36/365206",
    journal = "J. Phys. A",
    volume = "45",
    pages = "365206",
    year = "2012"
}

@article{Seibold:2020ywq,
    author = "Seibold, Fiona K. and Van Tongeren, Stijn J. and Zimmermann, Yannik",
    title = "{The twisted story of worldsheet scattering in $\eta$-deformed AdS$_5$ $\times$ S$^5$}",
    eprint = "2007.09136",
    archivePrefix = "arXiv",
    primaryClass = "hep-th",
    doi = "10.1007/JHEP12(2020)043",
    journal = "JHEP",
    volume = "12",
    pages = "043",
    year = "2020"
}

@article{Beisert:2014qba,
    author = "Beisert, Niklas and Broedel, Johannes and Rosso, Matteo",
    title = "{On Yangian-invariant regularization of deformed on-shell diagrams in N = 4 super-Yang-Mills theory}",
    eprint = "1401.7274",
    archivePrefix = "arXiv",
    primaryClass = "hep-th",
    doi = "10.1088/1751-8113/47/36/365402",
    journal = "J. Phys. A",
    volume = "47",
    pages = "365402",
    year = "2014"
}

@article{Rej:2005qt,
    author = "Rej, Adam and Serban, Didina and Staudacher, Matthias",
    title = "{Planar N = 4 gauge theory and the Hubbard model}",
    eprint = "hep-th/0512077",
    archivePrefix = "arXiv",
    reportNumber = "AEI-2005-164, SPHT-T05-190, NSF-KITP-05-84",
    doi = "10.1088/1126-6708/2006/03/018",
    journal = "JHEP",
    volume = "03",
    pages = "018",
    year = "2006"
}

@article {Tolstoy:1992,
    author = "Tolstoy, V. N. and Khoroshkin, S. M.",
    title = {Universal {$R$}-matrix for quantized nontwisted affine {L}ie algebras},
    journal = {Funktsional'nyi Analiz i ego Prilozheniya},
    volume = {26},
    year = {1992},
    number = {1},
    pages = {85--88},
    issn = {0374-1990,2305-2899},
    doi = {10.1007/BF01077085},
}

@article{Khoroshkin:1991,
author = {S. M. Khoroshkin and V. N. Tolstoy},
title = {Universal $R$-matrix for quantized (super)algebras},
volume = {141},
journal = {Communications in Mathematical Physics},
number = {3},
publisher = {Springer},
pages = {599 -- 617},
year = {1991},
}

@article{Heckenberger:2007ry,
    author = "Heckenberger, Istvan and Spill, Fabian and Torrielli, Alessandro and Yamane, Hiroyuki",
    title = "{Drinfeld second realization of the quantum affine superalgebras of D(1)(2,1;x) via the Weyl groupoid}",
    eprint = "0705.1071",
    archivePrefix = "arXiv",
    primaryClass = "math.QA",
    reportNumber = "MIT-CTP-3835, HU-EP-07-15",
    journal = "Publ. Res. Inst. Math. Sci. Kyoto B",
    volume = "8",
    pages = "171",
    year = "2008"
}

@article{Beisert:2018zxs,
    author = "Beisert, Niklas and Garus, Aleksander and Rosso, Matteo",
    title = "{Yangian Symmetry for the Action of Planar N = 4 Super Yang-Mills and N = 6 Super Chern-Simons Theories}",
    eprint = "1803.06310",
    archivePrefix = "arXiv",
    primaryClass = "hep-th",
    reportNumber = "HU-EP-18/08, HU-EP-18-08",
    doi = "10.1103/PhysRevD.98.046006",
    journal = "Phys. Rev. D",
    volume = "98",
    number = "4",
    pages = "046006",
    year = "2018"
}

@inproceedings{Dolan:2004ps,
    author = "Dolan, Louise and Nappi, Chiara R. and Witten, Edward",
    title = "{Yangian symmetry in D = 4 superconformal Yang-Mills theory}",
    booktitle = "{3rd International Symposium on Quantum Theory and Symmetries}",
    eprint = "hep-th/0401243",
    archivePrefix = "arXiv",
    doi = "10.1142/9789812702340_0036",
    pages = "300--315",
    month = "1",
    year = "2004"
}

@book{Schlichenmaier+2014,
url = {https://doi.org/10.1515/9783110279641},
title = {Krichever--Novikov Type Algebras},
subtitle = {Theory and Applications},
author = {Martin Schlichenmaier},
publisher = {De Gruyter},
address = {Berlin, Boston},
doi = {doi:10.1515/9783110279641},
isbn = {9783110279641},
year = {2014},
lastchecked = {2025-08-12}
}

@ARTICLE{1995q.alg.....6005E,
       author = {Etingof, Pavel and Kazhdan, David},
        title = "{Quantization of Lie bialgebras, I}",
      journal = {Selecta Mathematica},
         year = 1996,
       volume = 2,
        pages = {1-41},
          doi = {10.1007/BF01587938},
archivePrefix = {arXiv},
       eprint = {q-alg/9506005},
 primaryClass = {math.QA}
}

@ARTICLE{1997q.alg.....1038E,
       author = {Etingof, Pavel and Kazhdan, David},
        title = "{Quantization of Lie bialgebras, II}",
      journal = {Selecta Mathematica},
         year = 1998,
       volume = 4,
        pages = {213},
          doi = {10.1007/s000290050030},
archivePrefix = {arXiv},
       eprint = {q-alg/9701038},
 primaryClass = {math.QA}
}

@ARTICLE{1996q.alg....10030E,
       author = {{Etingof}, Pavel and {Kazhdan}, David},
        title = "{Quantization of Lie bialgebras, III}",
      journal = {Selecta Mathematica},
         year = 1998,
       volume = 4,
        pages = {233},
          doi = {10.1007/s000290050031},
archivePrefix = {arXiv},
       eprint = {q-alg/9610030},
 primaryClass = {math.QA}
}

@article{krichever1987algebras,
  title={Algebras of Virasoro type, Riemann surfaces and structures of the theory of solitons},
  author={Krichever, Igor Moiseevich and Novikov, Sergei Petrovich},
  journal={Functional Analysis and Its Applications},
  volume={21},
  number={2},
  pages={126--142},
  year={1987},
  doi = {10.1007/BF01077803},
}

@article{Cherednik:1984vvp,
    author = "Cherednik, I. V.",
    title = "{Factorizing Particles on a Half Line and Root Systems}",
    doi = "10.1007/BF01038545",
    journal = "Theor. Math. Phys.",
    volume = "61",
    pages = "977--983",
    year = "1984"
}

@article{Sklyanin:1988yz,
    author = "Sklyanin, E. K.",
    title = "{Boundary Conditions for Integrable Quantum Systems}",
    reportNumber = "E-11-86",
    doi = "10.1088/0305-4470/21/10/015",
    journal = "J. Phys. A",
    volume = "21",
    pages = "2375--2389",
    year = "1988"
}

@article{Dorey:2007xn,
    author = "Dorey, Nick and Hofman, Diego M. and Maldacena, Juan Martin",
    title = "{On the Singularities of the Magnon S-matrix}",
    eprint = "hep-th/0703104",
    archivePrefix = "arXiv",
    doi = "10.1103/PhysRevD.76.025011",
    journal = "Phys. Rev. D",
    volume = "76",
    pages = "025011",
    year = "2007"
}

\end{bibtex}
\fi

\bibliographystyle{nb}
\bibliography{\jobname}

\begin{thebibliography}{10}
\providecommand{\href}[2]{#2}
\providecommand{\arxivref}[2]{\href{http://arxiv.org/abs/#1}{#2}}
\providecommand{\doiref}[2]{\href{http://dx.doi.org/#1}{#2}}
\providecommand{\nbbstauthor}[1]{#1}
\providecommand{\nbbstjournal}[1]{\textsf{#1}}
\providecommand{\nbbsttitle}[1]{\textit{#1}}
\providecommand{\nbbsturl}[1]{\texttt{#1}}
\providecommand{\nbbsteprint}[1]{\texttt{#1}}
\providecommand{\nbbststyle}{\raggedright\small\parskip0pt}
\nbbststyle

\bibitem{Beisert:2006qh}
\nbbstauthor{N.~Beisert},
\nbbsttitle{``{The Analytic Bethe Ansatz for a Chain with Centrally Extended
  su(2$/$2) Symmetry}''},
\nbbstjournal{\doiref{10.1088/1742-5468/2007/01/P01017}{J.~Stat.~Mech.~0701,~P01017~(2007)}},
\nbbsteprint{\arxivref{nlin/0610017}{nlin/0610017}}.

\bibitem{Essler:2005aa}
\nbbstauthor{F.~H.~L.~Essler, H.~Frahm, F.~Göhmann, A.~Klümper and
  V.~E.~Korepin},
\nbbsttitle{``The one-dimensional Hubbard model''},
Cambridge University Press (2005),
Cambridge, UK.

\bibitem{Beisert:2010jr}
\nbbstauthor{N.~Beisert et~al.},
\nbbsttitle{``{Review of AdS/CFT Integrability: An Overview}''},
\nbbstjournal{\doiref{10.1007/s11005-011-0529-2}{Lett.~Math.~Phys.~99,~3~(2012)}},
\nbbsteprint{\arxivref{1012.3982}{arxiv:1012.3982}}.

\bibitem{Drinfel'd:1985}
\nbbstauthor{V.~G.~Drinfel'd},
\nbbsttitle{``Hopf algebras and the quantum Yang–Baxter equation''},
\nbbstjournal{Sov.~Math.~Dokl.~32,~254~(1985)}.

\bibitem{Drinfel'd:1988}
\nbbstauthor{V.~G.~Drinfel'd},
\nbbsttitle{``Quantum groups''},
\nbbstjournal{\doiref{10.1007/BF01247086}{J.~Sov.~Math.~41,~898~(1988)}}.

\bibitem{Shastry:1986bb}
\nbbstauthor{B.~S.~Shastry},
\nbbsttitle{``Exact Integrability of the One-Dimensional Hubbard Model''},
\nbbstjournal{\doiref{10.1103/PhysRevLett.56.2453}{Phys.~Rev.~Lett.~56,~2453~(1986)}}.

\bibitem{Beisert:2005tm}
\nbbstauthor{N.~Beisert},
\nbbsttitle{``{The SU(2$/$2) dynamic S-matrix}''},
\nbbstjournal{\doiref{10.4310/ATMP.2008.v12.n5.a1}{Adv.~Theor.~Math.~Phys.~12,~945~(2008)}},
\nbbsteprint{\arxivref{hep-th/0511082}{hep-th/0511082}}.

\bibitem{Janik:2006dc}
\nbbstauthor{R.~A.~Janik},
\nbbsttitle{``{The AdS$_5$ $\times$ S$^5$ superstring worldsheet S-matrix and
  crossing symmetry}''},
\nbbstjournal{\doiref{10.1103/PhysRevD.73.086006}{Phys.~Rev.~D~73,~086006~(2006)}},
\nbbsteprint{\arxivref{hep-th/0603038}{hep-th/0603038}}.

\bibitem{Beisert:2006ez}
\nbbstauthor{N.~Beisert, B.~Eden and M.~Staudacher},
\nbbsttitle{``{Transcendentality and Crossing}''},
\nbbstjournal{\doiref{10.1088/1742-5468/2007/01/P01021}{J.~Stat.~Mech.~0701,~P01021~(2007)}},
\nbbsteprint{\arxivref{hep-th/0610251}{hep-th/0610251}}.

\bibitem{Beisert:2006ib}
\nbbstauthor{N.~Beisert, R.~Hernández and E.~López},
\nbbsttitle{``{A Crossing-symmetric phase for AdS$_5$ $\times$ S$^5$
  strings}''},
\nbbstjournal{\doiref{10.1088/1126-6708/2006/11/070}{JHEP~0611,~070~(2006)}},
\nbbsteprint{\arxivref{hep-th/0609044}{hep-th/0609044}}.

\bibitem{Dorey:2007xn}
\nbbstauthor{N.~Dorey, D.~M.~Hofman and J.~M.~Maldacena},
\nbbsttitle{``{On the Singularities of the Magnon S-matrix}''},
\nbbstjournal{\doiref{10.1103/PhysRevD.76.025011}{Phys.~Rev.~D~76,~025011~(2007)}},
\nbbsteprint{\arxivref{hep-th/0703104}{hep-th/0703104}}.

\bibitem{Klose:2006zd}
\nbbstauthor{T.~Klose, T.~McLoughlin, R.~Roiban and K.~Zarembo},
\nbbsttitle{``{Worldsheet scattering in AdS$_5$ $\times$ S$^5$}''},
\nbbstjournal{\doiref{10.1088/1126-6708/2007/03/094}{JHEP~0703,~094~(2007)}},
\nbbsteprint{\arxivref{hep-th/0611169}{hep-th/0611169}}.

\bibitem{Torrielli:2007mc}
\nbbstauthor{A.~Torrielli},
\nbbsttitle{``{Classical r-matrix of the su(2$/$2) SYM spin-chain}''},
\nbbstjournal{\doiref{10.1103/PhysRevD.75.105020}{Phys.~Rev.~D~75,~105020~(2007)}},
\nbbsteprint{\arxivref{hep-th/0701281}{hep-th/0701281}}.

\bibitem{Moriyama:2007jt}
\nbbstauthor{S.~Moriyama and A.~Torrielli},
\nbbsttitle{``{A Yangian double for the AdS/CFT classical r-matrix}''},
\nbbstjournal{\doiref{10.1088/1126-6708/2007/06/083}{JHEP~0706,~083~(2007)}},
\nbbsteprint{\arxivref{0706.0884}{arxiv:0706.0884}}.

\bibitem{Beisert:2007ty}
\nbbstauthor{N.~Beisert and F.~Spill},
\nbbsttitle{``The Classical r-matrix of AdS/CFT and its Lie Bialgebra
  Structure''},
\nbbstjournal{\doiref{10.1007/s00220-008-0578-2}{Commun.~Math.~Phys.~285,~537~(2009)}},
\nbbsteprint{\arxivref{0708.1762}{arxiv:0708.1762}}.

\bibitem{Beisert:2022vnc}
\nbbstauthor{N.~Beisert and E.~Im},
\nbbsttitle{``{Classical Lie Bialgebras for AdS/CFT Integrability by
  Contraction and Reduction}''},
\nbbstjournal{\doiref{10.21468/SciPostPhys.14.6.157}{SciPost~Phys.~14,~157~(2023)}},
\nbbsteprint{\arxivref{2210.11150}{arxiv:2210.11150}}.

\bibitem{1995q.alg.....6005E}
\nbbstauthor{P.~Etingof and D.~Kazhdan},
\nbbsttitle{``{Quantization of Lie bialgebras, I}''},
\nbbstjournal{\doiref{10.1007/BF01587938}{Selecta~Mathematica~2,~1~(1996)}},
\nbbsteprint{\arxivref{q-alg/9506005}{q-alg/9506005}}.

\bibitem{1997q.alg.....1038E}
\nbbstauthor{P.~Etingof and D.~Kazhdan},
\nbbsttitle{``{Quantization of Lie bialgebras, II}''},
\nbbstjournal{\doiref{10.1007/s000290050030}{Selecta~Mathematica~4,~213~(1998)}},
\nbbsteprint{\arxivref{q-alg/9701038}{q-alg/9701038}}.

\bibitem{1996q.alg....10030E}
\nbbstauthor{P.~{Etingof} and D.~{Kazhdan}},
\nbbsttitle{``{Quantization of Lie bialgebras, III}''},
\nbbstjournal{\doiref{10.1007/s000290050031}{Selecta~Mathematica~4,~233~(1998)}},
\nbbsteprint{\arxivref{q-alg/9610030}{q-alg/9610030}}.

\bibitem{Beisert:2010kk}
\nbbstauthor{N.~Beisert},
\nbbsttitle{``The Classical Trigonometric r-Matrix for the Quantum-Deformed
  Hubbard Chain''},
\nbbstjournal{\doiref{10.1088/1751-8113/44/26/265202}{J.~Phys.~A~44,~265202~(2011)}},
\nbbsteprint{\arxivref{1002.1097}{arxiv:1002.1097}}.

\bibitem{Beisert:2024rke}
\nbbstauthor{N.~Beisert and E.~Im},
\nbbsttitle{``{Affine Classical Lie Bialgebras for AdS/CFT Integrability}''},
\nbbsteprint{\arxivref{2401.10327}{arxiv:2401.10327}}.

\bibitem{Young:2007wd}
\nbbstauthor{C.~A.~S.~Young},
\nbbsttitle{``{q-deformed supersymmetry and dynamic magnon representations}''},
\nbbstjournal{\doiref{10.1088/1751-8113/40/30/033}{J.~Phys.~A~40,~9165~(2007)}},
\nbbsteprint{\arxivref{0704.2069}{arxiv:0704.2069}}.

\bibitem{Borsato:2017lpf}
\nbbstauthor{R.~Borsato and A.~Torrielli},
\nbbsttitle{``{$q$-Poincaré supersymmetry in AdS$_5$ / CFT$_4$}''},
\nbbstjournal{\doiref{10.1016/j.nuclphysb.2018.01.017}{Nucl.~Phys.~B~928,~321~(2018)}},
\nbbsteprint{\arxivref{1706.10265}{arxiv:1706.10265}}.

\bibitem{krichever1987algebras}
\nbbstauthor{I.~M.~Krichever and S.~P.~Novikov},
\nbbsttitle{``Algebras of Virasoro type, Riemann surfaces and structures of the
  theory of solitons''},
\nbbstjournal{\doiref{10.1007/BF01077803}{Functional~Analysis~and~Its~Applications~21,~126~(1987)}}.

\bibitem{Schlichenmaier+2014}
\nbbstauthor{M.~Schlichenmaier},
\nbbsttitle{``Krichever--Novikov Type Algebras''},
De Gruyter (2014),
Berlin, Boston.

\bibitem{Nahm:1977tg}
\nbbstauthor{W.~Nahm},
\nbbsttitle{``{Supersymmetries and their Representations}''},
\nbbstjournal{\doiref{10.1016/0550-3213(78)90218-3}{Nucl.~Phys.~B~135,~149~(1978)}}.

\bibitem{Bremner:1994}
\nbbstauthor{M.~Bremner},
\nbbsttitle{``Generalized Affine Kac-Moody Lie Algebras Over Localizations of
  the Polynomial Ring in One Variable''},
\nbbstjournal{\doiref{10.4153/CMB-1994-004-8}{Can.~Math.~Bull.~37,~21–28~(1994)}}.

\bibitem{Cherednik:1984vvp}
\nbbstauthor{I.~V.~Cherednik},
\nbbsttitle{``{Factorizing Particles on a Half Line and Root Systems}''},
\nbbstjournal{\doiref{10.1007/BF01038545}{Theor.~Math.~Phys.~61,~977~(1984)}}.

\bibitem{Sklyanin:1988yz}
\nbbstauthor{E.~K.~Sklyanin},
\nbbsttitle{``{Boundary Conditions for Integrable Quantum Systems}''},
\nbbstjournal{\doiref{10.1088/0305-4470/21/10/015}{J.~Phys.~A~21,~2375~(1988)}}.

\end{thebibliography}

\end{document}